\documentclass[twocolumn,prd,superscriptaddress,preprintnumbers,floatfix]{revtex4}[10pt]
\pdfoutput=1
\usepackage{amsmath,amssymb,graphicx} %use drftcite in draftmode
\usepackage{epsf,verbatim}
\usepackage{pstricks}
\usepackage{hyperref}
\usepackage{comment}
\usepackage{cancel}

\usepackage{array}
\usepackage{rotating}
\usepackage{multirow}

\newcommand{\pt}{p_{\rm T}}

\newcommand{\squishlist}{
 \begin{list}{$\bullet$}
  { \setlength{\itemsep}{0pt}
     \setlength{\parsep}{3pt}
     \setlength{\topsep}{3pt}
     \setlength{\partopsep}{0pt}
     \setlength{\leftmargin}{1.5em}
     \setlength{\labelwidth}{1em}
     \setlength{\labelsep}{0.5em} } }

\newcommand{\squishlisttwo}{
 \begin{list}{$\bullet$}
  { \setlength{\itemsep}{0pt}
     \setlength{\parsep}{0pt}
    \setlength{\topsep}{0pt}
    \setlength{\partopsep}{0pt}
    \setlength{\leftmargin}{2em}
    \setlength{\labelwidth}{1.5em}
    \setlength{\labelsep}{0.5em} } }

\newcommand{\squishend}{
  \end{list}  }

\newcommand{\ifb}{\mathrm{fb}^{-1}}

\newcommand{\f}{\frac}

\newcommand{\gev}{~\mathrm{GeV}}

\newcommand{\fref}[1]{Fig.~\ref{f.#1}}
\newcommand{\eref}[1]{Eq.~(\ref{e.#1})}

\newcommand{\sref}[1]{Section \ref{s.#1}}
\newcommand{\ssref}[1]{Section \ref{ss.#1}}

\newcommand{\cref}[1]{Chapter \ref{c.#1}}

\newcommand{\bi}{\begin{itemize} \itemsep=0.75mm}
\newcommand{\ei}{\end{itemize}}
\newcommand{\benum}{\begin{enumerate} \itemsep=0.75mm}
\newcommand{\eenum}{\end{enumerate}}
\newcommand{\ben}{\begin{equation}}
\newcommand{\een}{\end{equation}}
\newcommand{\be}{\begin{equation*}}
\newcommand{\ee}{\end{equation*}}

\newcommand{\fval}{0.40}
\newcommand{\ttbar}{t\bar{t}}
\newcommand{\gtt}{{\tilde g}}

\begin{document}

\preprint{YITP-SB-12-39}

\title{Boosted Multijet Resonances and New Color-Flow Variables}

\author{David Curtin}
\affiliation{C. N. Yang Institute for Theoretical Physics, Stony Brook University, 
 Stony Brook, NY 11794, U.S.A.}

 \author{Rouven Essig}
\affiliation{C. N. Yang Institute for Theoretical Physics, Stony Brook University, 
 Stony Brook, NY 11794, U.S.A.}
 
 \author{Brian Shuve}
 \affiliation{Perimeter Institute for Theoretical Physics, 31 Caroline St.~N, Waterloo, Ontario N2L 2Y5, Canada.}
    \affiliation{Department of Physics \& Astronomy, McMaster University, 1280 Main St.~W, Hamilton, Ontario  L8S 4L8, Canada.}

\begin{abstract}
We use modern jet-substructure techniques to propose LHC searches for multijet-resonance signals without leptons or missing energy. Ê
We focus on three-jet resonances produced by $R$-parity-violating decays of boosted gluinos,
showing that shape analyses searching for a mass peak can probe such gluinos up to masses of $\sim 750$ GeV (650 GeV) 
with $20\,\,\mathrm{fb}^{-1}$ ($5\,\,\mathrm{fb}^{-1}$) at the LHC at 8 TeV.  
This complements existing search strategies, which also include counting methods that are inherently more prone to systematic 
uncertainties. Ê
Since $R$-parity-violating gluinos lighter than all squarks hadronize before decaying, we introduce new color-flow variables, Ê
``radial pull'' and ``axis contraction'', which are sensitive to the color structure of the $R$-hadron's decay. 
The former measures the inward pull of subjets in a fat jet, while the latter quantifies the inward drift of the $N$-subjettiness axes 
when changing the distance measure.  
We show that they can dramatically improve the discrimination of a boosted gluino signal versus QCD, $t\bar t$, 
and combinatoric background for $m_{\tilde g}\sim m_t$. 
Cuts on axis contraction also noticeably improve the resonance shape for heavy gluinos with $m_{\tilde g} \gtrsim 500 \gev$. 
With minor adaptations, these variables could  find application in substructure searches for particles in different color representations or with other decay topologies. 
We also compare how several different Monte Carlo generators model the high-multiplicity QCD background. 
This provides evidence that the discriminating power of our color-flow observables are robust, and provides useful guidance for future substructure studies.
\end{abstract}

\maketitle

 \setcounter{equation}{0} \setcounter{footnote}{0}

 %%%%%%%%%%%%%%%%%%%%%%%%%%%%%%%%%%%%%%%%%%%%%%%%%%%%%%%%%%%%%%%%%%%%%%
%%%%%%%%%%%%%%%%%%%%%%%%%%%%%%%%%%%%%%%%%%%%%%%%%%%%%%%%%%%%%%%%%%%%%%%
\section{Introduction}
\label{s.intro} 
%%%%%%%%%%%%%%%%%%%%%%%%%%%%%%%%%%%%%%%%%%%%%%%%%%%%%%%%%%%%%%%%%%%%%%
%%%%%%%%%%%%%%%%%%%%%%%%%%%%%%%%%%%%%%%%%%%%%%%%%%%%%%%%%%%%%%%%%%%%%

The Large Hadron Collider (LHC) is setting ever-more stringent constraints on many Beyond-Standard Model (BSM) theories.  
The most constrained BSM theories are those that produce large amounts of missing transverse energy (MET)
and/or leptons \cite{ATLASnote:2012a, 2012mfa, ATLASnote:2012b, CMSnote:2011b, 2012ar, CMSnote:2011c, Chatrchyan:2012sa, Aad:2011cw, 2012he}.  
However, there are many theories that do not feature these signatures. 
One of the more experimentally challenging signals are jets~\cite{Sterman:1977wj} with no MET or leptons,  
for which the background from ordinary Quantum Chromodynamics (QCD) processes is prodigious.  

\begin{figure}
\begin{center}\vspace{4mm}
\includegraphics[width=0.43\textwidth]{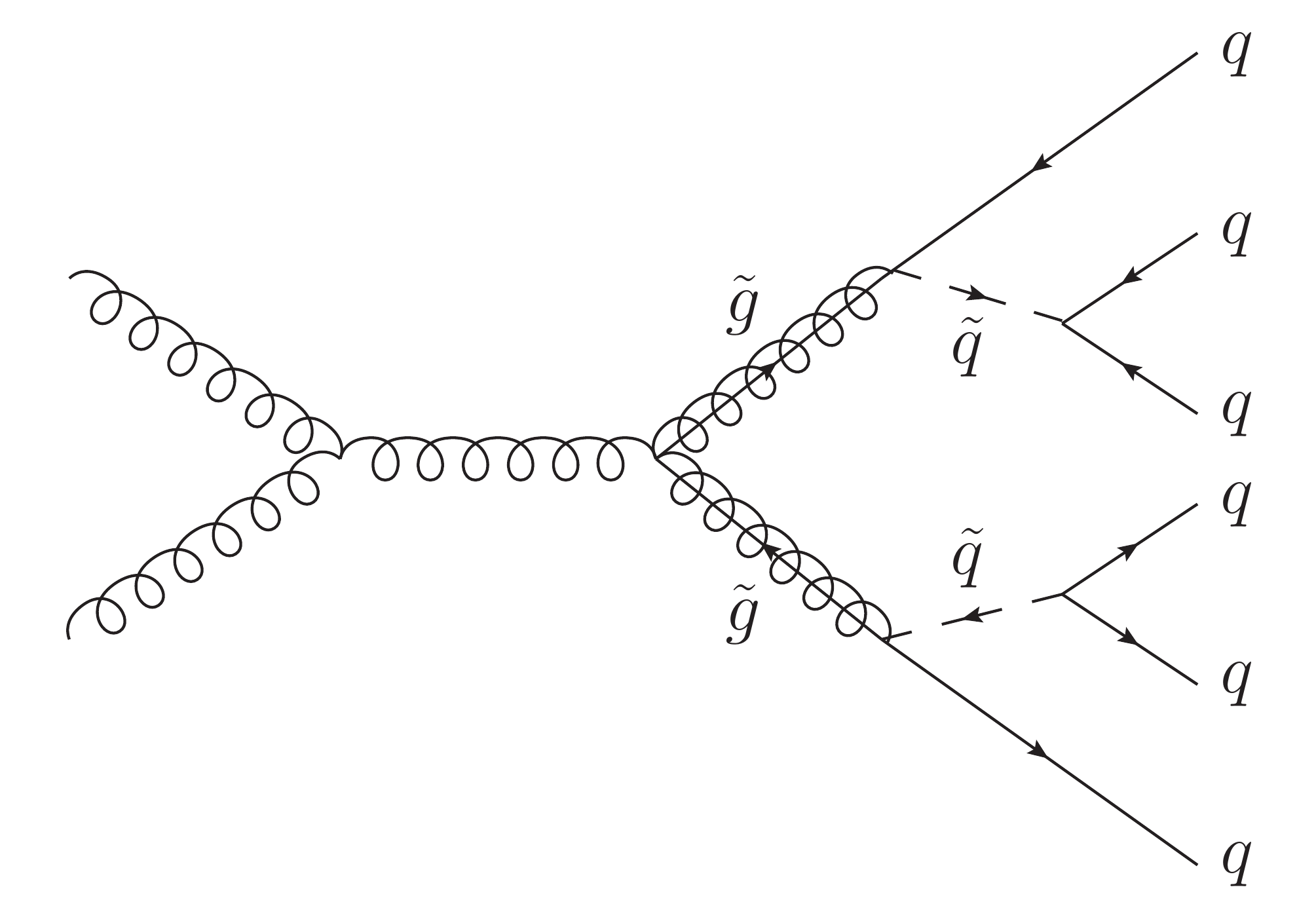}
\end{center}
\caption{The signal studied in this paper: pair production of gluinos, with each decaying to three quarks via an intermediate off-shell squark. 
}
\label{f.gluinodecay}
\end{figure}

The particular signal that we study in this paper can be phrased in terms of a simplified model.   
In addition to the Standard Model particle content, we consider a colored particle that is pair-produced and decays to three 
light-flavored quarks~\cite{Chivukula:1990di,Chivukula:1991zk,Farhi:1979zx,Marciano:1980zf,Frampton:1987ut,Frampton:1987dn,rouventhesis}~\footnote{We leave decays to heavy-flavored quarks, which are in principle easier to detect, to future work.}.  
A useful benchmark model for this scenario is the Minimal Supersymmetric (SUSY) Standard Model with baryon-number 
violating $R$-parity violation (RPV)~\cite{RPVoriginal, RPVreview} and a gluino as the lightest supersymmetric particle (LSP): 
the colored particles are gluinos, which each decay to a quark and an off-shell squark that decays 
to two quarks via an RPV coupling (see Fig.~\ref{f.gluinodecay}). 
This leads to a six-jet signal, with two three-jet resonances from the two decaying gluinos.
For other examples of multi-jet searches, see \cite{Alwall:2008ve,Kilic:2008pm,Kilic:2010et,Alves:2011wf,Tavares:2011zg,Gross:2012bz,Hook:2012fd}.

If leptons and neutrinos appear in RPV cascade decays, bounds on superpartner masses can still be $\sim 1$~TeV~\cite{rpvmsugra}.  
However, the bounds are weaker if the signal is entirely hadronic, which we consider here.
In particular, for a gluino LSP decaying to three jets, a model-independent bound on exotic color octets from LEP excludes gluino masses below 51 GeV \cite{LEPrpvgluinobound}, while searches at CDF and CMS have only excluded gluinos with masses in the 
range $\sim 77-144$ GeV~\cite{TEVATRONrpvgluinobound}, $\sim 200-280$~GeV~\cite{CMSrpvgluinobound36}, and 
$\sim 280 - 450$~GeV~\cite{CMSrpvgluinobound5}.  A recent ATLAS search has closed the gap between $144-200$ GeV and excludes gluinos up to 660 GeV~\cite{ATLASrpvgluinobound5}.  
The weakened bounds relative to $R$-parity-conserving SUSY make RPV an attractive possibility for natural 
SUSY models~\cite{Allanach:2012vj,Brust:2011tb,Brust:2012uf,Evans:2012bf}. 

The six-jet signal is challenging to see due to the large QCD background.  
A further difficulty is the large combinatorial ambiguity in correctly identifying the three jets from each gluino.  The method in~\cite{rouventhesis,TEVATRONrpvgluinobound,CMSrpvgluinobound36, CMSrpvgluinobound5} uses a correlation between the sum of the transverse momentum ($p_{T,jjj}$) and the invariant mass ($M_{jjj}$) of three jets to select a phase space region with a high number of signal relative to combinatorial and QCD background events.  Even so $S/B$ often ends up being very small, motivating the investigation of complementary analysis methods. 

The ATLAS study~\cite{ATLASrpvgluinobound5} exploits the fact that the jets from gluino decays tend to have similar $\pt$, while QCD six-jet events exhibit a $\pt$ hierarchy. Counting the number of events on the high-end tail of the sixth-hardest-jet-$\pt$ distribution excludes RPV gluinos up to 660~GeV~\cite{ATLASrpvgluinobound5}, but does require an extremely reliable understanding of the background normalization.  
It is thus desirable to cross-check this result using an orthogonal search channel with a nearly independent set of systematic uncertainties 
(as also suggested in~\cite{ATLASrpvgluinobound5}).

This motivates the focus of our paper: we examine the decay and radiation pattern produced by two  \emph{boosted} gluinos, whose decay products tend to be collimated and fall into the same region of the detector, producing two hard \emph{fat jets}. The study of boosted gluinos allows for a reconstruction of $m_{\tilde g}$ in the fat-jet invariant mass distribution, giving a cleaner and more robust signal than other methods, and allowing a shape analysis to extract the gluino mass peak. We study boosted gluinos using jet-substructure techniques~\cite{nsubjettiness, Stewart:2010tn, nsubjettinessminaxes, pull, Bassetto:1984ik, dipolarity, girth, WtagRcores, quarkgluontag, planarflowtoptag, BDRStagger, jetpruning, filteringfortag, trimming, filterforhiggs, YsplitterATLAS, Soper:2011cr, Butterworth:2007ke, Almeida:2010pa, Jankowiak:2011qa, Ellis:2012sn, Butterworth:2002tt, Jankowiak:2012na, Salam:2009jx, TWtoptagger, Hopkinstoptagger}, which have matured enormously in recent years and are being verified experimentally~\cite{ATLASnote:2012a, ATLASnote:2012b, CMSnote:2011b, CMSnote:2011c, Abazov:2011vh, ATLASsubstructurecomparison, Aad:2011kq, Aad:2012jf, CMSnote:2011a, Aaltonen:2011pg, ATLASnote:2011a, Chatrchyan:2012zt, Seymour:1993mx}  
(for some recent reviews see~\cite{boost2010, toptagreview, boost2011}).  
We achieve the best signal sensitivity using $N$-subjettiness \cite{nsubjettiness, nsubjettinessminaxes} to isolate three-pronged fat jets, requiring two high-$\pt$ fat jets with similar masses, and vetoing events with a large sub-jet $p_{\rm T}$-hierarchy. 
In addition, we introduce two new color-flow variables, \emph{radial pull} and \emph{axis contraction}. {\it Radial pull} is based on the {\it pull} variable defined in~\cite{pull}, while \emph{axis contraction} 
exploits the shift in the minimizing axes of $N$-subjettiness when changing the distance measure.   
Both variables help distinguish the \emph{signal's} QCD radiation pattern when compared with the QCD background. 
The background radiation pattern has been simulated with several Monte Carlo programs.  

The ATLAS study~\cite{ATLASrpvgluinobound5} examined boosted gluinos and excluded $m_{\tilde g}\sim100-300$~GeV 
(light compared to $\sqrt{s}$), but we show 
that a boosted search is viable at the 8 TeV LHC (LHC8) up to  
$m_{\tilde g} \lesssim 750$~GeV, where $\sigma_{\tilde g\tilde g} \sim 0.2$~pb and the boosted fraction is only $\mathcal{O}$(few \%).
This shows that looking for relatively heavy gluinos in the boosted regime, first proposed for the Tevatron~\cite{tevatronboostedgluinos}, 
carries over to the LHC in spite of the smaller boosted fraction and production cross section of a $pp$ collider relative to a $p\bar p$ collider.  
We also define a search for top-mass gluinos ($m_{\tilde g}\sim m_t$) with spectacular background discrimination that would improve 
on the ATLAS limit.  This is relevant for other simplified models.

 The new variables, \emph{radial pull} and \emph{axis contraction}, are designed to measure the distribution of the soft 
QCD radiation pattern inside each boosted-gluino fat jet. They should generalize to other examples of boosted jet 
studies~\cite{futurework}, for example hadronic RPV decays of neutralino LSPs (as in \cite{Butterworth:2009qa}). 
Their use in our study relies on the fact that gluinos decaying via an off-shell squark generically live longer than the hadronization 
time-scale, so that they form a color-singlet $R$-hadron before decaying.  
This leaves a measurable soft QCD radiation pattern within each boosted fat jet, 
which differs from a beam-connected color octet like that for an un-hadronized gluino, combinatorics background, a boosted hadronic top quark, or QCD 
background. 
This effect of \emph{color-connected} jets has been previously studied for a color singlet (such as $W$ or $h$) 
decaying into two jets forming a color dipole \cite{Bassetto:1984ik,Bjorken:1991xr,Bjorken:1992er,Fletcher:1993ij,Barger:1994zq,pull,dipolarity,girth}, and Tevatron experimental results demonstrated its viability~\cite{Abazov:2011vh}; 
however, color flow has to our knowledge never been investigated for decaying $R$-hadrons or non-dipole configurations. 

In \S\ref{s.rhadron}, we discuss the color flow in the gluino $R$-hadron decay and introduce variables sensitive to the radiation pattern. 
\S\ref{s.montecarlo} describes our Monte Carlo (MC) generation;  \S\ref{s.analysis} details the substructure variables that distinguish boosted gluino decays from background, and show the results for searches for heavy ($m_{\tilde g}\gtrsim500$ GeV) and top-mass gluinos.  
\S\ref{s.conclusions} contains our conclusions.  
A brief review of RPV gluinos is in Appenix \ref{s.review}. 
A detailed Monte Carlo comparison for the QCD background of substructure and color-flow observables is described in Appendix \ref{s.MCcomparison} to ensure robustness of our results.

 %%%%%%%%%%%%%%%%%%%%%%%%%%%%%%%%%%%%%%%%%%%%%%%%%%%%%%%%%%%%%%%%%%%%%%
%%%%%%%%%%%%%%%%%%%%%%%%%%%%%%%%%%%%%%%%%%%%%%%%%%%%%%%%%%%%%%%%%%%%%%%
\section{Novel probes of color flow}
\label{s.rhadron}  
%%%%%%%%%%%%%%%%%%%%%%%%%%%%%%%%%%%%%%%%%%%%%%%%%%%%%%%%%%%%%%%%%%%%%%
%%%%%%%%%%%%%%%%%%%%%%%%%%%%%%%%%%%%%%%%%%%%%%%%%%%%%%%%%%%%%%%%%%%%%
%
\begin{figure}
\begin{center}\hspace*{-3mm}
\begin{tabular}{c}
\includegraphics[width=0.45\textwidth]{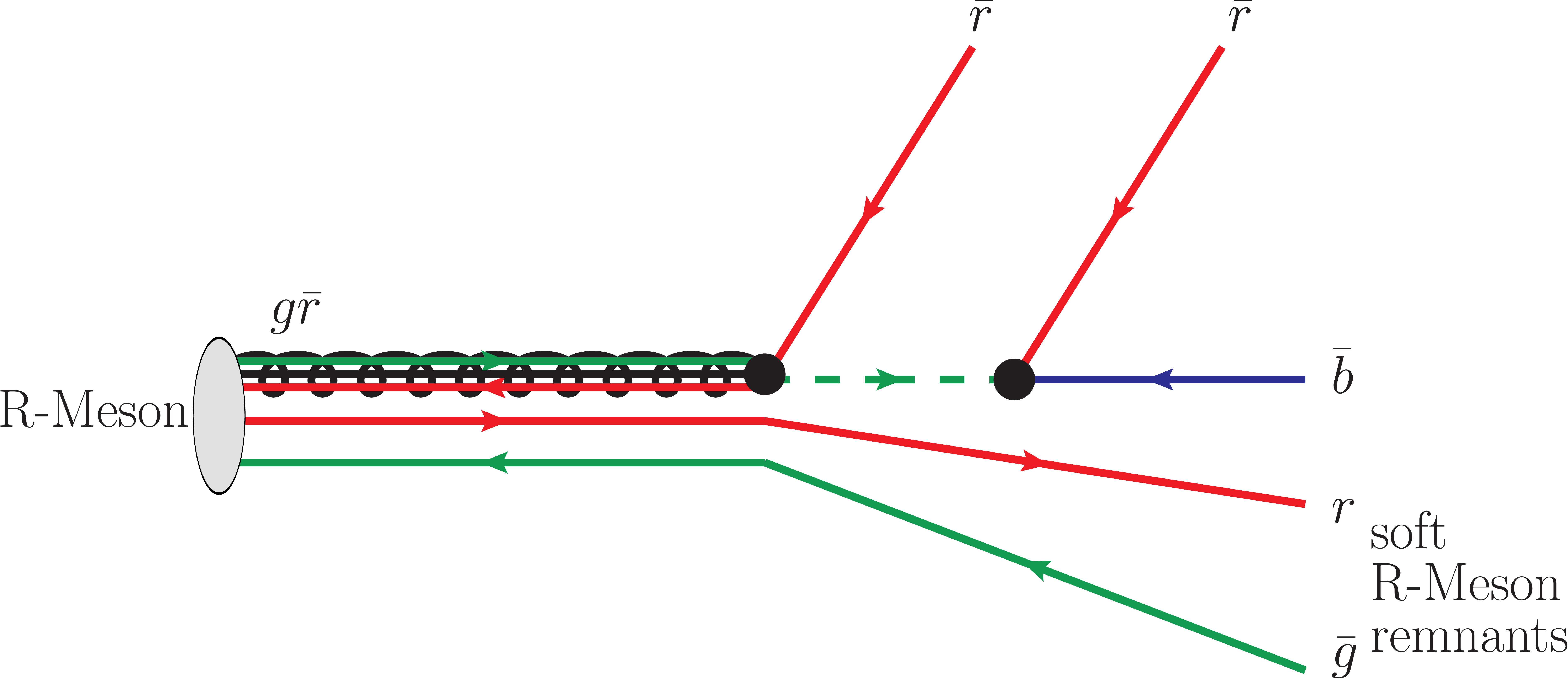}
\\
(a)
\\
\includegraphics[width=0.45\textwidth]{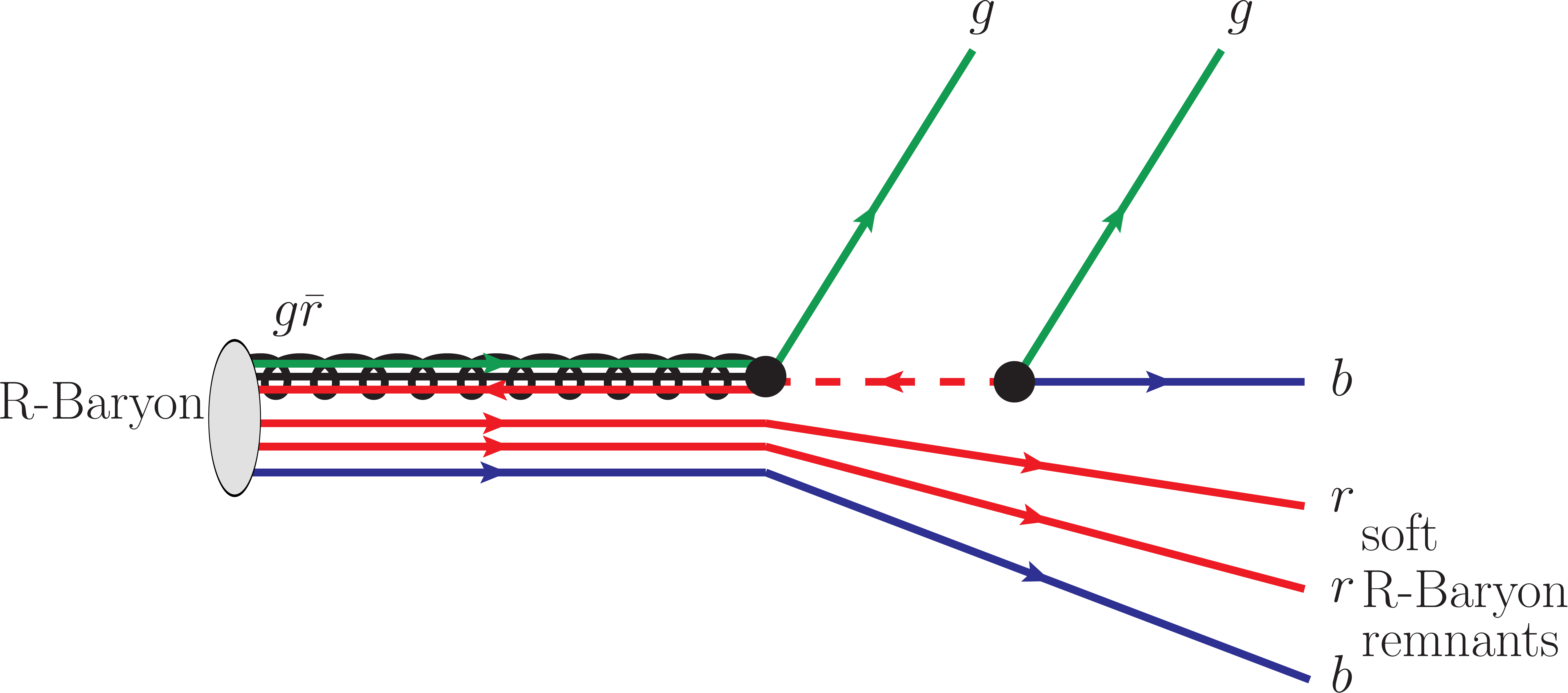}
\\
(b)
\\
\includegraphics[width=0.45\textwidth]{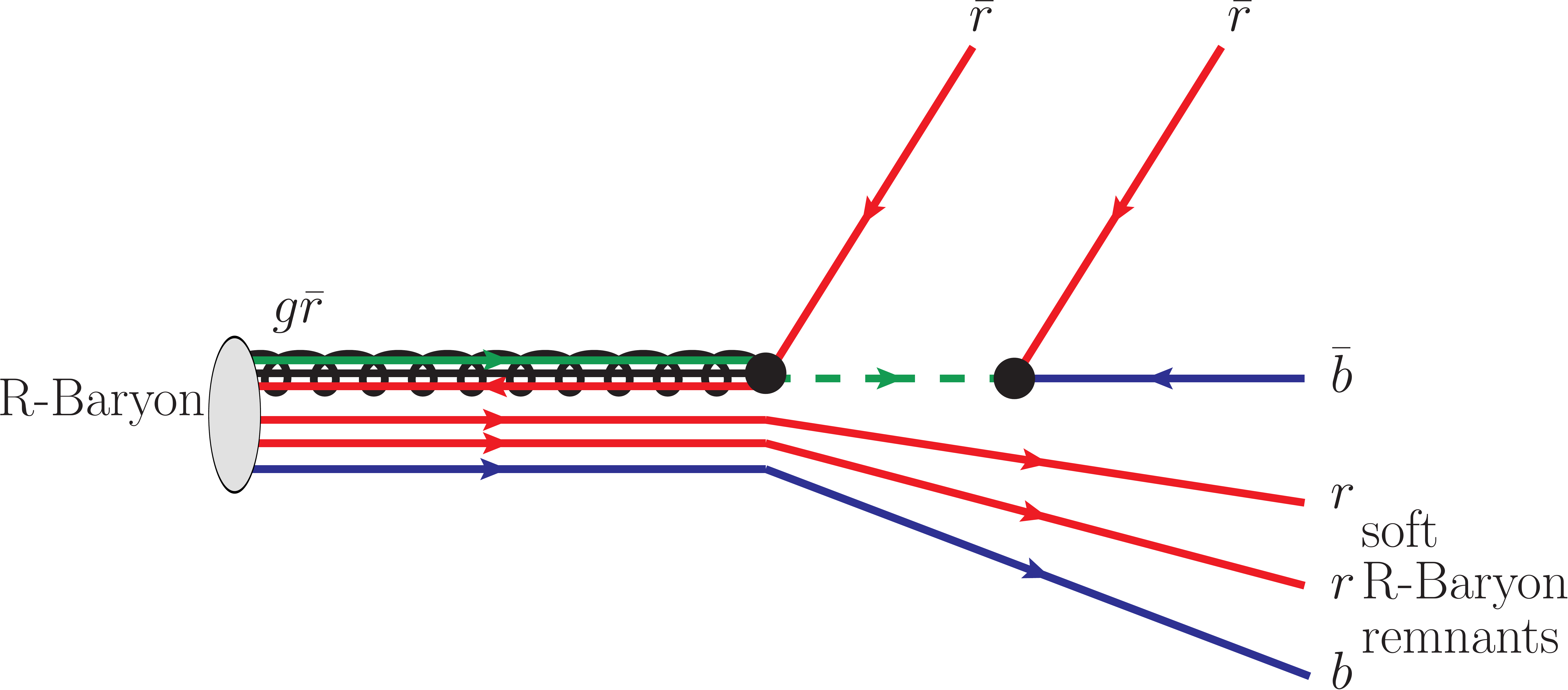}
\\
(c)
\end{tabular}
\end{center}
\caption{(a) The color flow in  $R$-meson $(\tilde g qq)$ decay due to RPV gluino decay via an intermediate off-shell squark. (b) and (c): The two physically distinct possibilities for color flow in  $R$-baryon $(\tilde g qqq)$ decay. Forward/backward arrows indicate color/anticolor, which are also indicated with $r,g,b$.}
\label{f.Rhadron}
\end{figure}
\begin{figure*}
\begin{center}
\includegraphics[width=0.4\textwidth]{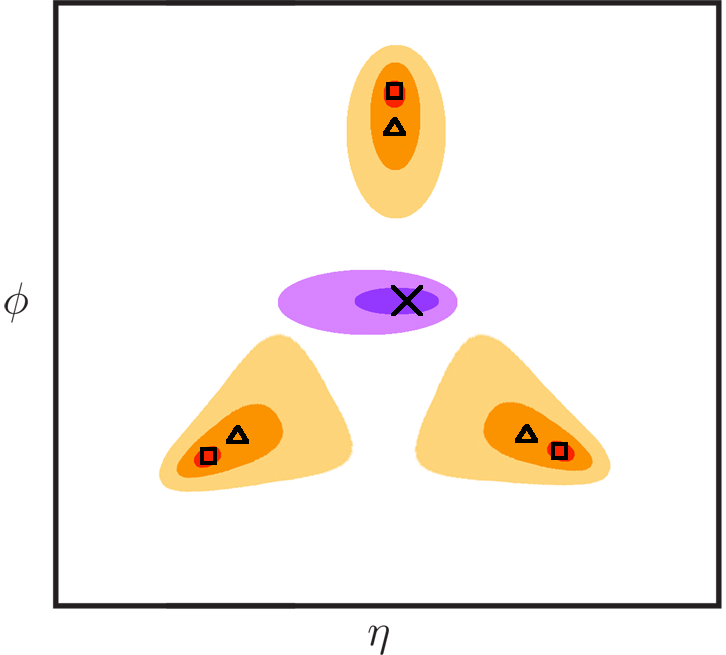}
\phantom{blablabl}
\includegraphics[width=0.4\textwidth]{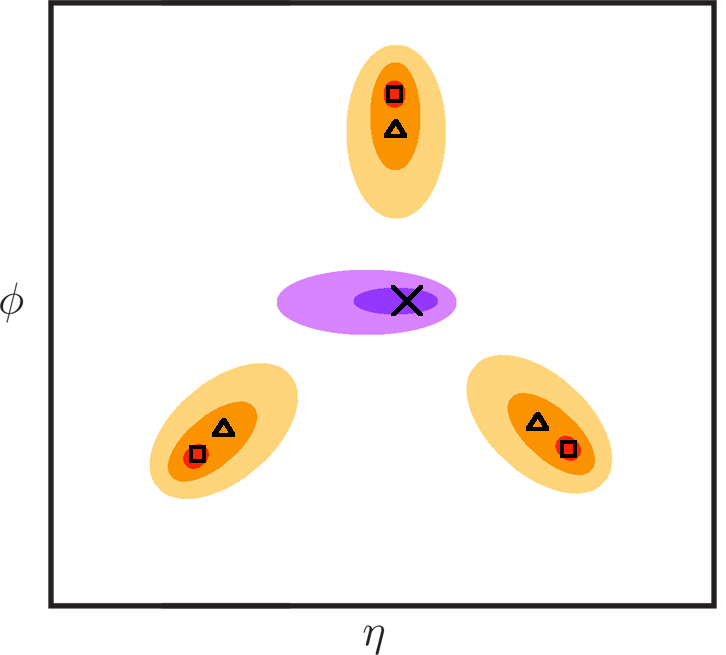}
\end{center}\vspace{-3mm}
\caption{Exaggerated schematic representation of the hypothesized radiation pattern in the $\eta-\phi$ plane produced by the $R$-hadron decays shown in \fref{Rhadron}.  The radiation pattern from the shower \& hadronization of the gluino is shown in purple at the center. The left-ward slant indicates (in this example) a color-connection to the beam, but this radiation pattern is extremely soft, hard to detect, and susceptible to uncertainties in the $R$-hadron formation model. The centers of the hard jets from RPV gluino decay are shown in red, with orange indicating the shape of the soft radiation pattern resulting from a \emph{second} round of showering \& hadronization of the three quarks produced in the gluino decay. 
Squares and triangles indicate the location of axes minimizing the N-subjettiness 
variables $\tau_3^{\beta = 1}$ and $\tau_3^{\beta = 2}$, respectively, and the cross marks the fat-jet centroid, i.e.~the original $R$-hadron direction.  
There are two possibilities for the radiation pattern: the \emph{left} plot corresponds to the $R$-meson and the $R$-baryon in 
Fig.~\ref{f.Rhadron}(a) and (b), respectively; the \emph{right} plot only occurs for the $R$-baryon in Fig.~\ref{f.Rhadron}(c) and is of limited importance, since $R$-baryons make up only $\sim 1\%$ of produced $R$-hadrons~\cite{pythiaRhadrondiscussion}. 
}
\label{f.colorflowcartoon}
\end{figure*}

Gluinos that decay to three quarks through an off-shell squark and the baryon-violating RPV coupling (which appears in the superpotential as $W\supset \lambda'' \bar{u}\bar{d}\bar{d}$)
generically have a lifetime that is prompt but longer than the hadronization scale --- see review in Appendix \ref{s.review}. 
(We do not consider gluino decays originating from a displaced vertex or occurring outside the detector.)
This means they first form color-singlet $R$-hadrons before decaying to three jets.  
The pattern produced by the radiation from these signal jets is different from QCD background jets or from jets that 
originate from a particle that is a color fundamental (like a top quark) or octet (like an un-hadronized gluino decaying to three jets).  
In this section, we first give an intuitive explanation for the radiation pattern before introducing two new variables, 
\emph{radial pull} and \emph{axis contraction}, that attempt to quantify this.  
The new variables should, with slight adaptations, also be useful to distinguish background jets from jets originating from other particles 
like the Higgs or $W$-boson, which we study elsewhere~\cite{futurework}.  

\subsection{Color Connections in $R$-hadron Decays}
\label{ss:colorconnection}

The soft radiation pattern of a jet produced during hadronization will depend on how the color of the parent quark 
or gluon is connected to the color of the other quarks or gluons in the event.  
Put simply, the radiation pattern of a jet will be, on average, slightly `pulled' towards other jets (or the beam) to which it is 
color-connected. 
Furthermore, gluons have a larger color charge than quarks and will produce on average a somewhat wider radiation pattern. 
A number of variables make use of this to, for example, tag dipole-pairs of jets and distinguish quarks from gluons \cite{WtagRcores, dipolarity, pull, girth, quarkgluontag}. 

The color flow of gluino pair production is identical to that of gluon pair production, with gluinos being color-connected to the proton remnants in the beam. Therefore, if the gluinos were stable, we would expect the $R$-hadrons to be surrounded by a soft radiation pattern not unlike that of gluon jets.  
However, for decaying $R$-hadrons the situation is more interesting. \fref{Rhadron} shows the color flow of a decaying $R$-meson $\tilde g qq$ and $R$-baryon $\tilde gqqq$. Of the three hard jets from the decaying $R$-meson, one forms a `mesonic' singlet with one of the $R$-meson remnant spectator quarks, while two form a `baryonic' singlet with the other one of remnant quarks (a similar pattern of color connection with the remnants holds for decaying gluinoballs $\tilde g g$). For $R$-baryons, there are two distinct possibilities: two baryonic singlets, one formed out of one hard jet and two spectators and the other formed out of two hard jets and one spectator; or three mesonic singlets incorporating one hard quark each. 

\begin{figure*}[t]
\begin{center}
\begin{tabular}{m{20mm}m{\fval\textwidth}m{3mm}m{\fval\textwidth}}
\includegraphics[width=15mm]{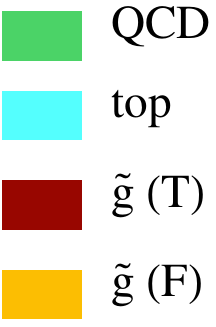}
&
\includegraphics[width=\fval\textwidth]{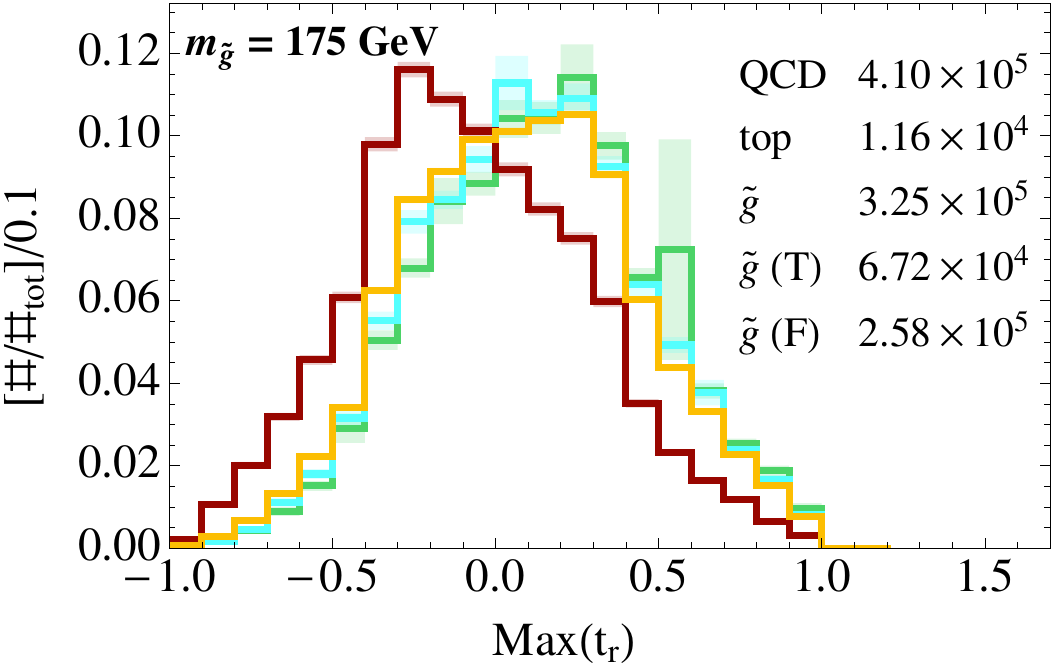}
&&
\includegraphics[width=\fval\textwidth]{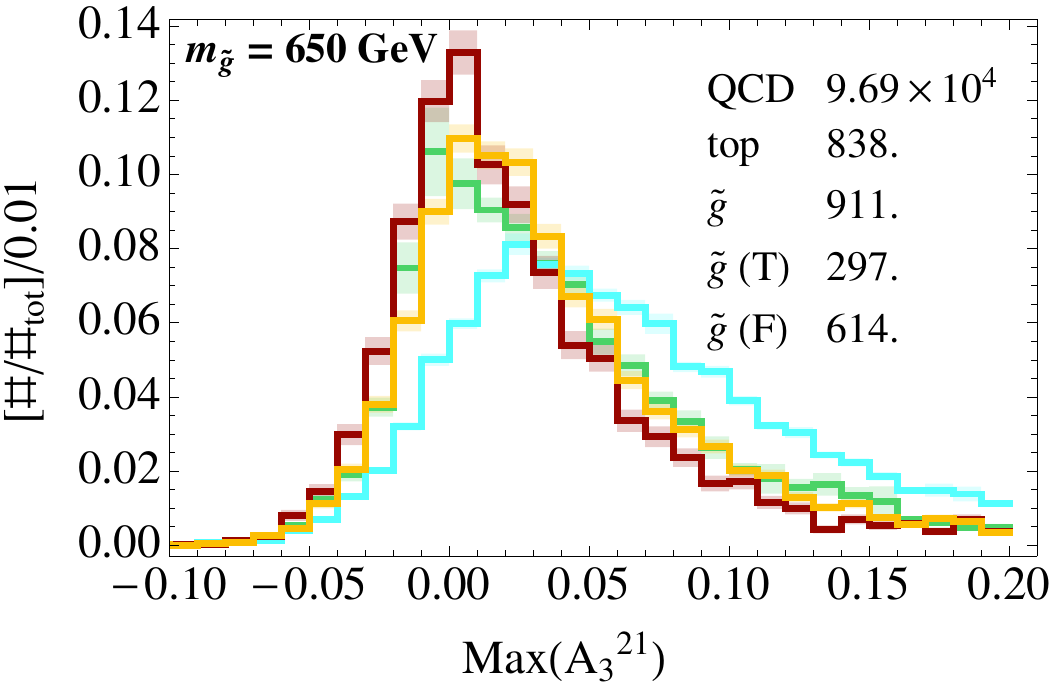}
\end{tabular}
\end{center}
\caption{{\it Left:} Normalized distributions of the new color-flow variable \emph{radial pull} ($t_{\rm r}$, Eq.~(\ref{e.radialpull}))  
for a gluino $R$-hadron with $m_\gtt = 175$~GeV 
decaying to three jets via RPV (red and orange); QCD (green) 
and $\ttbar$ (cyan) backgrounds are also shown. The signal is divided into two components:  red (orange) corresponds to events in which the hardest two fat jets are aligned (are not aligned) 
within $\Delta R = 0.3$ of the two gluino $R$-hadrons at the MC truth level.  
The red distribution is dominated by events where both fat jets reconstruct a decaying gluino, and thus constitute ``good'' (T) signal events, whereas the orange distribution shows events where the fat jets do not reconstruct a decaying gluino and are ``bad'' (F) signal events.  
For each event with two hard fat jets, the larger of the two radial pulls is shown in the histogram. 
The inset table shows the absolute sizes of the different samples, normalized to the number of expected events at LHC8 with 20 $\ifb$. 
Basic generator- and trigger-level cuts have been included, same as for \fref{cutflow.light} (a). Error bars indicate MC statistical uncertainty.  \vspace*{-1mm}\\
{\it Right:} Same as plot on left, but now showing the normalized distribution of the new color-flow variable \emph{axis-contraction}
($A^{\beta \beta'}_N$ with $\beta=2, \beta'=1, N=3$) for $m_\gtt = 650$~GeV.  An equivalent set of cuts has been applied, same as for \fref{cutflow.heavy} (a). 
In both cases the distribution of the good signal events differs markedly from the other (background) distributions.  
}
\label{f.color}
\end{figure*}

The resulting soft radiation pattern expected in the decay of a \emph{boosted} $R$-hadron is shown in \emph{strongly exaggerated} schematic form in \fref{colorflowcartoon}. The soft radiation field (purple region) of the \emph{first} round of hadronization (during 
$R$-hadron formation) is contained in the center of the fat jet, and might contain evidence of the initial gluino's color connection to the beam, indicated here by its left-ward slant. However, this radiation field is  extremely soft, as well as subject to the possible shortcomings of $R$-hadronization models, and it will not be our focus. Superimposed on this initial radiation field is the result of the $R$-hadron decay, and a \emph{second} round of hadronization to resolve the color connections amongst its decay products. It is in the radiation fields of the three hard subjets of the fat jet that it may be possible to find evidence of the \emph{overall singlet nature} of the $R$-hadron: there should be an overall pull towards the center, and possibly between two fat jets that form a baryonic singlet with part of the remnant. While there is substantial overlap in the signal and background distributions of color-flow variables, we do find that aggressive cuts on color-flow variables that isolate events with the patterns shown in Fig.~\ref{f.colorflowcartoon} are very helpful in improving signal-to-background ratios for RPV gluino searches. 
Any such variable can also be viewed as a generalization of existing dipole taggers, and could also find application elsewhere.

\subsection{Radial Pull}

The distribution of the radiation field of a jet can be measured with a variable called \emph{pull}~\cite{pull}, defined as
\begin{equation}
\label{e.pull}
\vec t = \sum_i \frac{p_T^i |r_i|}{p_T^\mathrm{jet}} \vec r_i, \phantom{blabl} {\textrm{ \it (pull)}}
\end{equation}
where $\vec r$ is a vector in the $\eta-\phi$ plane pointing from the jet axis to the $i^\mathrm{th}$ jet constituent, and 
$p_T^\mathrm{jet}$ ($p_T^i$) is the transverse momentum of the jet ($i^\mathrm{th}$ jet constituent). The direction of the pull vector indicates an overall slant in the jet's radiation distribution, while the magnitude contains limited information \cite{pull}. 

For a fat jet with $N$ subjets, we can calculate the pull vector for each subjet  and combine them in a quantity that we call {\it radial pull}, defined as 
\begin{equation}
\label{e.radialpull}
t_{\rm r} = \f{1}{N}\,{\displaystyle\sum_{j=1}^N \hat{t}_j \cdot \hat{n}_j}\,, \phantom{blabl} {\textrm{ \it (radial pull)}}
\end{equation}
where $\hat t_j$ is the pull vector (normalized to unity) for the $j$-th subjet and $\hat n_j$ is the unit vector from 
the fat-jet center to the subjet axis, i.e.~the difference between the subjet's and fat-jet three-momentum 
vector, normalized to unity.  
By definition, $t_{\rm r}\in [-1,1]$.

\begin{figure}[h!]
\begin{center}
\includegraphics[width=\fval\textwidth]{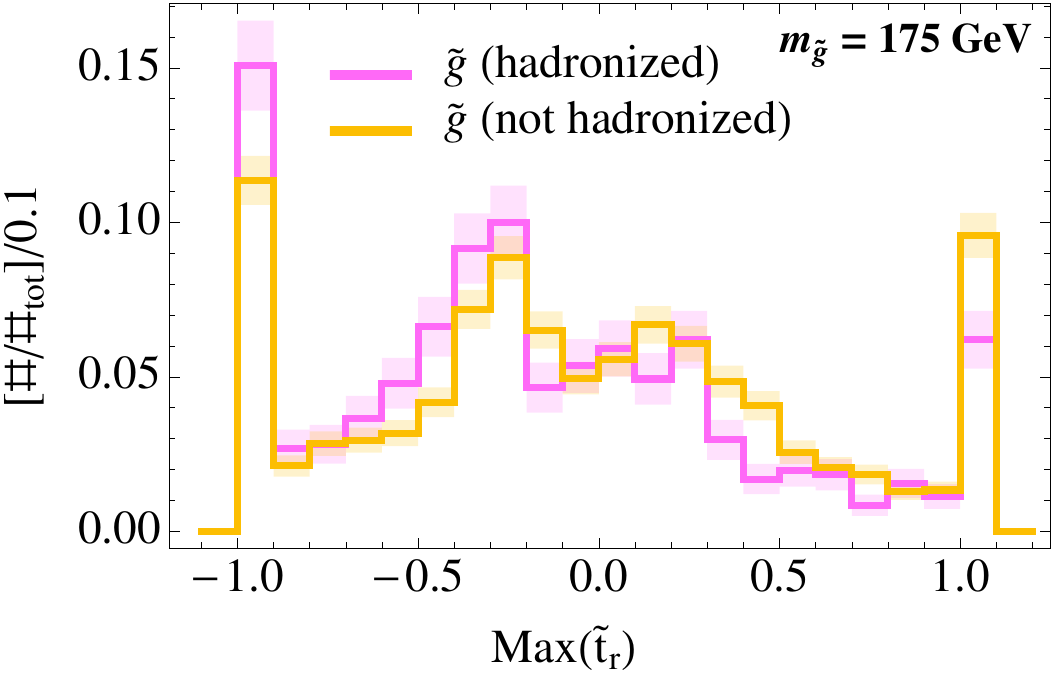}
\end{center}
\caption{Comparing the normalized distributions of $\mathrm{max}(\tilde t_{\rm r})$ (maximized over the two hardest fat jets in the event) for 175 GeV gluinos with and without formation of $R$-hadrons before decay. The kinematic cuts (though not the radial-pull cut) of the top-mass gluino analysis outlined in \ssref{topmass} have been applied. Comparing the first and last bin occupations readily distinguishes between gluinos that either form or do not form $R$-hadrons before decaying.
}
\label{f.had.vs.nohad}
\end{figure}

Radial pull characterizes the degrees to which a fat jet's subjets are color-connected to the fat-jet center. Radial pull is expected to be closer to $-1$ 
for the decay of a color-singlet $R$-hadron located at the fat-jet center (see Fig.~\ref{f.colorflowcartoon}), whereas fat jets from QCD are expected to have components with color connections to the beam and other jets in the event, yielding a radial pull closer to $+1$. 
This is confirmed by MC simulations of the signal and background (in case of the background, with several different generators, see \sref{montecarlo} and Appendix \ref{s.MCcomparison}).
Fig.~\ref{f.color} (left) shows the normalized distribution of $t_{\rm r}$ for a gluino with $m_{\tilde g} = 175$~GeV (red and orange) and QCD (green) 
and $\ttbar$ (cyan) backgrounds (basic generator and trigger level cuts have been included).
We divide the signal into two components:  red (orange) in Fig.~\ref{f.color} corresponds to those events in which the hardest 
two fat jets line up (do not line up) within $\Delta R = 0.3$ with the three-momentum of the two gluino $R$-hadrons at the MC truth level.  
The red distribution is thus enriched with events where two fat jets originate from a decaying gluino, and thus constitute ``good'' (T) signal events.  
We see that the good signal events peak at $t_{\rm r}$ values closer to $-1$ and have a distinct distribution from the ``bad'' (F) signal events and from QCD 
and $\ttbar$ backgrounds.

We also tried an alternative definition of radial pull, 
\begin{equation}
\label{e.radialpull2}
\tilde{t}_r = \f{\displaystyle\sum_{j=1}^N \vec{t}_j \cdot \hat{n}_j}{\displaystyle\sum_{j=1}^N |\vec{t}_j \cdot \hat{n}_j|}\,. \phantom{b} {\textrm{ \it (radial pull, alternative def.)}}
\end{equation}
This variable is most sensitive to the dominant subjet pull, and so $\tilde t_{\rm r}$ is peaked strongly 
at $\pm 1$.  We find that it works less well in distinguishing signal from background compared to the definition in~Eq.~(\ref{e.radialpull}).  However, 
Eq.~(\ref{e.radialpull2}) is better at distinguishing radiation patterns from hadronized gluinos ($R$-hadrons) vs.~unhadronized gluinos. This is evident in Fig.~\ref{f.had.vs.nohad}, where a ratio between the +1 and -1 bin distinguishes clearly between decaying hadronized and un-hadronized gluinos.
It is possible that for other types of signal Eq.~(\ref{e.radialpull2}) may be more useful, but 
in this paper, we  use Eq.~(\ref{e.radialpull}).

\subsection{Axis Contraction}
\label{ss.axiscontraction}

A second variable to measure the QCD radiation pattern of a boosted heavy particle exploits previously unexplored 
properties of $N$-subjettiness, $\tau_N^\beta$ \cite{nsubjettiness}, which we first review. 
$\tau_N^\beta$ is based on the variable $N$-jettiness~\cite{Stewart:2010tn}, and is the average of the $\pt$'s of jet-constituent particles, 
weighted by the distance to a set of $N$ axes, 
\begin{eqnarray}
\label{e.nsubjettiness}
\tau_N^\beta &\equiv& \frac{1}{d_0}\sum_i\,p_{\mathrm{T}\,i}\mathrm{min}\left[(\Delta R_{1,i})^\beta,\ldots,(\Delta R_{N,i})^\beta\right],\\
d_0 &=& \sum_i\,p_{\mathrm{T}\,i}\,R_0^\beta,
\end{eqnarray}
where $R_0$ is the fat-jet radius and the sum is over all jet constituents $i$. $\tau_N^\beta$ characterizes how well the radiation in the jet is aligned along these $N$ chosen axes, which are labelled by $a = 1, \ldots , N$. $\Delta R_{a,i}$ is the distance in the $\eta-\phi$ plane between axis $a$ and constituent $i$, and the axes are chosen in each instance to minimize the value of $\tau_N^\beta$ \cite{nsubjettinessminaxes} unless indicated otherwise. (This is implemented in the \texttt{FastJet} $N$-subjettiness plug-in \cite{nsubjettinessminaxes}  and provides substantially better discriminating power in boosted particle taggers than when the $N$-subjettiness axes are chosen using a traditional jet clustering algorithm such as ~anti-$k_{\rm T}$) \cite{nsubjettiness}.) $\beta$ determines the weighting in the sum of radiation far from the axes; typically $\beta\sim 1-2$. 

By comparing values of $\tau_N$ for different $N$, it is possible to characterize the number of subjets in the event. For example, if $\tau_{21} \equiv \tau_2/\tau_1 \ll 1$, the radiation in the jet is clustered around two separate axes within the jet, implying 
that the jet contains two prominent subjets. By contrast, if $\tau_{21} \sim 1$, then the radiation is distributed fairly evenly around the central jet axis and is not well characterized by two subjets. The ratio $\tau_{21}$ is useful for tagging jets from the decay of a boosted $W$-boson, whereas the ratio $\tau_{32} \equiv \tau_3/\tau_2$ can isolate jets with three distinct subjets, which is 
useful to tag boosted top quarks~\cite{boost2010, boost2011, nsubjettiness, nsubjettinessminaxes} as well as boosted RPV gluino decays, as 
we show below.

Varying the parameter $\beta$ changes the sensitivity of $\tau_N^{\beta}$ to radiation far from the $N$-subjettiness axes, and
can therefore probe the shape of the radiation inside the fat jet. This provides a sensitivity not just to the \emph{number} of `hard subjet-like structures' in a fat jet (which might be accessed by reclustering) but also to the \emph{shape} of their radiation patterns. Traditionally, this has been exploited in top-taggers \cite{nsubjettiness, nsubjettinessminaxes}  by setting $\beta \approx 1$ to make $N$-subjettiness hone in on very tight radiation centers.

A novel way of exploiting this shape sensitivity is to study the \emph{$\beta$-variance of the axes which minimize $\tau_N^\beta$}. (The idea of varying the parameters of event shape variables to extract additional information was first proposed in~\cite{georgeeventshape}.) 
Consider a fat jet with $N$ well-defined subjets. We denote the $a = 1, \ldots, N$ axes which minimize $\tau_N^\beta$ by 
\begin{equation}
\label{e.axis}
\vec R_{a,N}^\beta \equiv (\eta_{a,N}^\beta, \phi_{a,N}^\beta)\,.
\end{equation}
If the soft radiation field of a subjet associated with axis $a$ is skewed towards one side, then the axis
 is shifted away from the center of the subjet, and this shift is more pronounced for higher values of $\beta$. 
Therefore, for example, the vector  
 \begin{equation}
 \label{e.axisshift}
\Delta \vec R_{a,N}^{\beta \beta'} \equiv  \vec R_{a,N}^\beta - \vec R_{a,N}^{\beta'}, \phantom{bbbbbb} \mathrm{\it (axis\ pull)}
 \end{equation}
 which we call \emph{axis pull}, should point in the direction of the skew for $\beta>\beta'$.  This typically points in the same direction as the pull (\ref{e.pull}).

For the radiation pattern from gluinos, shown in Fig.~\ref{f.colorflowcartoon}, the skew of the radiation field should lie
between the subjet axis and the fat jet center. To capture this effect, we define a scalar quantity called \emph{axis contraction} as 
\begin{equation}
\label{e.axiscontraction}
A^{\beta \beta'}_N = \frac
{\displaystyle \sum_{a=1}^N \left|\vec R_{a,N}^\beta - \vec R_\mathrm{cen} \right|}
{\displaystyle \sum_{a=1}^N \left| \vec R_{a,N}^{\beta'} - \vec R_\mathrm{cen} \right|} - 1\,, \phantom{bla} {\textrm{ \it (axis contraction)}}
\end{equation}
where $\vec R_\mathrm{cen}$ marks the fat jet centroid, i.e.~the fat jet momentum  in the $\eta-\phi$ plane, 
and $\beta>\beta'$.  For our application we choose $(\beta, \beta') = (2,1)$.  

If the $\tau_N^\beta$ axes shift towards (away) from the centroid of the fat jet as $\beta$ is changed from 1 to 2, then 
$A^{\beta\beta'}_N$  is smaller (larger) than 0. We therefore expect that for boosted RPV gluinos, $A^{21}_3$ is on average \emph{smaller} for \emph{signal} fat jets that contain all of a gluino's decay products than for fat jets from kinematically identical QCD 
or $t \bar t$ backgrounds, or signal fat jets that do not contain the decay products of a single gluino.  
This is again confirmed by MC simulations of the signal and background, which are described in detail in \S\ref{s.montecarlo}.  
For light gluinos, axis contraction performs comparably to radial pull, though the latter does have slightly better discrimination power. However, axis contraction is more suitable when cutting conservatively and wanting to preserve the largest amount of signal, as opposed to achieving maximum signal purity. This makes it more suitable for the heavy gluino case.
Fig.~\ref{f.color} (right) shows the normalized distributions of $A^{21}_3$ for a gluino with $m_{\tilde g} = 650$~GeV (red and orange) 
and QCD (green) and $\ttbar$ (cyan) backgrounds (basic generator and trigger level cuts have been included).  The discrimination power of the variable for this heavy gluinos becomes more apparent as additional cuts are applied, see Fig.~\ref{f.cutflow.heavy}(g).

Note that the small numerical range of the axis contraction variable over our event samples is not indicative of any unrealistic detector resolution required to observe its variation, but is just a matter of how the variable is normalized. Furthermore, the shapes of the jet-$p_T$ distributions are very different at maximal vs minimal values of axis contraction (or radial pull). These differences are easily distinguished by the calorimeter and particle tracker.

The idea of using a shift in the $N$-subjettiness axes under a change of $\beta$ to probe color flow is very general. 
One could imagine using $A_2^{21}$ as a dipole tagger, for example. This will be investigated  further in~\cite{futurework}.

  %%%%%%%%%%%%%%%%%%%%%%%%%%%%%%%%%%%%%%%%%%%%%%%%%%%%%%%%%%%%%%%%%%%%%%
%%%%%%%%%%%%%%%%%%%%%%%%%%%%%%%%%%%%%%%%%%%%%%%%%%%%%%%%%%%%%%%%%%%%%%%
\section{Monte Carlo Event Generation}
\label{s.montecarlo}
%%%%%%%%%%%%%%%%%%%%%%%%%%%%%%%%%%%%%%%%%%%%%%%%%%%%%%%%%%%%%%%%%%%%%%
%%%%%%%%%%%%%%%%%%%%%%%%%%%%%%%%%%%%%%%%%%%%%%%%%%%%%%%%%%%%%%%%%%%%%

We simulated gluino pair production, $R$-hadron formation, and RPV decay with a developmental version of \texttt{Pythia 8.165} \cite{pythia6manual, pythia8RPV, pythia6RPV, pythia8} \footnote{We thank Torbj\"orn Sj\"ostrand and Peter Skands for their quick implementation 
in \texttt{Pythia} of a baryon-violating RPV decay after $R$-hadron formation, and for supplying us with a developmental version of \texttt{Pythia} that included this feature.  Their change is now implemented in the publicly available \texttt{Pythia 8.170}.}. 
This was done for  two ranges: ``top-mass'' gluinos with $m_{\tilde g} = 175 \gev$ and heavy gluinos with $m_{\tilde g} = 500, 550,\ldots 1000 \gev$. The number of unweighted signal events simulated for the heavy (light) gluino analysis was $\sim 10^5-10^6$ 
($\sim 4\times 10^7$) per mass point. All  distributions were reweighted to correspond to a given luminosity at  LHC8, usually 20 $\ifb$. Samples without $R$-hadron formation before the decay were also generated for comparison. All signal cross sections were calculated at NLO in \texttt{Prospino} \cite{prospino}. 

We used \texttt{Sherpa 1.4.0} \cite{sherpa, sherpaother} to generate fully matched QCD (with 2-6 hard jets from the Leading-Order matrix element) and $t \bar t$ background samples (fully hadronic decays with $\le 2$ additional hard jets), with additional jets produced by the shower.  Event generation was weighted to adequately sample high-multiplicity and high-$p_T$ events. 
We generated separate backgrounds for the heavy and the top-mass gluino searches:
\begin{itemize}
\item {\it Heavy gluino ($m_{\tilde g}\gtrsim 500$~GeV) search:} The relevant trigger is $H_{\rm T} \gtrsim 850$ GeV along with at least one jet of $\pt>200$ GeV~\cite{private.communication.1}, while the analysis itself requires two hard fat jets. Accordingly, we required background events at generator level to have two fat jets (anti-$k_{\rm T}$, $R=1.5$) each with $p_{\rm T}>500$ GeV, along with one thin jet (anti-$k_{\rm T}$, $R=0.4$) 
with $p_{\rm T}>200$ GeV.  All generator-level cuts act at parton-level, and thresholds are correspondingly conservative compared to later cuts on the fully showered event. Because efficiencies of subsequent cuts are low, we generated a sample of $\sim50$ million QCD events and $\sim10$ million $t\bar t$ events.  We find that the dominant background by far is QCD. Top backgrounds have a much smaller cross section and tend to produce fat jets that lie around $m_t\sim175$ GeV, well below the gluino mass scale under study. 
\item {\it Top-mass gluino search:} This search uses a trigger requiring six thin jets (anti-$k_{\rm T}$, $R=0.4$) with $p_{\rm T}\gtrsim 60$ GeV for full efficiency. Therefore background events were required at generator level to have two fat jets (anti-$k_{\rm T}$, $R=1.5$) with $p_{\rm T}>150$ GeV, as well as at least four thin jets with $p_{\rm T}>40$ GeV. Nevertheless, the poor trigger and cut efficiencies required the generation of around 100 million QCD events. The top, for which we generated $\sim 20$ million events, is subdominant to QCD for most of our subsequent chain of cuts, but can dominate after the final color-flow cuts. 
\end{itemize}

\begin{figure}
\begin{center}
\includegraphics[width=\fval\textwidth]{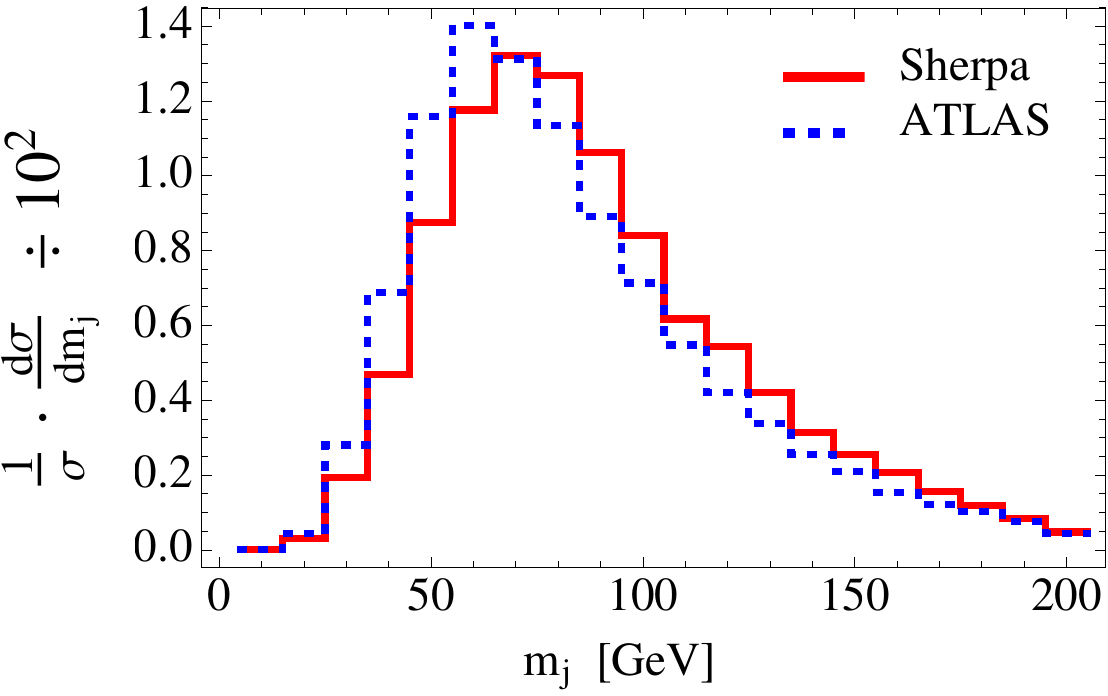}\\
\includegraphics[width=\fval\textwidth]{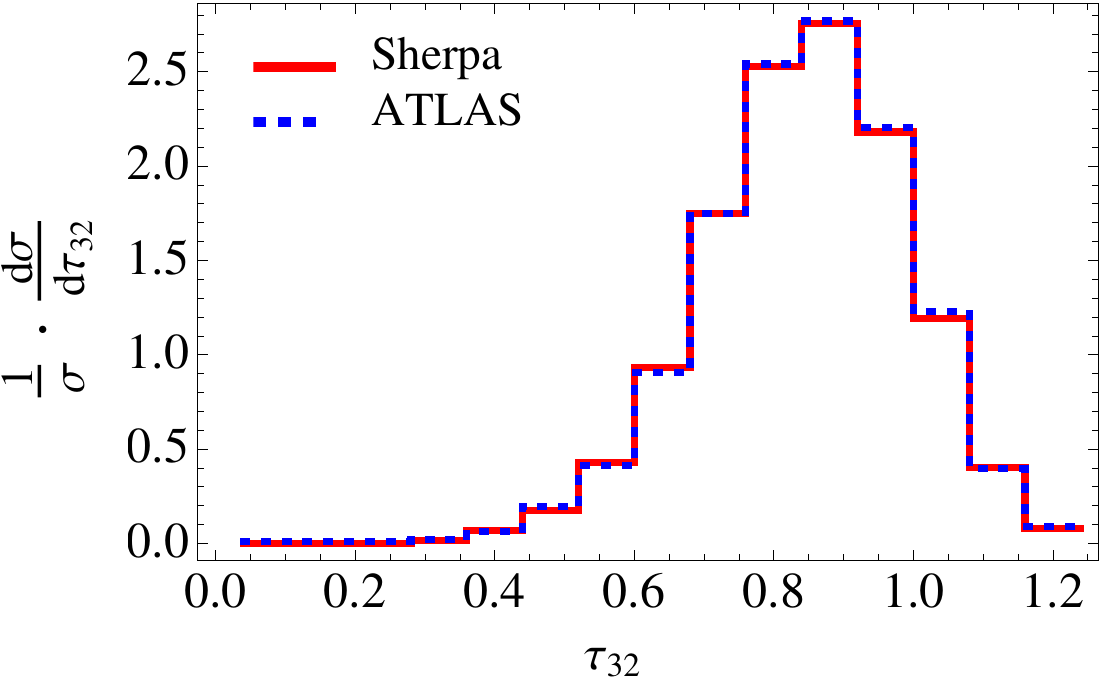}
\end{center}
\caption{Comparison of  300 GeV$<\pt<400$ GeV, $R=1.0$ anti-$k_{\rm T}$ fat jets from Sherpa and the ATLAS 7 TeV, $35\,\,\mathrm{pb}^{-1}$ analysis~\cite{ATLASsubstructurecomparison}. The plots are normalized to the same area. Top: Jet-mass distribution. Bottom: $N$-subjettiness ratio $\tau_{32}\equiv\tau_3/\tau_2$ distribution.
}
\label{f.ATLAScompare}
\end{figure}

As a check of the backgrounds generated with \texttt{Sherpa}, we also considered backgrounds generated with independent matrix element generators and showering programs: \texttt{POWHEG 1.0 + Pythia 6.4.27} and \texttt{POWHEG 1.0 + Pythia 8.1.65} for QCD, and \texttt{Madgraph 5.1.5 + Pythia 6.4.27} and \texttt{Madgraph 5.1.5 + Pythia 8.1.65} for $t\bar t$. The details of the MC comparison are in Appendix \ref{s.MCcomparison}. Overall, the different event generators are consistent with one another for jet-substructure and color-flow observables, although the tails of the distributions could differ in size by up to a factor of 10. Nevertheless, the broad conclusions of the efficacy of jet substructure and the discriminating power of color-flow observables are supported by all analyses. 
For the rest of this paper we compare the signal to backgrounds generated with \texttt{Sherpa}, since it is the only generator which provides fully matched event samples with up to 6 jets at matrix-element level as well as supporting weighted event generation to fully sample the tails of distributions.
Ultimately, experimental study of color-flow observables will be needed to determine which MC program has the best agreement with data, but the comparison from Appendix \ref{s.MCcomparison} give a rough estimate of the dependence of generator effects on our observables. Because hadronization and RPV decay of gluinos is unique to \texttt{Pythia 8}, we were unfortunately unable to similarly estimate the generator-dependence of the signal observables.

All events are analyzed in our own \texttt{FastJet 3.0.2} \cite{fastjet}-based code, interfaced with the $N$-subjettiness plugin \cite{nsubjettinessminaxes} to calculate $\tau_N^\beta$ and find minimizing axes.  Histograms were combined and reweighted in \texttt{Mathematica} to the number of expected events at LHC8 for a given integrated luminosity. For a histogram bin containing events with weight $\{w_i\}$, the MC statistical error on the sum $\sum_i w_i$ is $\sqrt{\sum_i w_i^2}$, which we include in our distributions. We generated enough events to ensure that this statistical error is no larger than Poisson error corresponding to the actual expected number of LHC8 events, and that the distributions are as smooth as possible. 
Systematic errors are not included in our predictions, since the number of events surviving all the cuts is generally so small that Poisson uncertainties dominate. 
Detector effects are not explicitly modeled, but are unlikely to be a limiting factor since we only use the central part of the event ($|\eta| < 2.5$), our fat jets are extremely hard ($p_T > 600 \gev$) and low statistics necessitate a large bin size for fat-jet-mass histograms anyway, precluding the use of any unrealistic mass resolution. Furthermore, as mentioned in the previous section, our color-flow variables do not rely on any unrealistic detector resolution to achieve their distinguishing power. 

All signal and background samples include the underlying event and are generated using default parton distribution functions. To verify their validity, we applied several cross-checks to our background samples. We were able to reproduce the QCD 6-jet distributions from \cite{CMSrpvgluinobound36} as well as jet-mass and $N$-subjettiness distributions from \cite{ATLASsubstructurecomparison} (see Fig.~\ref{f.ATLAScompare}). While the shape matched well across the entire range of all distributions (particularly for $N$-subjettiness), the normalization was half as large as the data. Therefore, we scaled up the cross section of our entire QCD background sample (supplied at LO by \texttt{Sherpa}) by a $K$-factor of 2. It was unnecessary to similarly re-scale the top background: it is negligible for the heavy gluino search,  and while it is the dominant background after color cuts are applied in the top-mass gluino search, it is nevertheless eliminated by an additional $b$-veto. 

Only $\sim 10\%$ of top-mass gluinos and $\sim1-4\%$ of heavy gluinos with a mass of $500 - 1000 \gev$ are boosted enough for all the decay products to potentially end up in two fat jets. Since our search attempts to reconstruct a resonance in the fat jet mass spectrum, the large majority of signal events has to be treated as combinatorics background. Therefore, we divide the signal sample into two groups: ``good'' (T) signal, which is enriched with boosted gluinos by requiring that the two hardest fat jets are within $\Delta R = 0.3$ of the truth-level gluinos, and ``bad'' (F) signal, i.e. the rest. This allows us to explicitly demonstrate that our cuts eliminate the QCD and $t \bar t$ as well as the combinatorics background.

 %%%%%%%%%%%%%%%%%%%%%%%%%%%%%%%%%%%%%%%%%%%%%%%%%%%%%%%%%%%%%%%%%%%%%%
%%%%%%%%%%%%%%%%%%%%%%%%%%%%%%%%%%%%%%%%%%%%%%%%%%%%%%%%%%%%%%%%%%%%%%%
\section{Analysis}
\label{s.analysis} 
%%%%%%%%%%%%%%%%%%%%%%%%%%%%%%%%%%%%%%%%%%%%%%%%%%%%%%%%%%%%%%%%%%%%%%
%%%%%%%%%%%%%%%%%%%%%%%%%%%%%%%%%%%%%%%%%%%%%%%%%%%%%%%%%%%%%%%%%%%%%

This section details our analysis strategy and results. 

%%%%%%%%%%%%%%%%%%%%%%%%%%%%%%%%%%%%%%%%%%%%%%%%%%%%%%%%%%%%%%%%%%%%%%
%%%%%%%%%%%%%%%%%%%%%%%%%%%%%%%%%%%%%%%%%%%%%%%%%%%%%%%%%%%%%%%%%%%%%%%
\subsection{Strategy}
\label{ss.strategy} 
%%%%%%%%%%%%%%%%%%%%%%%%%%%%%%%%%%%%%%%%%%%%%%%%%%%%%%%%%%%%%%%%%%%%%%
%%%%%%%%%%%%%%%%%%%%%%%%%%%%%%%%%%%%%%%%%%%%%%%%%%%%%%%%%%%%%%%%%%%%%

We  discuss here the set of variables used to distinguish signal from background.  
Our aim is to remove as much QCD background as possible and have the gluino resonance be clearly distinguishable from the background.  In particular, we do not rely on a counting analysis, as this is more susceptible to uncertainties in the background normalization.  

In addition to the generator- and trigger-level cuts discussed in \S~\ref{s.montecarlo}, 
we select events with the following properties: 
\benum
\item Two hard fat jets of size $\Delta R = 1.5$, clustered with the anti-$k_T$ algorithm~\cite{Cacciari:2008gp}.  
Each fat jet  contains the three jets from a gluino decay if the gluino is boosted enough.  We always  consider only the two hardest fat jets in each event, and in many cases cut on the larger/smaller value of a variable evaluated for each jet (indicated with max/min).
\item Each fat jet should be three-pronged.  We use the \emph{N-subjettiness} variable $\tau_N^\beta$ defined in Eq.~(\ref{e.nsubjettiness}) to select events with small values of 
\ben\label{e.nsubjettinesscut}
\tau_{32} \equiv \tau^{\beta = 1}_3 / \tau^{\beta = 1}_2 \,.
\een
\item The two signal fat jets should have comparable invariant masses, since both reconstruct the gluino resonance.  
We thus select events with small values of the 
\emph{jet-mass symmetry} parameter 
\ben\label{e.masssymmetry}
s_{\rm m} \equiv \frac{|m_1-m_2|}{(m_1+m_2)/2}\,.
\een
\item  Each fat jet should contain three hard subjets with similar $\pt$, since the QCD background jets tend to have
asymmetric $\pt$'s.  We thus select events where the \emph{subjet hierarchy} 
parameter 
\ben\label{e.subjethierarchy}
h_{31} \equiv \frac{p_{\rm T3}}{p_{\rm{T1}}}\,,
\een
is close to one.  Here $p_{\rm{T}\,i}$ is the $i$-th subjet found by reclustering each fat jet constituents with anti-$k_T$, $R=0.4$. This variable is similar to the flow variable 
in~\cite{Falkowski:2010hi}.  Versions of both $s_{\rm m}$ and $h_{31}$ have been used in~\cite{tevatronboostedgluinos}.
\item
The radiation pattern within each fat jet should be that of a color-singlet $R$-hadron.  We thus select events with 
constraints on the new variables \emph{radial pull} or \emph{axis contraction} defined in Eqs.~(\ref{e.radialpull}) and (\ref{e.axiscontraction}), respectively.  For radial pull, we use subjets obtained by reclustering the fat-jet constituents with anti-$k_T$, R = 0.4, the same as for $h_{31}$. 
\eenum

We also evaluated other substructure variables like \emph{girth}~\cite{girth, quarkgluontag} and 
\emph{planar flow}~\cite{planarflowtoptag, planarflow}, but they did not add any discriminating power in our case. 

%%%%%%%%%%%%%%%%%%%%%%%%%%%%%%%%%%%%%%%%%%%%%%%%%%%%%%%%%%%%%%%%%%%%%%
%%%%%%%%%%%%%%%%%%%%%%%%%%%%%%%%%%%%%%%%%%%%%%%%%%%%%%%%%%%%%%%%%%%%%%%
\subsection{Pile-up Considerations \& Effect of \\Jet-Grooming}
\label{ss.pileup} 
%%%%%%%%%%%%%%%%%%%%%%%%%%%%%%%%%%%%%%%%%%%%%%%%%%%%%%%%%%%%%%%%%%%%%%
%%%%%%%%%%%%%%%%%%%%%%%%%%%%%%%%%%%%%%%%%%%%%%%%%%%%%%%%%%%%%%%%%%%%%

In the high luminosity environment of the LHC, the effects of pile-up (PU) can distort the spectrum of jet observables such as its 
mass and $N$-subjettiness.
The effects of pile-up can be largely  eliminated using one of several jet-grooming techniques \cite{BDRStagger, jetpruning, filteringfortag, trimming,filterforhiggs, Ellis:2009me}. While we do not explicitly include the effects of jet-grooming in our main analysis, we verify here that an analysis based on variables outlined in  \S\ref{ss.strategy} give results that are applicable when jet grooming is included.

To see the effects of jet grooming, we trim \cite{trimming} our samples by re-clustering fat jets into subjets of $R=0.4$ using the $k_{\rm T}$ algorithm and discarding all subjets with $p_{\mathrm{T\,subjet}}/p_{\mathrm{T\,jet}}<f_{\rm cut}$. We take $f_{\rm cut}=0.02$ ($f_{\rm cut}=0.05$) for the heavy (top-mass) gluino analysis, with the precise value chosen to eliminate most of the effects of PU based on a simple model with mean PU energy of 12 GeV per unit area and intra-event fluctuations of 3 GeV per unit area (characteristic of PU with 20 primary vertices per event) \cite{pileupboost}. 
We choose a smaller $f_{\rm cut}$ for heavier gluinos than for $m_{\tilde g}\sim m_t$, since PU remains fixed even as the fat-jet $p_{\rm T}$ increases for heavier gluinos. These parameters are comparable to those used by~\cite{ATLASrpvgluinobound5}. 

Trimming does alter the distributions of some of our kinematic variables; in particular, it tends to shift $N$-subjettiness to lower values. However, we find that it is always possible to choose values of the cuts on $\tau_{32}$, $s_{\rm m}$, and $h_{31}$ such that the overall signal and background efficiencies are essentially the same  for trimmed and untrimmed samples, and so the outcomes of the ungroomed analyses remain the same  when grooming is included. In fact, it is possible in some instances to achieve a background acceptance  that is lower for the trimmed sample while maintaining the same signal efficiency; this makes our ungroomed analysis somewhat conservative, although we leave an optimization of jet grooming parameters to future work.

Our color-flow variables, however, must be calculated on the jets prior to grooming, because grooming can remove the soft radiation within and between subjets that distinguishes different color flows. To mitigate the effects of PU, it is preferable to include soft radiation over the smallest possible area. We find that, in most events, the values of the color-flow variables are dominated by radiation close to the subjet center, and we suggest that including only soft radiation within $R=0.4$ of each subjet axis would reduce contamination from PU (this is similar to the method of PU mitigation used in  \cite{dipolarity}). Furthermore, we find that the scale of radiation driving the color-flow variables is $\sim12-30$ GeV (and sometimes higher for heavy gluinos), which is above the characteristic scale of intra-event fluctuations in PU ($\sim3$ GeV for 20 primary vertices \cite{pileupboost}). We do not, however, include PU in our MC samples, and we leave it to the experimental collaborations to study precisely how color-flow variables are affected by PU and to confirm their utility.

\newcommand{\tilt}[1]{\begin{sideways}#1\end{sideways}}
\newcommand{\cutentryfirst}[2]{\begin{tabular}{r}$#1$\\$#2$\end{tabular}}
\newcommand{\cutentry}[2]{\begin{tabular}{r}$#1$\\$#2$\%\end{tabular}}
\newcommand{\cutentryTF}[6]{\begin{tabular}{r}$#1 \ (#2)$ \\$#4 \ (#5 )$ \% \end{tabular}}
\newcommand{\cutentrylast}[3]{\begin{tabular}{r} $#1$ \\ $#2$\% \\ $S/B = #3$ \end{tabular}}
\newcommand{\cutentrylastTF}[7]{\begin{tabular}{r}$#1 \ (#2)$ \\$#4 \ (#5 )$ \% \\ $S/B = #7$
\end{tabular}} 
\newcommand{\gluinoheader}[1]{\begin{tabular}{r}Gluinos\\ $[#1]$\end{tabular}}
\begin{table*}
\footnotesize
\vspace*{-3mm}
\begin{center}
\hspace*{-1mm}
\begin{tabular}{|l|r||r|r|r|r||r|r|r||r|r|}
\hline
\multicolumn{2}{|c||}{}
&
\multicolumn{7}{c||}{Common Cuts}
&
\multicolumn{2}{c|}{Optimized Cuts}
\\
\cline{3-11}
\multicolumn{2}{|m{1cm}||}{Analysis}
&
\tilt{generator} \tilt{level} \tilt{cuts}
&
\tilt{Trigger:} \tilt{$H_T > 850 \gev$,} \tilt{$p_T^{j1} > 250 \gev$}
& 
\tilt{two fat jets} \tilt{with} \tilt{$p_T > 600 \gev$}
&
\tilt{$\mathrm{max}(\tau_{32}) < 0.7$}
&
\tilt{$\mathrm{max}(\tau_{32}) < 0.5$}
&
\tilt{$\mathrm{max}(s_{\rm m}) < 0.1$}
&
\tilt{$\mathrm{min}(h_{31}) > 0.2$}
&
\tilt{fat jet}
\tilt{ $\mathrm{min}(p_T) > p_T^{max}$}
&
\tilt{axis} 
\tilt{contraction cut}
\tilt{ $\mathrm{max}(A_3^{21})$}
\tilt{$ < A^{max}$}
\\ \hline \hline

\multirow{3}{*}{\tilt{$m_{\tilde g} = 500 \gev$}}
& QCD 
& \cutentryfirst{4.7 \times 10^6}{-}
& \cutentry{4.7 \times 10^6}{99.9}
& \cutentry{1.6 \times 10^6}{34}
& \cutentry{9.7 \times 10^4}{6.1}
& \cutentry{2.1 \times 10^3}{2.2}
& \cutentry{380}{18}
& \cutentry{88}{23}

&\cutentry{88}{100}
&\cutentry{37 \pm 3}{42}
\\ 

\cline{2-11}
& top &
 \cutentryfirst{6.9 \times 10^3}{-}
& \cutentry{6.8 \times 10^3}{99}
& \cutentry{2.4 \times 10^3}{35}
& \cutentry{840}{35}
& \cutentry{50}{6.0}
& \cutentry{13}{26}
& \cutentry{0.56}{4.3}

& \cutentry{0.56}{100}
& \cutentry{0.14 \pm 0.03}{25}
\\

\cline{2-11}
&
\gluinoheader{4.1\mathrm{pb}}
& \cutentryfirst{8.3\times 10^4}{-}
&  \cutentry{5.5 \times 10^4}{66}
& \cutentry{5.9 \times 10^3}{11}
& \cutentry{2.6 \times 10^3}{44}
& \cutentryTF{310}{162}{150} {12}{14}{11}
& \cutentryTF{110}{82}{31}{36}{50}{21}
& \cutentryTF{69}{52}{17}{61}{64}{54}
& \cutentryTF{69}{52}{17}{100}{100}{100}
& \cutentrylastTF{51 \pm 4}{41 \pm 4}{9.9 \pm2.0}{73}{78}{59}{1.4}
\\ \hline \hline

\multirow{3}{*}{\tilt{$m_{\tilde g} = 550 \gev$}}
& QCD 
& \cutentryfirst{4.7 \times 10^6}{-}
& \cutentry{4.7 \times 10^6}{99.9}
& \cutentry{1.6 \times 10^6}{34}
& \cutentry{9.7 \times 10^4}{6.1}
& \cutentry{2.1 \times 10^3}{2.2}
& \cutentry{380}{18}
& \cutentry{88}{23}

&\cutentry{88}{100}
&\cutentry{37 \pm 3}{42}
\\ 

\cline{2-11}
& top &
 \cutentryfirst{6.9 \times 10^3}{-}
& \cutentry{6.8 \times 10^3}{99}
& \cutentry{2.4 \times 10^3}{35}
& \cutentry{840}{35}
& \cutentry{50}{6.0}
& \cutentry{13}{26}
& \cutentry{0.56}{4.3}

& \cutentry{0.56}{100}
& \cutentry{0.14 \pm 0.02}{25}
\\

\cline{2-11}
&
\gluinoheader{2.2\mathrm{pb}}
& \cutentryfirst{4.3 \times 10^4}{-}
&  \cutentry{3.3 \times 10^4}{76}
&  \cutentry{4.0 \times 10^3}{12}
&  \cutentry{1.8 \times 10^3}{44}
& \cutentryTF{230}{111}{115}{13}{15}{11}
& \cutentryTF{78}{52}{26}{34}{47}{22}
& \cutentryTF{47}{32}{15}{60}{61}{59}
& \cutentryTF{47}{32}{15}{100}{100}{100}
& \cutentrylastTF{32 \pm 2}{24\pm2}{8.3 \pm 1.4}{68}{75}{55}{0.87}
\\ \hline \hline

\multirow{3}{*}{\tilt{$m_{\tilde g} = 600 \gev$}}
& QCD 
& \cutentryfirst{4.7 \times 10^6}{-}
& \cutentry{4.7 \times 10^6}{99.9}
& \cutentry{1.6 \times 10^6}{34}
& \cutentry{9.7 \times 10^4}{6.1}
& \cutentry{2.1 \times 10^3}{2.2}
& \cutentry{380}{18}
& \cutentry{88}{23}

&\cutentry{53}{61}
&\cutentry{23 \pm 2}{44}
\\ 

\cline{2-11}
& top &
 \cutentryfirst{6.9 \times 10^3}{-}
& \cutentry{6.8 \times 10^3}{99}
& \cutentry{2.4 \times 10^3}{35}
& \cutentry{840}{35}
& \cutentry{50}{6.0}
& \cutentry{13}{26}
& \cutentry{0.56}{4.3}

& \cutentry{0.25}{45}
& \cutentry{0.080 \pm 0.020}{31}
\\

\cline{2-11}
&
\gluinoheader{1.2 \mathrm{pb}}
& \cutentryfirst{2.4 \times 10^4}{-}
&  \cutentry{2.0 \times 10^4}{84}
&  \cutentry{2.9 \times 10^3}{15}
&  \cutentry{1.3 \times 10^3}{44}
& \cutentryTF{170}{67}{100}{14}{14}{13}
& \cutentryTF{53}{29}{24}{31}{43}{23}
& \cutentryTF{33}{18}{16}{63}{60}{67}
& \cutentryTF{26}{16}{11}{78}{89}{67}
& \cutentrylastTF{17 \pm 1}{12 \pm 1}{4.7 \pm 0.7}{63}{76}{45}{0.71}
\\ \hline \hline

\multirow{3}{*}{\tilt{$m_{\tilde g} = 650 \gev$}}
& QCD 
& \cutentryfirst{4.7 \times 10^6}{-}
& \cutentry{4.7 \times 10^6}{99.9}
& \cutentry{1.6 \times 10^6}{34}
& \cutentry{9.7 \times 10^4}{6.1}
& \cutentry{2.1 \times 10^3}{2.2}
& \cutentry{380}{18}
& \cutentry{88}{23}

&\cutentry{35}{40}
&\cutentry{21 \pm 1}{59}
\\ 

\cline{2-11}
& top &
 \cutentryfirst{6.9 \times 10^3}{-}
& \cutentry{6.8 \times 10^3}{99}
& \cutentry{2.4 \times 10^3}{35}
& \cutentry{840}{35}
& \cutentry{50}{6.0}
& \cutentry{13}{26}
& \cutentry{0.56}{4.3}

& \cutentry{0.11}{20}
& \cutentry{0.049 \pm 0.009}{44}
\\

\cline{2-11}
&
\gluinoheader{ 0.65\mathrm{pb}}
& \cutentryfirst{1.3 \times 10^4}{-}
&  \cutentry{1.2 \times 10^4}{89}
&  \cutentry{2.1 \times 10^3}{18}
&  \cutentry{910}{43}
& \cutentryTF{120}{43}{77}{13}{15}{13}
& \cutentryTF{33}{16}{17}{27}{37}{22}
& \cutentryTF{22}{9.8}{12}{67}{62}{72}
& \cutentryTF{13}{7.6}{5.2}{58}{77}{43}
& \cutentrylastTF{11 \pm 1}{6.8 \pm 0.7}{3.8 \pm 0.5}{83}{90}{73}{0.51}
\\ \hline \hline

\multirow{3}{*}{\tilt{$m_{\tilde g} = 700 \gev$}}
& QCD 
& \cutentryfirst{4.7 \times 10^6}{-}
& \cutentry{4.7 \times 10^6}{99.9}
& \cutentry{1.6 \times 10^6}{34}
& \cutentry{9.7 \times 10^4}{6.1}
& \cutentry{2.1 \times 10^3}{2.2}
& \cutentry{380}{18}
& \cutentry{88}{23}

&\cutentry{22}{25}
&\cutentry{11 \pm 1}{50}
\\ 

\cline{2-11}
& top &
 \cutentryfirst{6.9 \times 10^3}{-}
& \cutentry{6.8 \times 10^3}{99}
& \cutentry{2.4 \times 10^3}{35}
& \cutentry{840}{35}
& \cutentry{50}{6.0}
& \cutentry{13}{26}
& \cutentry{0.56}{4.3}

& \cutentry{0.070}{13}
& \cutentry{0.018 \pm 0.006}{25}
\\

\cline{2-11}
&
\gluinoheader{0.38 \mathrm{pb}}
& \cutentryfirst{7.6 \times 10^3}{-}
&  \cutentry{7.1 \times 10^3}{93}
&  \cutentry{1.6 \times 10^3}{23}
&  \cutentry{700}{43}
& \cutentryTF{99}{29}{70}{14}{16}{14}
& \cutentryTF{26}{10}{15}{26}{36}{22}
& \cutentryTF{16}{6.4}{10}{64}{61}{66}
& \cutentryTF{7.4}{4.8}{2.6}{45}{75}{26}
& \cutentrylastTF{6.0 \pm 0.3}{4.2 \pm 0.2}{1.8 \pm 0.2}{82}{88}{69}{0.55}
\\ \hline \hline

\multirow{3}{*}{\tilt{$m_{\tilde g} = 750 \gev$}}
& QCD 
& \cutentryfirst{4.7 \times 10^6}{-}
& \cutentry{4.7 \times 10^6}{99.9}
& \cutentry{1.6 \times 10^6}{34}
& \cutentry{9.7 \times 10^4}{6.1}
& \cutentry{2.1 \times 10^3}{2.2}
& \cutentry{380}{18}
& \cutentry{88}{23}

&\cutentry{15}{17}
&\cutentry{7.6 \pm 0.8}{50}
\\ 

\cline{2-11}
& top &
 \cutentryfirst{6.9 \times 10^3}{-}
& \cutentry{6.8 \times 10^3}{99}
& \cutentry{2.4 \times 10^3}{35}
& \cutentry{840}{35}
& \cutentry{50}{6.0}
& \cutentry{13}{26}
& \cutentry{0.56}{4.3}

& \cutentry{0.029}{5.2}
& \cutentry{0.0062 \pm 0.002}{21}
\\

\cline{2-11}
&
\gluinoheader{0.23 \mathrm{pb}}
& \cutentryfirst{4.5 \times 10^3}{-}
&  \cutentry{4.3 \times 10^3}{95}
&  \cutentry{1.2 \times 10^3}{28}
&  \cutentry{510}{43}
& \cutentryTF{75}{18}{57}{15}{16}{14}
& \cutentryTF{19}{6.5}{12}{25}{35}{21}
& \cutentryTF{12}{4.0}{8.4}{67}{61}{70}
& \cutentryTF{4.1}{2.6}{1.5}{34}{66}{18}
& \cutentrylastTF{3.1}{2.1}{0.93}{74}{82}{60}{0.40}
\\ \hline \hline

\multirow{3}{*}{\tilt{$m_{\tilde g} = 800 \gev$}}
& QCD 
& \cutentryfirst{4.7 \times 10^6}{-}
& \cutentry{4.7 \times 10^6}{99.9}
& \cutentry{1.6 \times 10^6}{34}
& \cutentry{9.7 \times 10^4}{6.1}
& \cutentry{2.1 \times 10^3}{2.2}
& \cutentry{380}{18}
& \cutentry{88}{23}

&\cutentry{8.4}{9.5}
&\cutentry{6.0 \pm 0.5}{72}
\\ 

\cline{2-11}
& top &
 \cutentryfirst{6.9 \times 10^3}{-}
& \cutentry{6.8 \times 10^3}{99}
& \cutentry{2.4 \times 10^3}{35}
& \cutentry{840}{35}
& \cutentry{50}{6.0}
& \cutentry{13}{26}
& \cutentry{0.56}{4.3}

& \cutentry{0.022}{4.0}
& \cutentry{0.0029 \pm 0.0009}{72}
\\

\cline{2-11}
&
\gluinoheader{0.14 \mathrm{pb}}
& \cutentryfirst{2.7 \times 10^3}{-}
&  \cutentry{2.6 \times 10^3}{97}
&  \cutentry{890}{34}
&  \cutentry{370}{42}
& \cutentryTF{54}{11}{43}{15}{17}{14}
& \cutentryTF{13}{3.5}{9.7}{24}{30}{23}
& \cutentryTF{8.9}{2.3}{6.6}{67}{68}{67}
& \cutentryTF{2.3}{1.3}{0.91}{25}{57}{14}
& \cutentrylastTF{1.8 \pm 0.2}{1.1 \pm 0.1}{0.69 \pm 0.09}{81}{84}{76}{0.30}
\\ \hline \hline

\multirow{3}{*}{\tilt{$m_{\tilde g} = 850 \gev$}}
& QCD 
& \cutentryfirst{4.7 \times 10^6}{-}
& \cutentry{4.7 \times 10^6}{99.9}
& \cutentry{1.6 \times 10^6}{34}
& \cutentry{9.7 \times 10^4}{6.1}
& \cutentry{2.1 \times 10^3}{2.2}
& \cutentry{380}{18}
& \cutentry{88}{23}

&\cutentry{5.3}{6.0}
&\cutentry{3.5\pm 0.4}{66		}
\\ 

\cline{2-11}
& top &
 \cutentryfirst{6.9 \times 10^3}{-}
& \cutentry{6.8 \times 10^3}{99}
& \cutentry{2.4 \times 10^3}{35}
& \cutentry{840}{35}
& \cutentry{50}{6.0}
& \cutentry{13}{26}
& \cutentry{0.56}{4.3}

& \cutentry{0.012}{2.2}
& \cutentry{0.0013 \pm 0.0005}{11}
\\

\cline{2-11}
&
\gluinoheader{82 \mathrm{fb}}
& \cutentryfirst{1.7 \times 10^3}{-}
&  \cutentry{1.6 \times 10^3}{98}
&  \cutentry{640}{40}
&  \cutentry{260}{41}
& \cutentryTF{40}{6.7}{33}{15}{16}{15}
& \cutentryTF{9.2}{2.1}{7.2}{23}{31}{22}
& \cutentryTF{6.0}{1.3}{4.8}{66}{61}{67}
& \cutentryTF{1.1}{0.62}{0.45}{18}{49}{9.4}
& \cutentrylastTF{0.79 \pm 0.08}{0.51 \pm 0.06}{0.28 \pm 0.05}{74}{82}{63}{0.23}
\\ \hline \hline

\end{tabular}
\end{center} \vspace{-5mm}
\caption{Number of expected events at LHC8 with 20 $\ifb$ for signal and background after each cut for the heavy gluino analyses, and gluino pair production cross sections \cite{prospino} in square brackets. The generator-level cuts outlined in \sref{montecarlo} are only applied to the background samples. ``max'' and ``min'' in the cut variables applies to the two values obtained for the \emph{two hardest fat jets} in each event. The second line in each row shows the efficiency of each individual cut step. For signal, the numbers in brackets refer to the ``good'' combinatorially correct signal defined in \sref{montecarlo}. The last $p_T$ and axis-contraction cuts were optimized for each $m_{\tilde g}$, with the following thresholds \{$p_T^{max}, A^{max}$\}: 
$m_{\tilde g} = 500$: \{$600, 0.02$\}, 
$m_{\tilde g} = 550$: \{$ 600, 0.02$\}, 
$m_{\tilde g} = 600$: \{$ 600, 0.02$\}, 
$m_{\tilde g} = 650$: \{$ 700, 0.04$\}, 
$m_{\tilde g} = 700$: \{$ 750, 0.03$\}, 
$m_{\tilde g} = 750$: \{$ 800, 0.03$\}, 
$m_{\tilde g} = 800$: \{$ 850, 0.04$\}, 
$m_{\tilde g} = 850$: \{$ 900, 0.03$\}.
(Masses and momenta in GeV.)
}
\label{f.HEAVYcuts}
\end{table*}

\begin{figure*}
\newcommand{\upshift}{\vspace*{-26mm}}
\vspace*{-4mm}
\begin{center}
\hspace*{-3mm}
\begin{tabular}{m{17cm}m{3mm}m{5mm}}
\begin{tabular}{p{5mm}p{1mm}p{\fval\textwidth}p{4mm}p{5mm}p{1mm}p{\fval\textwidth}}
\upshift (a)&&
\includegraphics[width=\fval\textwidth]{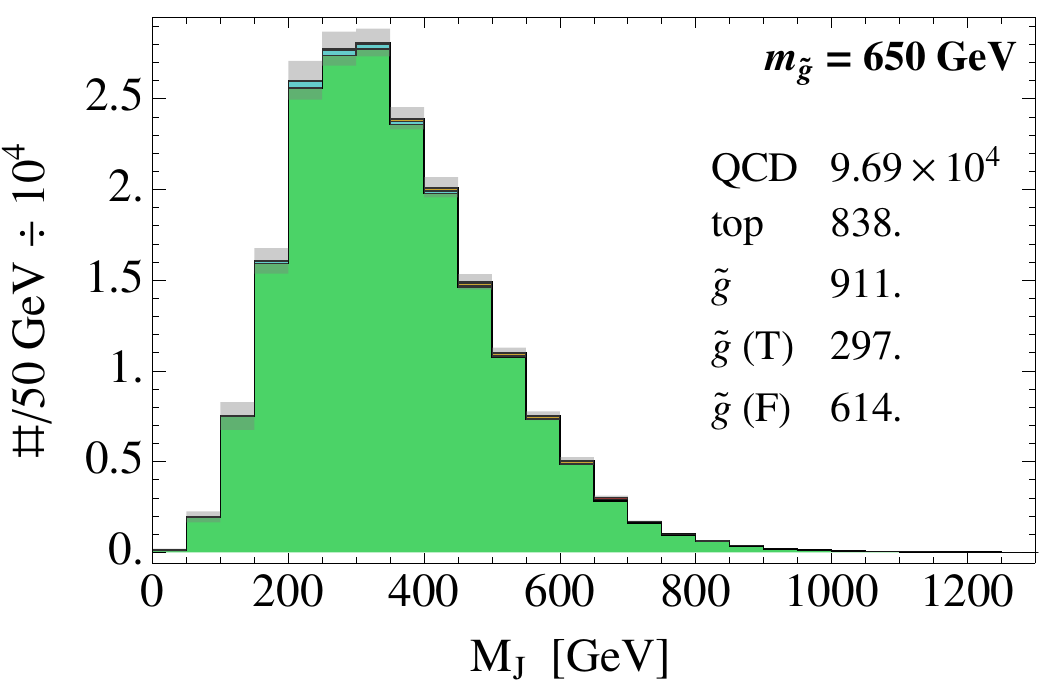}
&&
\upshift(b) &&
\includegraphics[width=\fval\textwidth]{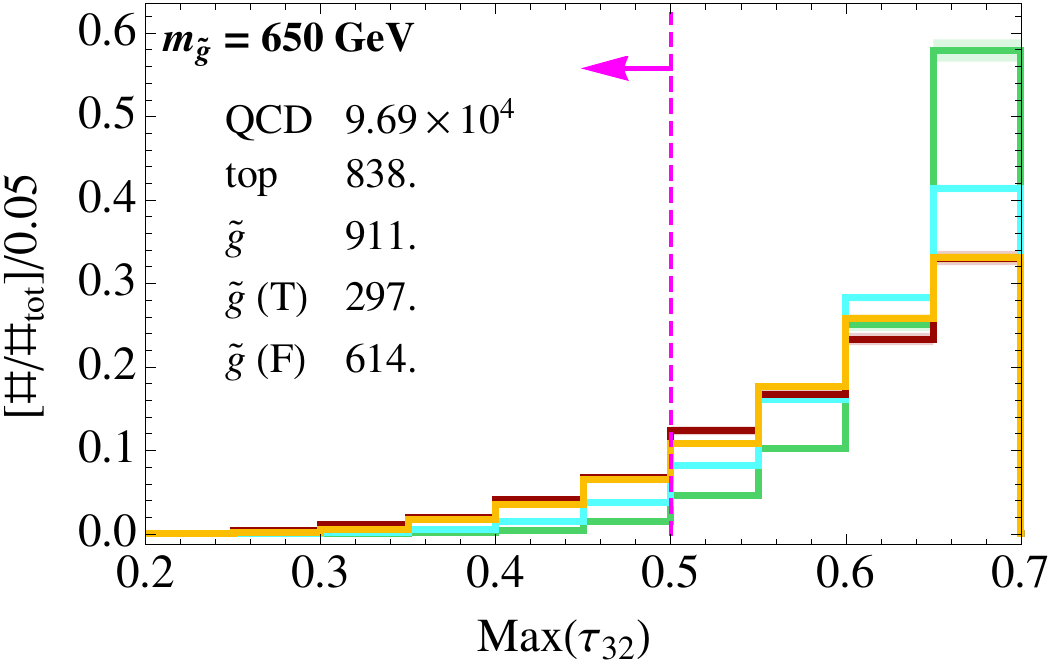}
 \vspace{2mm}
 \\
\upshift (c)&&
\includegraphics[width=\fval\textwidth]{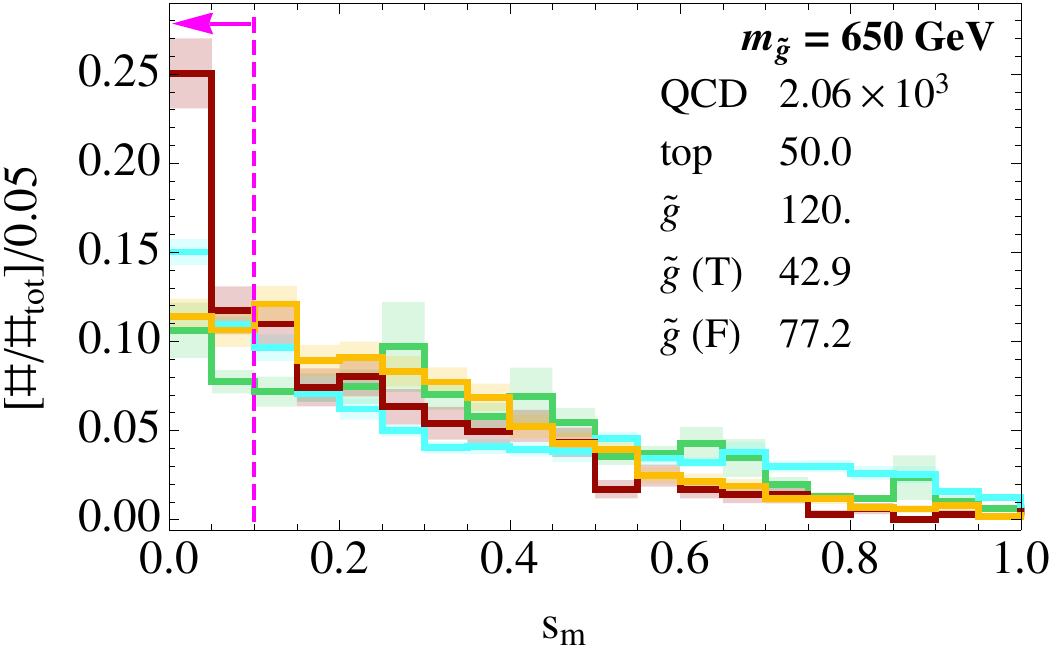}
&&
\upshift(d)&&
\includegraphics[width=\fval\textwidth]{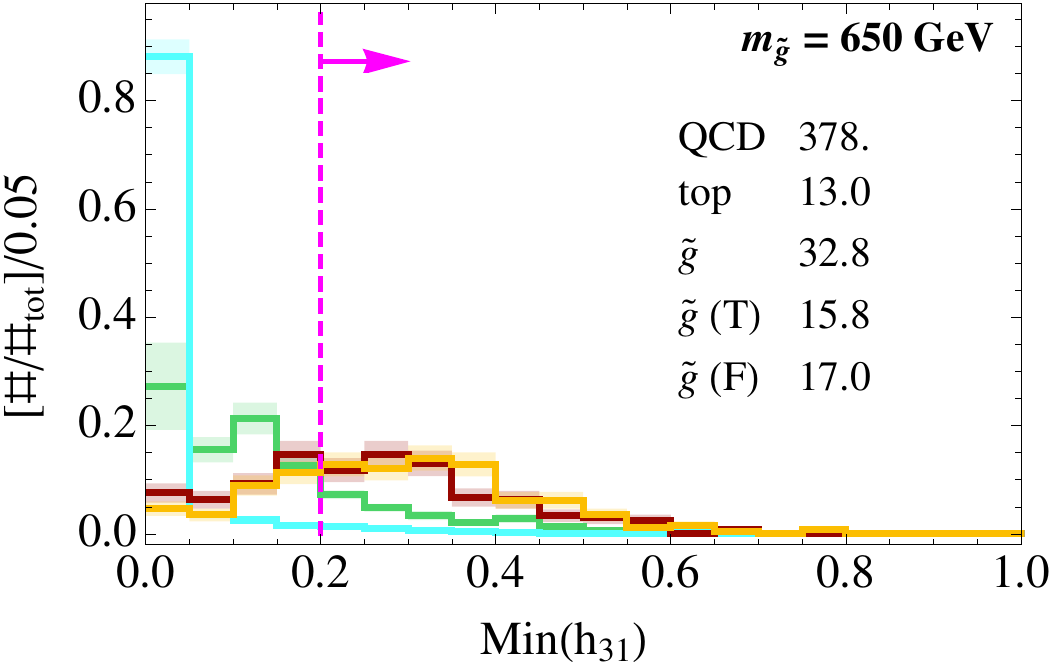}
 \vspace{2mm}
 \\
\upshift (e)&&
 \includegraphics[width=\fval\textwidth]{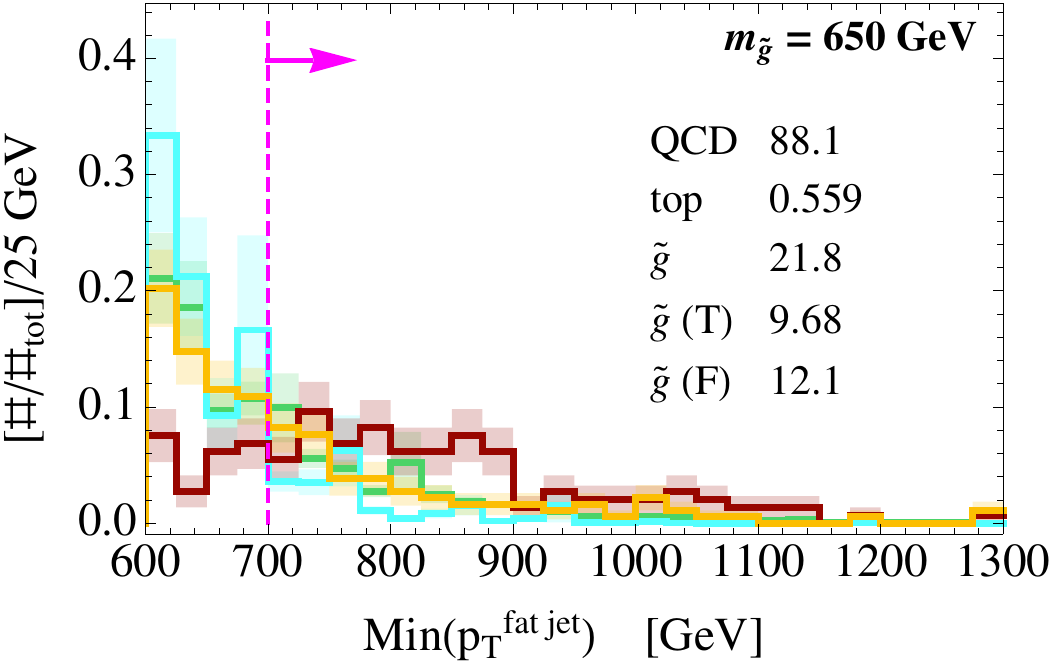}
&&
\upshift(f)&&
\includegraphics[width=\fval\textwidth]{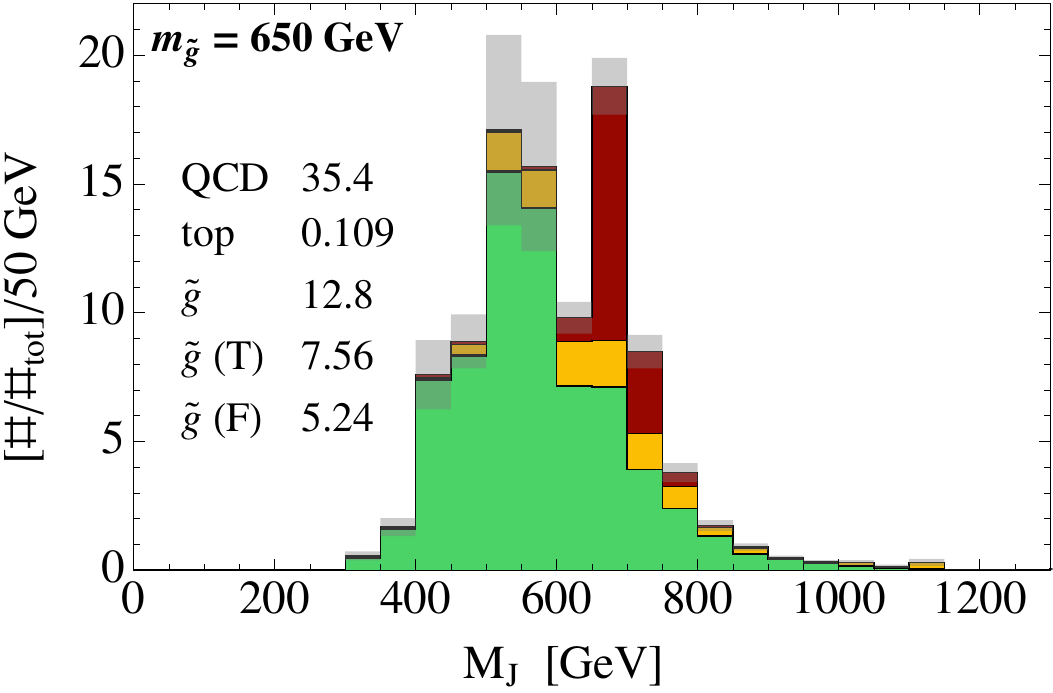}
 \vspace{1mm}
 \\
\upshift (g)  &&
 \includegraphics[width=\fval\textwidth]{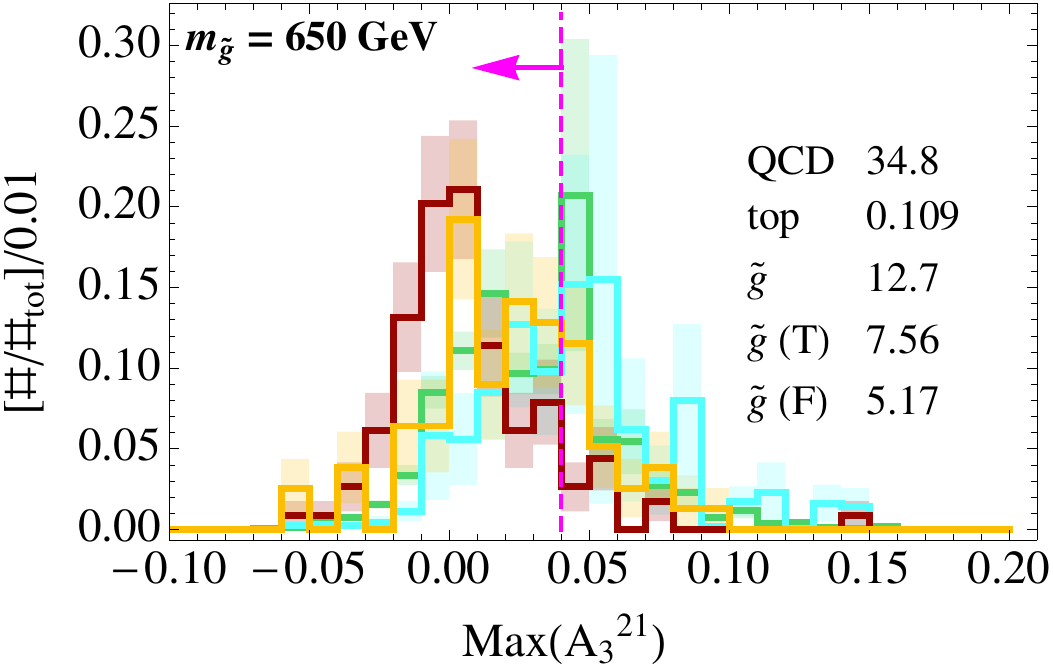}
&&
\upshift(h) &&
\includegraphics[width=\fval\textwidth]{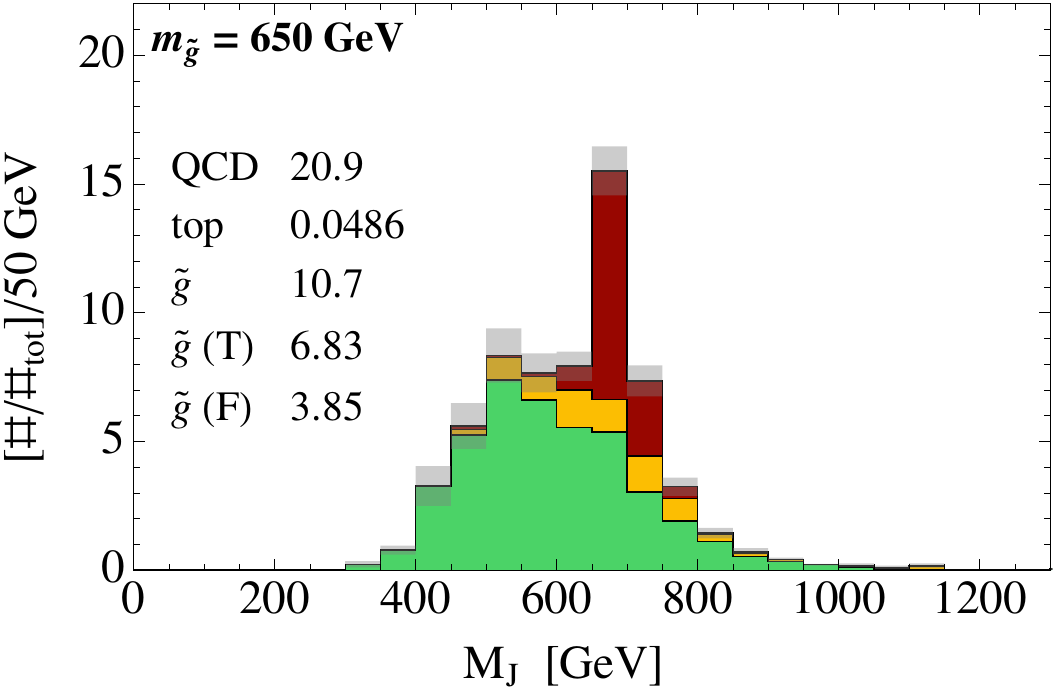}
\end{tabular}
&&
\begin{sideways}\includegraphics[height=0.55cm]{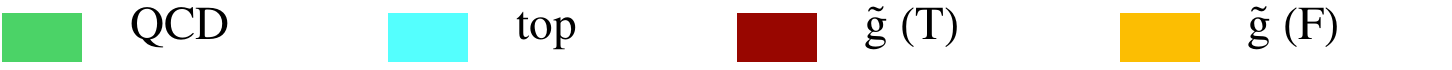}\end{sideways}
\end{tabular}
\end{center}
\vskip -0.6cm
\caption{
Distributions for kinematic and substructure variables at various stages in our chain of cuts 
for $m_{\tilde g} = 650$~GeV. Unstacked distributions are separately normalized to unity, while stacked distributions show actual number of events expected at LHC8 for 20 $\ifb$.
QCD ($\ttbar$) background is in green (cyan), while signal is in red (orange) for events whose fat-jet momenta are aligned 
within $\Delta R = 0.3$ of the gluino $R$-hadron momenta checked with MC truth. 
The shown distributions are 
(a) the invariant mass $M_J$ of \emph{both} hardest fat jets in each event (\emph{both} are counted in the histogram), for events containing two fat jets with $p_T > 600 \gev$ and $\tau_{32} < 0.7$,
(b) the maximum N-subjettiness $\tau_{32}$ (Eq.(\ref{e.nsubjettinesscut})) of the two fat jets, with the same cuts applied, 
(c) the jet-mass symmetry $s_{\rm m}$ (Eq.(\ref{e.masssymmetry})) after the cut max$(\tau_{32})<0.5$, 
(d) the minimum subjet hierarchy $h_{31}$ (Eq.(\ref{e.subjethierarchy})) of the two fat jets including the cut $s_{\rm m}<0.1$,
(e) the minimum fat-jet momentum $p_T^{\rm fat~jet}$  of the two fat jets including the cut min$(h_{31})>0.2$, 
(f) the resulting $M_J$, 
(g) the color variable {\it axis contraction} max($A_3^{21}$) (Eq.(\ref{e.axiscontraction})), and 
(h) the final $M_J$ including the cut max$(A_3^{21})<0.03$. 
MC statistical error is shown in faint-colored bands with the color corresponding to the respective signal or background component.  
In $M_J$ distributions, the gray error bands indicate the MC statistical error of the various components added in quadrature.  
A vertical magenta line indicates the applied cut, with the arrow pointing to the events that are kept.  
The inset shows the actual number of expected events at LHC8 with 20$\ifb$.
}
\label{f.cutflow.heavy}
\end{figure*}

 %%%%%%%%%%%%%%%%%%%%%%%%%%%%%%%%%%%%%%%%%%%%%%%%%%%%%%%%%%%%%%%%%%%%%%
%%%%%%%%%%%%%%%%%%%%%%%%%%%%%%%%%%%%%%%%%%%%%%%%%%%%%%%%%%%%%%%%%%%%%%%
\subsection{Results for $m_{\tilde g}\gtrsim 500 \gev$}
\label{ss.heavy} 
%%%%%%%%%%%%%%%%%%%%%%%%%%%%%%%%%%%%%%%%%%%%%%%%%%%%%%%%%%%%%%%%%%%%%%
%%%%%%%%%%%%%%%%%%%%%%%%%%%%%%%%%%%%%%%%%%%%%%%%%%%%%%%%%%%%%%%%%%%%%

\begin{figure*}
\begin{center}
\begin{tabular}{m{16cm}m{5mm}}
\begin{tabular}{ccc}
Without axis-contraction cut& & With axis-contraction cut\\\\
\includegraphics[width=\fval\textwidth]{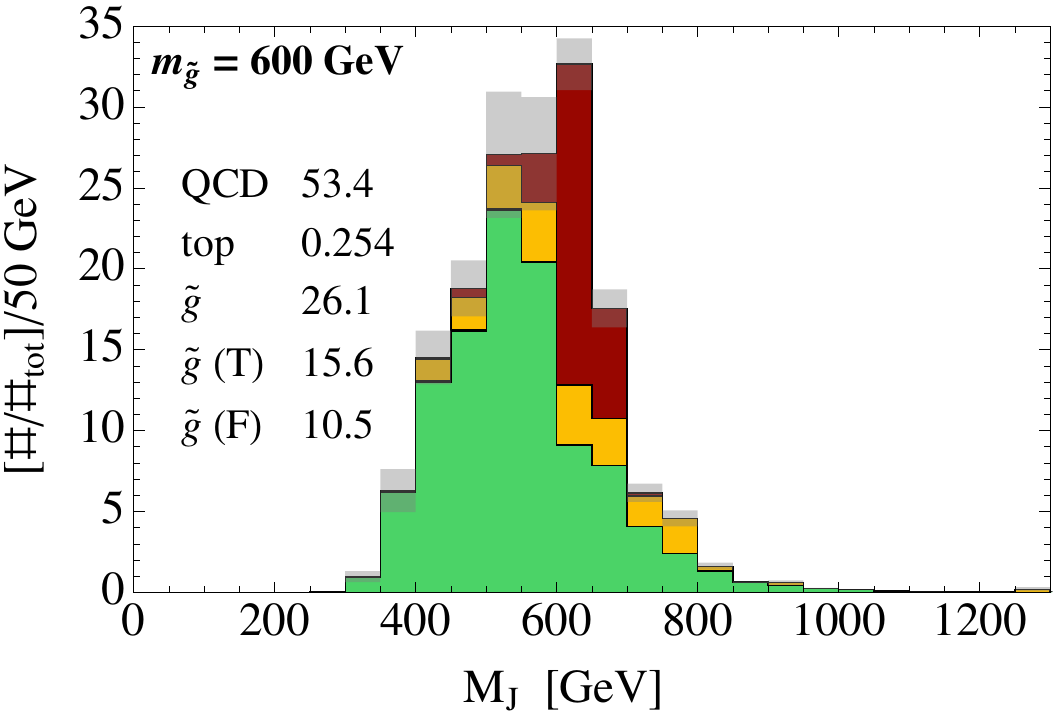}
\vspace{3mm}
&\phantom{blaba}&
\includegraphics[width=\fval\textwidth]{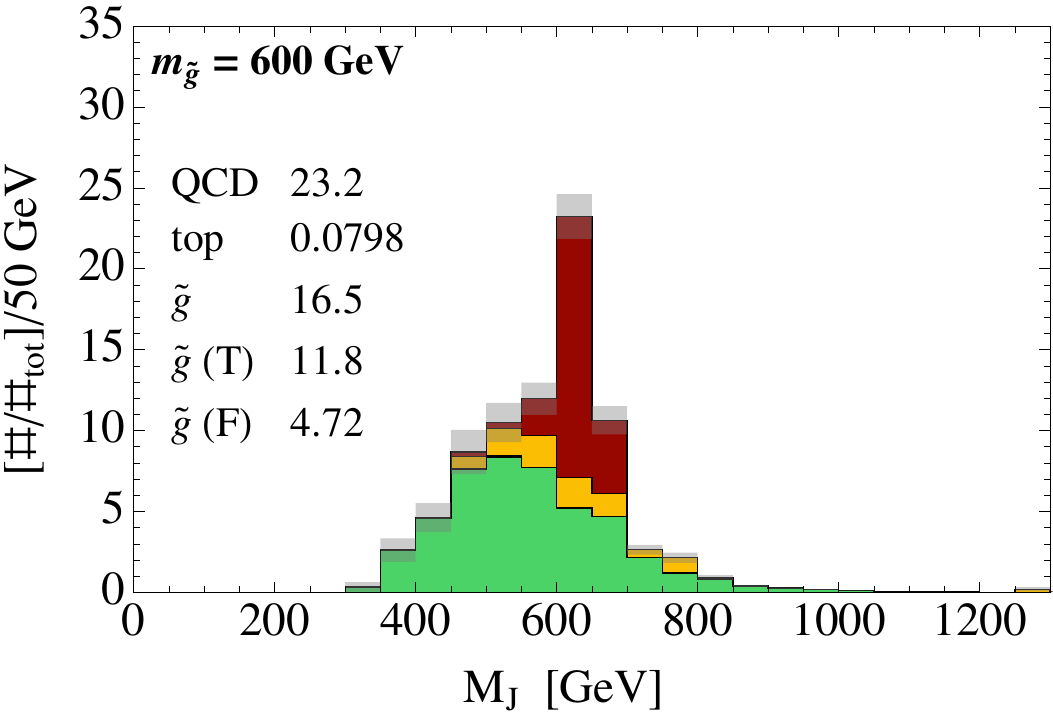}
\\
\includegraphics[width=\fval\textwidth]{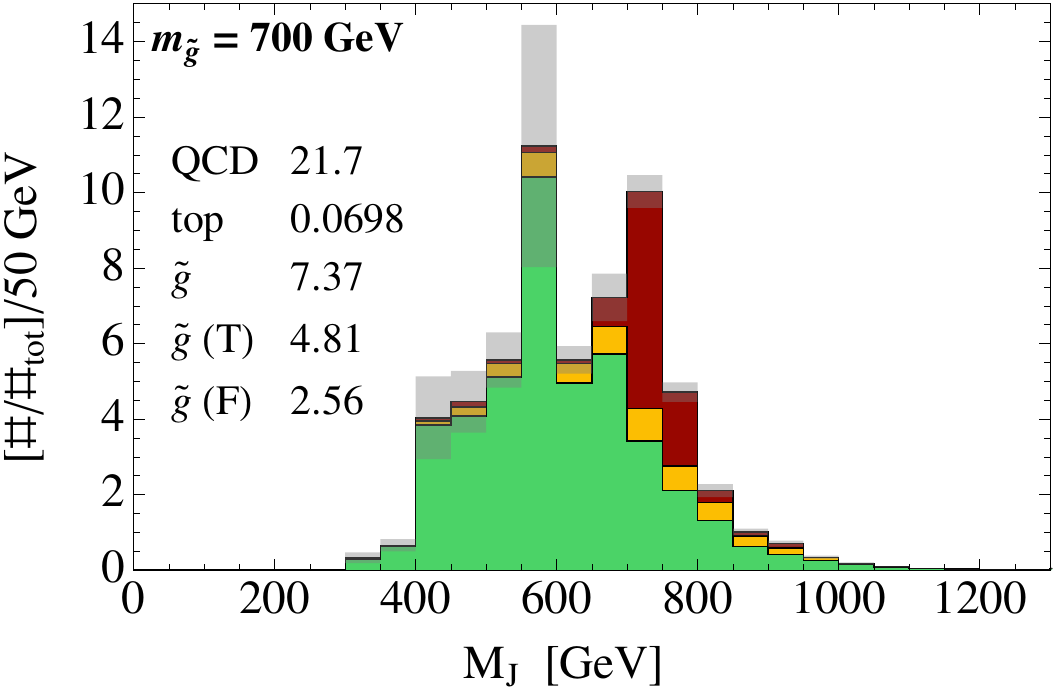}
\vspace{3mm}
&\phantom{blaba}&
\includegraphics[width=\fval\textwidth]{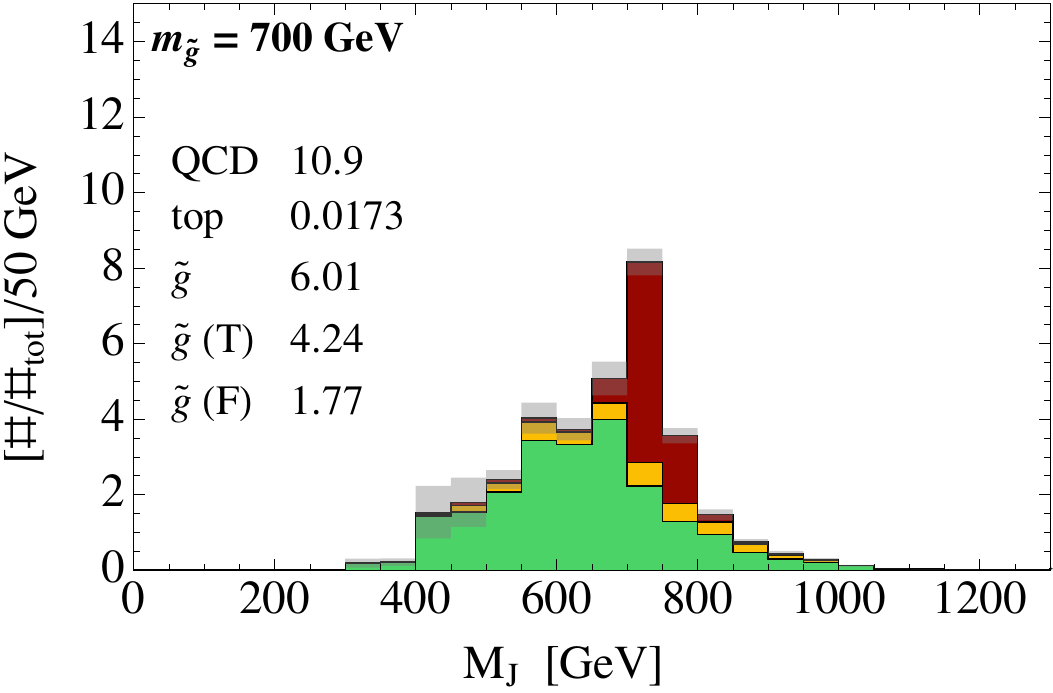}
\\
\includegraphics[width=\fval\textwidth]{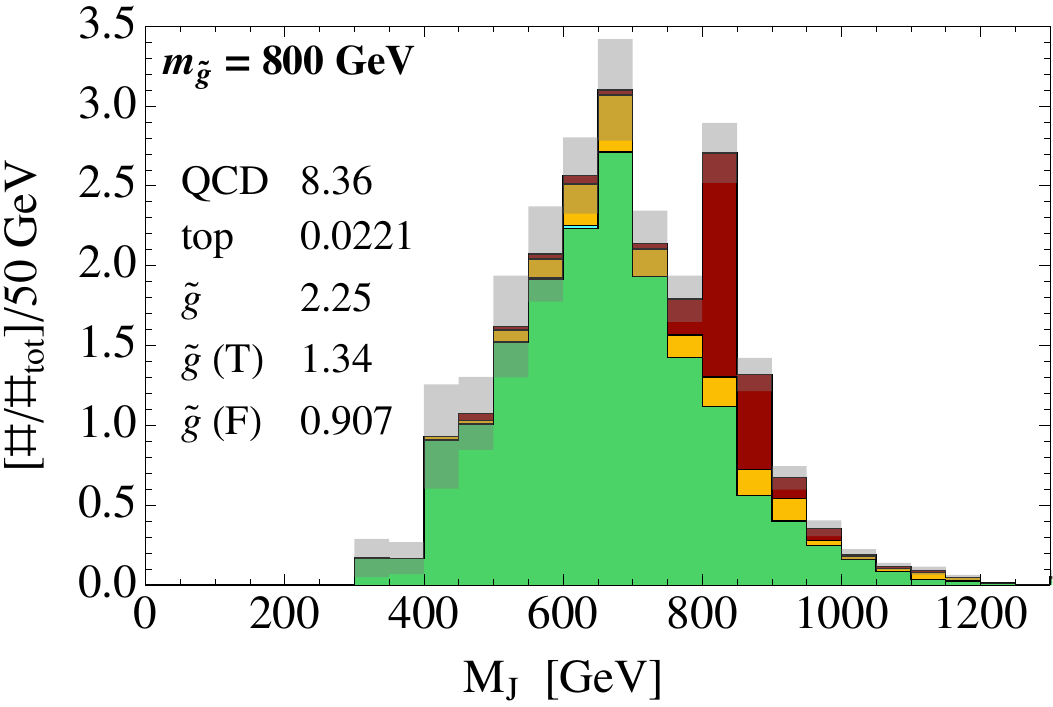}
\vspace{3mm}
&\phantom{blaba}&
\includegraphics[width=\fval\textwidth]{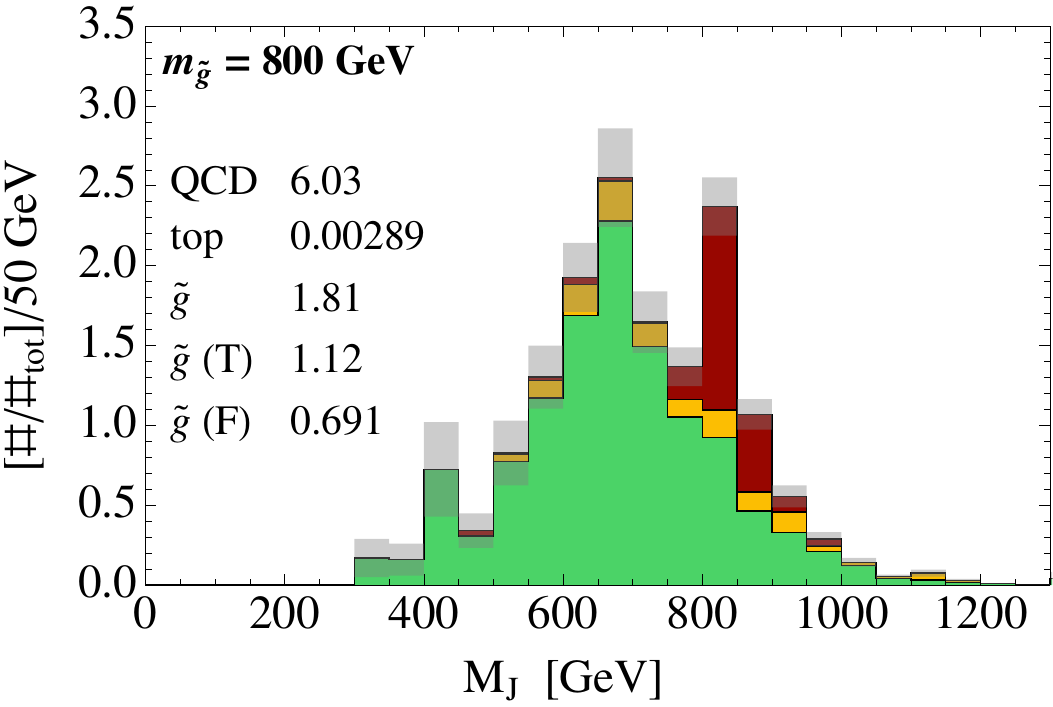}
\end{tabular}
&
\begin{sideways}\includegraphics[height=0.55cm]{legend1.pdf}\end{sideways}
\end{tabular}
\end{center}
\caption{
From top to bottom, stacked fat-jet-mass distributions of signal and background for $m_{\tilde g} = 600, 700, 800 \gev$, similar to \fref{cutflow.heavy} (f) and (h). Plots on the left include the final \emph{axis contraction} cut, while plots on the right do not.
}
\label{f.cutflow.all.heavy}
\end{figure*}

\begin{figure*}
\begin{center}
\includegraphics[width=\fval\textwidth]{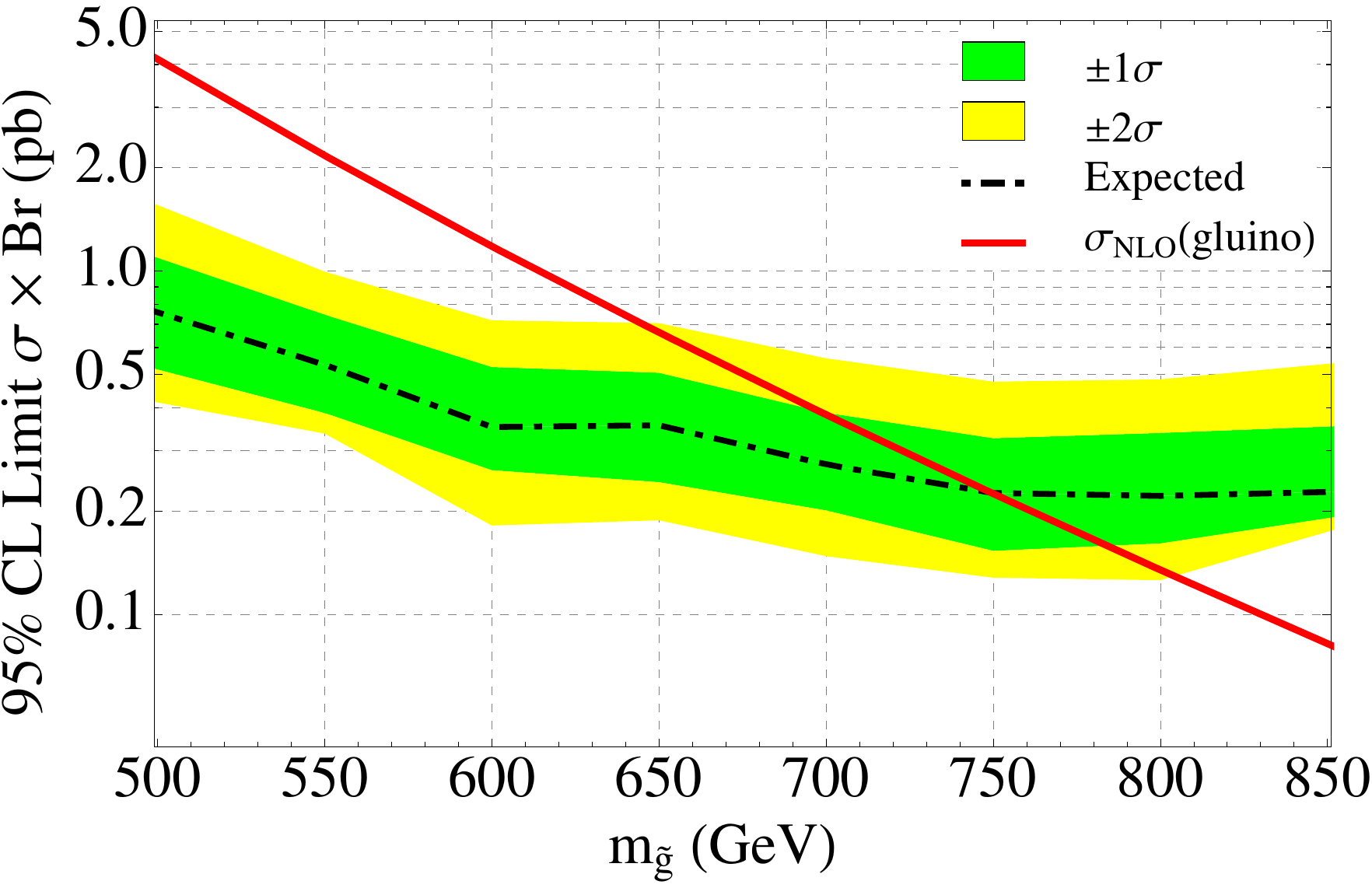}
\hspace{1cm}
\includegraphics[width=\fval\textwidth]{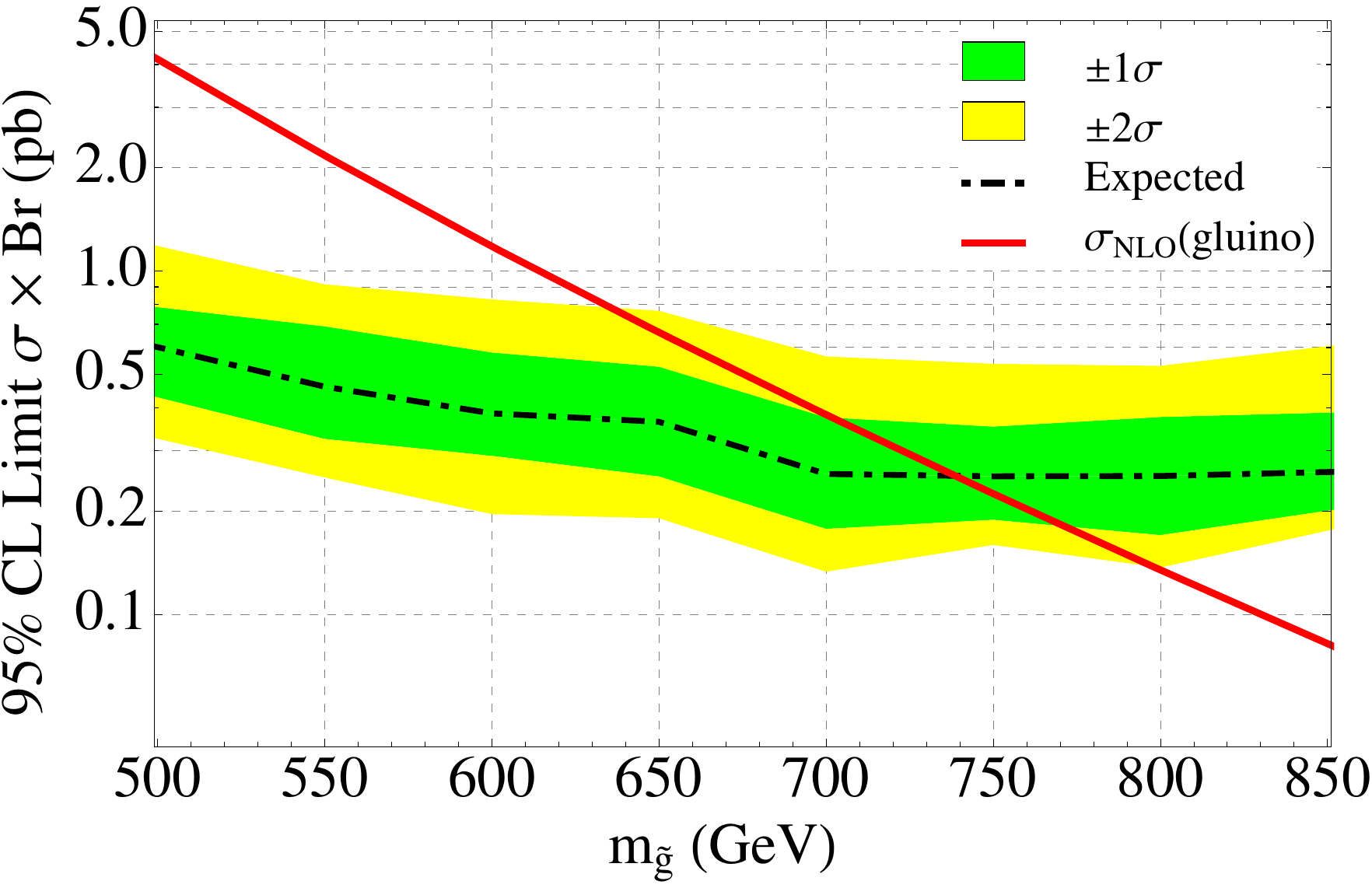}
\end{center}
\caption{Expected 95\% confidence-level cross section exclusion (and its $\pm1\sigma$ and $\pm 2 \sigma$ fluctuations) obtained with the $CL_\mathrm{s}$ method \cite{CLsmethod} for different gluino masses at LHC8 with $20~\ifb$ of data. The left (right) exclusions were derived without (with) color-flow cuts on axis contraction (Eq.~(\ref{e.axiscontraction})), with a floating background normalization in all fits to pseudodata. In both cases the reach is $\sim 750 \gev$, though the color-flow cuts improve exclusion at lower masses and improve the resonance shape. Fixing the background normalization increases the reach by $\sim 20 \gev$ (not shown). For $5 \ifb$, the reach is $\sim 650 \gev$ (not shown). 
}
\label{f.heavy.reach}
\end{figure*}

\renewcommand{\tilt}[1]{\begin{sideways}#1\end{sideways}}
\renewcommand{\cutentryfirst}[2]{\begin{tabular}{r}$#1$\\$#2$\end{tabular}}
\renewcommand{\cutentry}[2]{\begin{tabular}{r}$#1$\\$#2$\%\end{tabular}}
\renewcommand{\cutentryTF}[6]{\begin{tabular}{r}$#1$ \\ $(#2)$ \\$#4 \%$ \\ $(#5  \%)$\end{tabular}}
\renewcommand{\cutentrylast}[3]{\begin{tabular}{r} $#1$ \\ $#2$\% \\ $S/B = #3$ \end{tabular}}
\renewcommand{\cutentrylastTF}[7]{\begin{tabular}{r}$#1$\\ $(#2)$ \\$#4 \%$ \\ $(#5  \%$) \\ $S/B = #7$ \end{tabular}}
\renewcommand{\gluinoheader}[1]{\begin{tabular}{r}Gluinos\\ $[#1]$\end{tabular}}
\begin{table*}
\footnotesize
\begin{center}
\begin{tabular}{|l|r||r|r|r|r||r|r|r||r|r|}
\hline
\multicolumn{2}{|m{1cm}||}{Analysis}
&
\tilt{generator} \tilt{level} \tilt{cuts}
&
\tilt{Trigger:} \tilt{6 thin jets with} \tilt{$p_T > 60 \gev$}
& 
\tilt{two fat jets} \tilt{with} \tilt{$p_T > 200 \gev$}
&
\tilt{$\mathrm{max}(\tau_{32}) < 0.7$}
&
\tilt{$\mathrm{max}(\tau_{32}) < 0.5$}
&
\tilt{$\mathrm{max}(s_{\rm m}) < 0.1$}
&
\tilt{$\mathrm{min}(h_{31}) > 0.5$}
&
\tilt{radial pull cut} 
\tilt{$\mathrm{max}(t_{\rm r}) < -0.6$}
&
\tilt{$b$-veto}
\\ \hline \hline

\multirow{3}{*}{\tilt{$m_{\tilde g} = 175 \gev$}}
& QCD 
& \cutentryfirst{2.5 \times 10^8}{-}
& \cutentry{1.9 \times 10^6}{0.78}
& \cutentry{1.7 \times 10^6}{88}
& \cutentry{4.1 \times 10^5}{24}
& \cutentry{2.0 \times 10^4}{4.8}
& \cutentry{4.5 \times 10^3}{24}
& \cutentry{488}{11}
& \cutentry{0.86}{0.18}
& \cutentry{0.81 \pm 0.24}{94}
\\ 

\cline{2-11}
& top
& \cutentryfirst{1.1 \times 10^6}{-}
& \cutentry{2.6 \times 10^4}{2.4}
& \cutentry{2.3 \times 10^4}{88}
& \cutentry{1.2 \times 10^4}{51}
& \cutentry{2.2 \times 10^3}{19}
& \cutentry{771}{36}
& \cutentry{253}{33}
& \cutentry{5.6}{2.2}
& \cutentry{0.48 \pm 0.25}{8.6}
\\

\cline{2-11}
&
\gluinoheader{1.9 \mathrm{nb}}
& \cutentryfirst{3.7 \times 10^7}{-}
& \cutentry{1.1 \times 10^6}{2.9}
& \cutentry{9.5 \times 10^5}{90}
& \cutentry{3.3 \times 10^5}{34}
& \cutentryTF{2.0 \times 10^4}{6.6 \times 10^3}{1.3 \times 10^3} {6.1}{9.9}{5.2}
& \cutentryTF{6.1 \times 10^3}{3.2 \times 10^3}{2.8 \times 10^3} {30}{49}{21}
& \cutentryTF{1.5 \times 10^3}{1.0 \times 10^3}{460} {24}{31}{16}
& \cutentryTF{128}{118}{10} {8.7}{12}{2.3}
& \cutentrylastTF{120 \pm 17}{111 \pm 15}{9.8 \pm 6.6}{94}{94}{94}{93}
\\ \hline \hline

\end{tabular}
\end{center} \vspace{-3mm}
\caption{Number of expected events at LHC8 with 20 $\ifb$ for signal and \texttt{Sherpa} background after each performed cut for the top-mass gluino analysis, and gluino pair production cross section \cite{prospino} in square brackets. The generator-level cuts outlined in \sref{montecarlo} are only applied to the background samples. ``max'' and ``min'' in the cut variables applies to the two values obtained for the two hardest fat jet in each event. The percentage in each cell is the efficiency of that column's individual cut step. For signal, the numbers in brackets refer to the ``good'' combinatorially correct signal defined in \sref{montecarlo}. The cut chain after the second vertical double-line is illustrated in \fref{cutflow.light}. Note that the efficiency of the ``bad'' signal for the color-flow cut is only 2.3\%, compared to 12\% for the ``good'' signal. See Table \ref{f.LIGHTcompare} for a comparison to other Monte Carlo generators.
}
\label{f.LIGHTcuts}
\end{table*}

Table~\ref{f.HEAVYcuts} details our cuts, together with the signal and background efficiencies for $m_{\tilde g} \in [500, 850] \gev$ and the expected number of events at LHC8 at 20 $\ifb$. We note, firstly, that due to the generator-level cuts outlined in \sref{montecarlo}, the  event numbers for background samples can only be compared to actual data numbers after the third cut, which requires two fat jets with $p_T > 600 \gev$. Secondly, even though we classify the entire gluino sample as `signal', only the $\mathcal{O}$(few \%) boosted fraction that ends up in a fat jet that  reconstructs the gluino mass is of interest. Therefore, while the signal efficiency is extremely low, a more relevant figure-of-merit is the $\sim 10\%$ final acceptance of that boosted fraction. The cuts are extremely effective at reducing background from $\sim 10^6$ events after the first fat-jet $p_T$ cut to $\sim 10$ events after all other cuts, and we can reconstruct gluino resonances as heavy as $\sim 750 \gev$. 
Finally, the last two cuts have been optimized for each $m_{\tilde g}$, mostly to target the boosted fraction ($p_T\gtrsim m_{\tilde g}$) and make the cut on axis contraction \eref{axiscontraction} more conservative for larger $m_{\tilde g}$ where there is less (boosted) signal available.  Interestingly, for the conservative cuts necessary to retain high signal efficiency in the heavy gluino case, axis contraction  turns out to be slightly better-suited than radial pull, which is why we use the former in spite of the latter's superior signal separation on the tail of the distribution.

\fref{cutflow.heavy} illustrates the cut chain following the first fat-jet $p_T$ and $\tau_{32}$ cuts, for $m_{\tilde g} = 650 \gev$. As can be seen from the jet-mass distribution (a), the QCD background (green) completely dominates over both top background (cyan) and signal (split into `good' signal in red, where the hardest two fat jets are within $\Delta R = 0.3$ of the gluinos, and the rest in orange). Each unstacked histogram shows the unity-normalized distributions of a variable before cutting on it, for each signal and background component. The cut is indicated with a dashed magenta line, with events in the direction of the arrow kept for the next cut. (f) and (h) show the fat-jet-mass distribution before and after a conservative cut on axis contraction, shown in (g), is applied. (For a distribution of axis contraction at the same stage of cuts as (a), (b), see \fref{color} (b)). The color cut loses very little signal but reduces background by almost one-half, and significantly improves the shape of the distribution, making the gluino peak stand out very clearly. This is also illustrated in \fref{cutflow.all.heavy} for a few other gluino masses. According to the MC comparison in Appendix \ref{s.MCcomparison}, the cuts on jet-substructure and color-flow observables in the heavy gluino analysis do not vary substantially among different generators, and the analysis is expected to be consistent within a small factor.

To estimate our analysis' mass reach, we follow the maximum likelihood procedure for a shape analysis outlined in \cite{Cowan:2010js}, 
but using the $CL_\mathrm{s}$ method~\cite{CLsmethod}. This involves using our predictions for the final jet-mass distributions, e.g.~\fref{cutflow.all.heavy}, to produce large collections of pseudodata for the `background-only' and the `background + signal' hypothesis, where the signal is scaled by some overall signal strength $\mu$. (We ignore the statistical uncertainties of the MC prediction, as for such low event rates the 
Poisson fluctuations dominate.) By comparing the distributions of the $CL_\mathrm{s}$ test statistic computed for those pseudodata sets one can arrive at an expected signal strength exclusion, as well as the exclusion's $\pm 1\sigma$ and $\pm 2 \sigma$ fluctuations. The background normalization was allowed to float in each of these fits individually, making this a true shape analysis. 

The resulting expected cross-section exclusions are compared to the gluino production cross section at LHC8 with $20~\ifb$ in \fref{heavy.reach}, both with and without the axis-contraction cut. The expected reach is $\sim 750 \gev$ and could be as high as 800 GeV, surprisingly high for a fully boosted search.  With 5 $\ifb$ of LHC8 data the reach is $m_{\tilde g} \sim 650 \gev$, higher than the 
LHC7 CMS search~\cite{CMSrpvgluinobound5} and comparable to the ATLAS \emph{resolved} search~\cite{ATLASrpvgluinobound5}, which was a pure counting experiment. Since the boosted analysis has $S/B\sim1$ and is predominantly limited by statistics, 
more data should lead to better mass exclusion using the boosted analysis techniques discussed here.

Due to the small number of events surviving all the cuts, our estimate of the mass reach is not actually increased by the color-flow cuts, though the excluded cross section is increased for small $m_{\tilde g}$, where more events survive all cuts. Furthermore, the color-flow cuts do decrease the background at little cost to the signal and may be useful to control systematic uncertainties on the background. 

Exclusion could be improved by studying the fat-jet-mass distribution in control samples where no signal events are expected; a similar procedure was applied to the three-jet-mass distribution by CMS~\cite{CMSrpvgluinobound36, CMSrpvgluinobound5}, effectively fixing the background normalization. This would increase our mass reach by $\sim 20 \gev$, the small improvement being indicative that our simple shape analysis already does well in fixing the background normalization.  

\begin{figure*}
\newcommand{\upshift}{\vspace*{-26mm}}
\vspace*{-4mm}
\begin{center}
\hspace*{-3mm}
\begin{tabular}{m{17cm}m{3mm}m{5mm}}
\begin{tabular}{p{5mm}p{1mm}p{\fval\textwidth}p{4mm}p{5mm}p{1mm}p{\fval\textwidth}}
\upshift (a)&&
\includegraphics[width=\fval\textwidth]{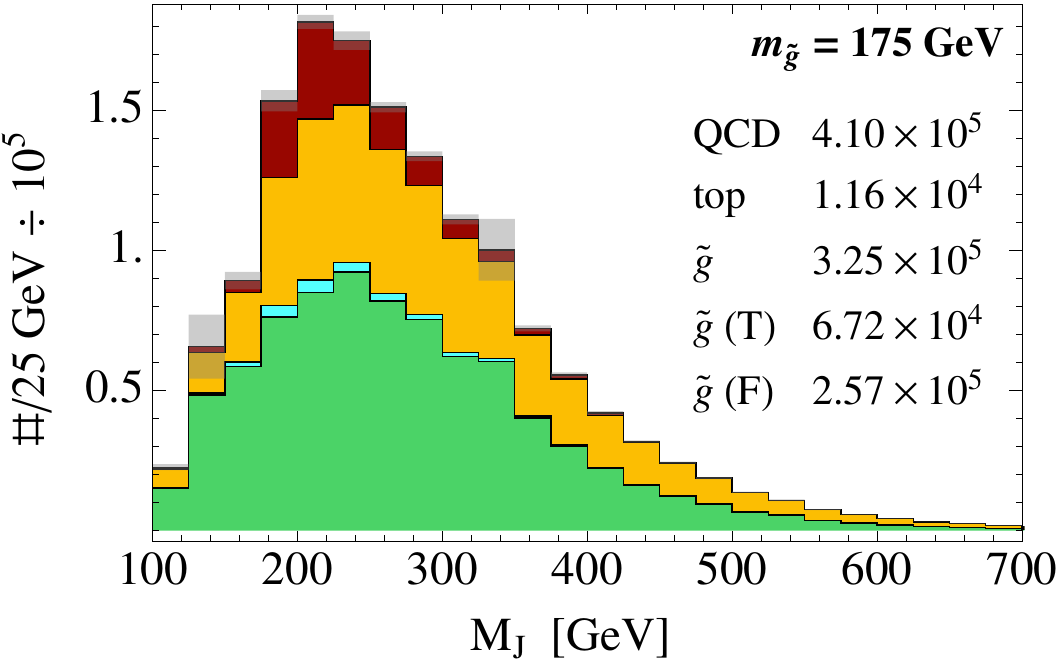}
&&
\upshift(b) &&
\includegraphics[width=\fval\textwidth]{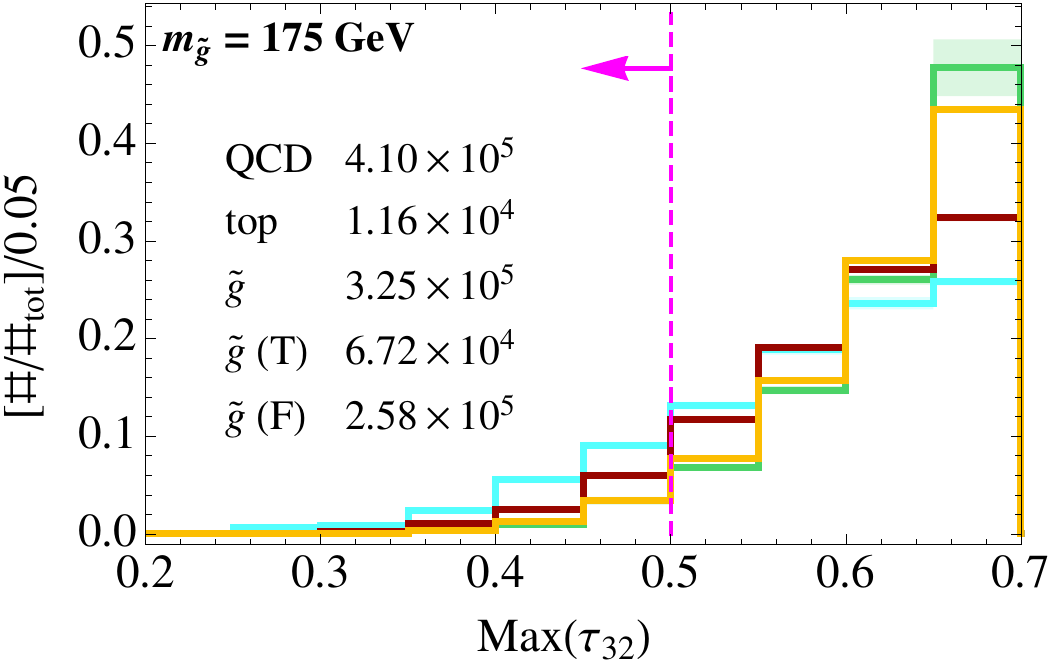}
 \vspace{2mm}
 \\
\upshift (c)&&
\includegraphics[width=\fval\textwidth]{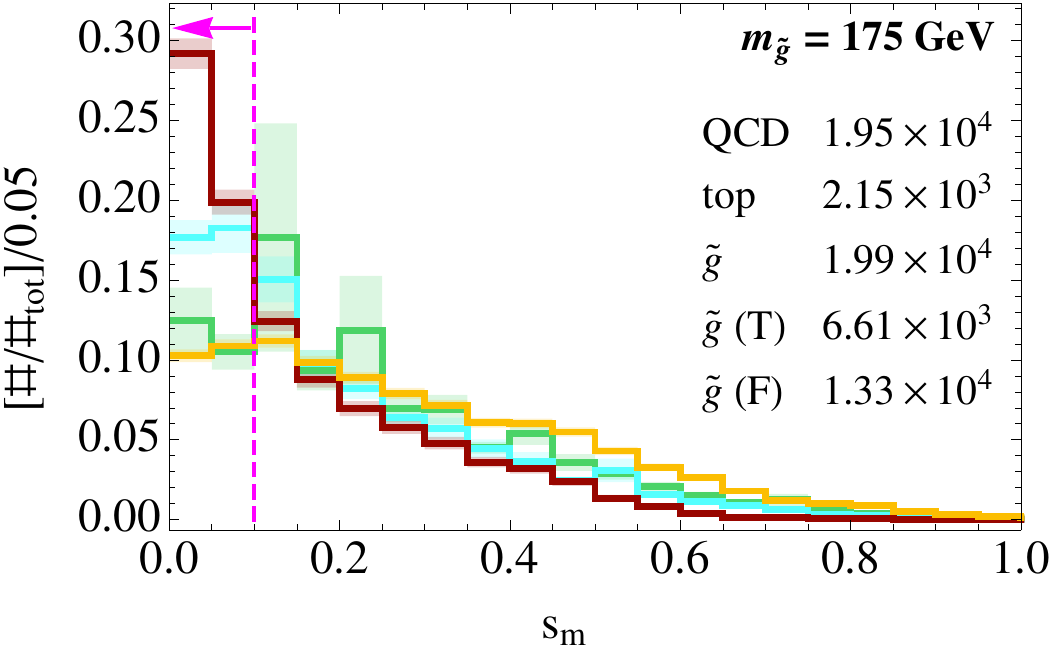}
&&
\upshift(d)&&
\includegraphics[width=\fval\textwidth]{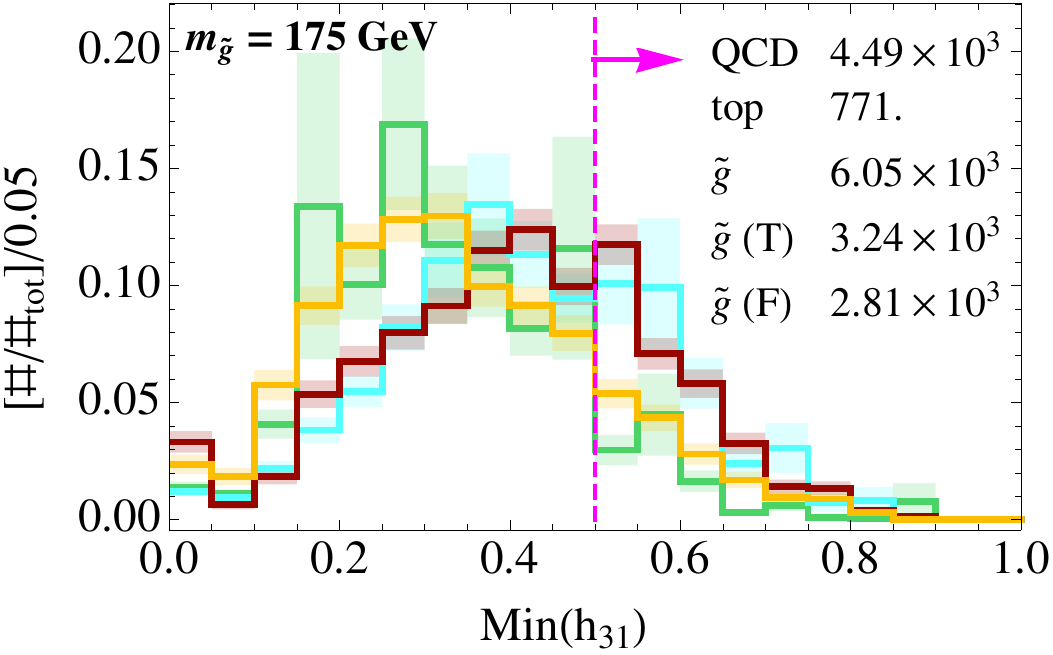}
 \vspace{2mm}
 \\
\upshift (e)&&
 \includegraphics[width=\fval\textwidth]{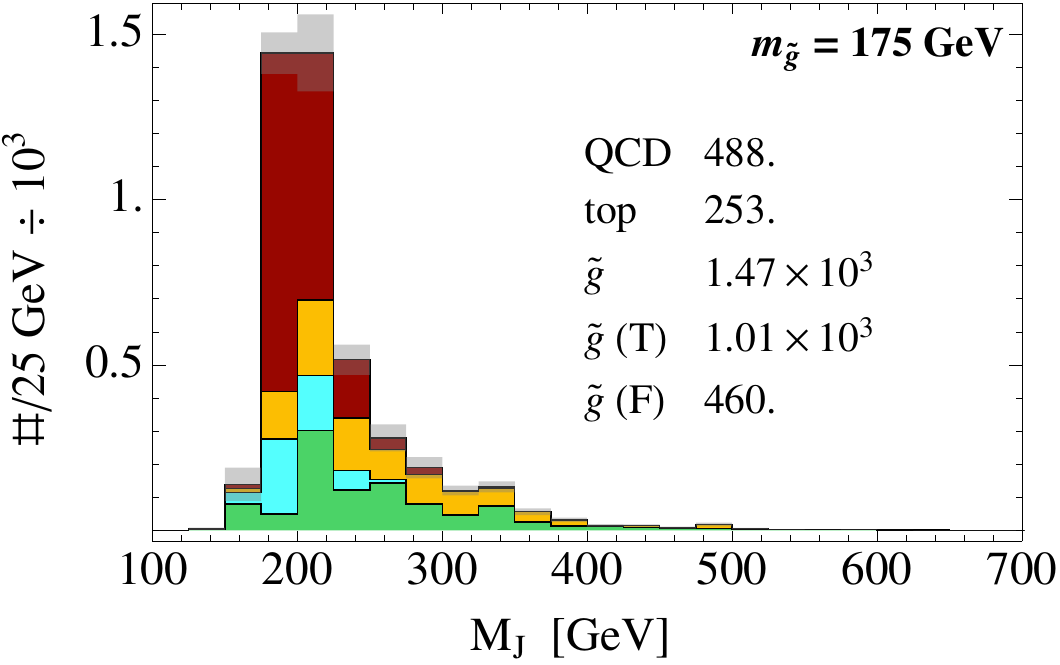}
&&
\upshift(f)&&
\includegraphics[width=\fval\textwidth]{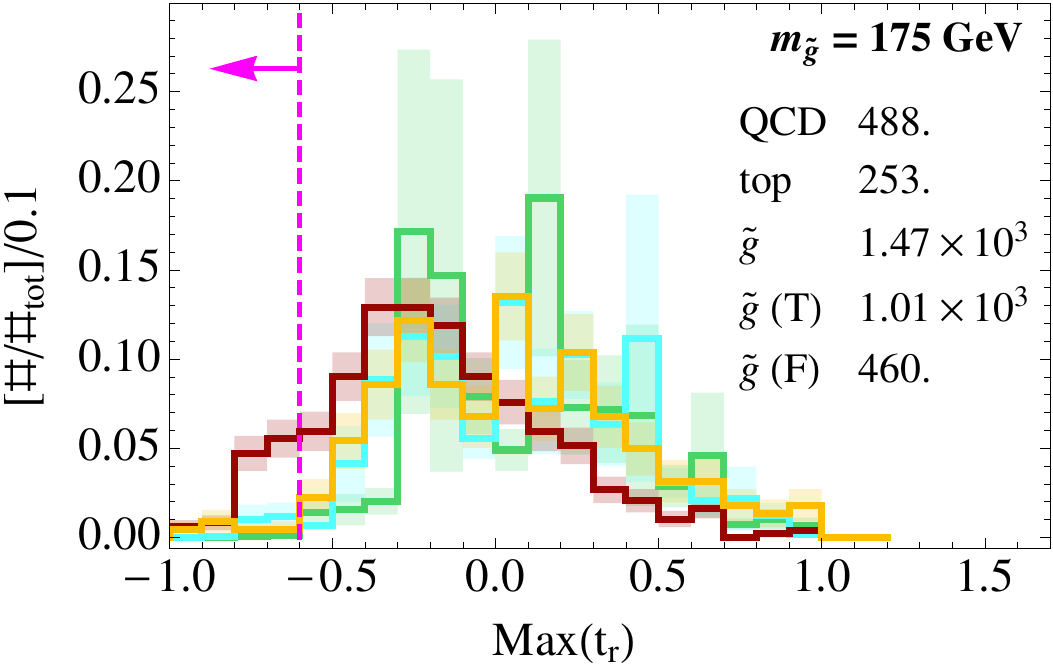}
 \vspace{1mm}
 \\
\upshift (g)  &&
 \includegraphics[width=\fval\textwidth]{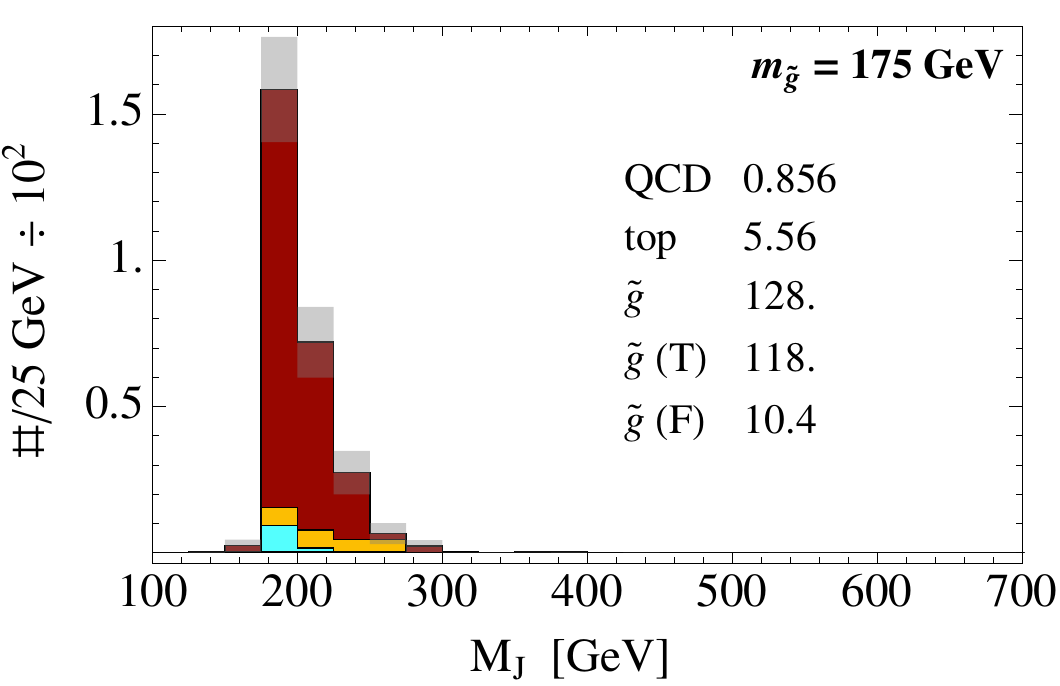}
&&
\upshift(h) &&
\includegraphics[width=\fval\textwidth]{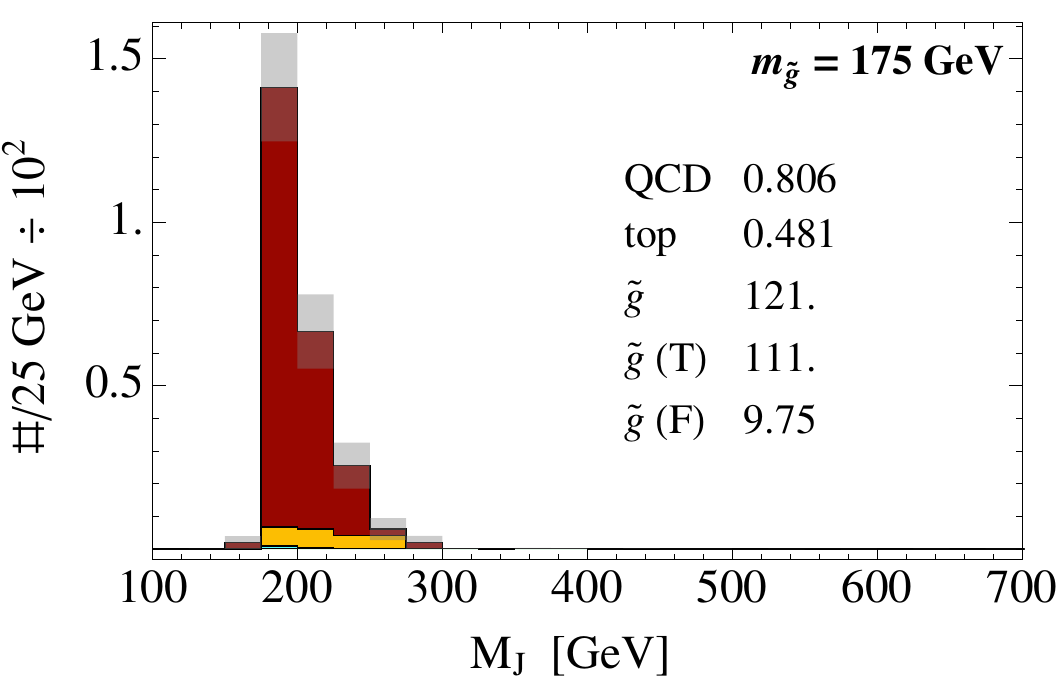}
\end{tabular}
&&
\begin{sideways}\includegraphics[height=0.55cm]{legend1.pdf}\end{sideways}
\end{tabular}
\end{center}
\vskip -0.4cm
\caption{
Distributions for kinematic and substructure variables at various stages in our chain of cuts 
for $m_{\tilde g} = 175$~GeV.  Unstacked distributions are separately normalized to unity, while stacked distributions show actual number of events expected at LHC8 for 20 $\ifb$.
QCD ($\ttbar$) background is in green (cyan), while signal is in red (orange) for events whose fat-jet momenta are aligned 
within $\Delta R = 0.3$ of the gluino $R$-hadron momenta checked with MC truth. 
The shown distributions are 
(a) the invariant mass $M_J$ of \emph{both} hardest fat jets in each event, for events passing the 6-jet (60 GeV) trigger and containing two fat jets with $p_T > 200 \gev$ and $\tau_{32} < 0.7$,
(b) the maximum N-subjettiness $\tau_{32}$ (Eq.(\ref{e.nsubjettinesscut})) of the two fat jets, with the same cuts applied, 
(c) the jet-mass symmetry $s_{\rm m}$ (Eq.(\ref{e.masssymmetry})) after also max$(\tau_{32})<0.5$, 
(d) the minimum subjet hierarchy $h_{31}$ (Eq.(\ref{e.subjethierarchy})) of the two fat jets including also $s_{\rm m}<0.1$,
(e) the resulting $M_J$ distribution,
(f) the color variable {\it radial pull} $\mathrm{max}(t_{\rm r})$ (Eq.(\ref{e.radialpull})), 
(g) the $M_J$ distribution after the radial pull cut, and 
(h) the same distribution after a $b$-veto.
MC statistical error is shown in faint-colored bands with the color corresponding to the respective signal or background component.  
In $M_J$ distributions, the gray error bands indicate the MC statistical error of the various components added in quadrature.  
A vertical magenta line indicates the applied cut, with the arrow pointing to the events that are kept.  
The inset shows the actual number of expected events at LHC8 with 20$\ifb$ for each signal and background component. For a comparison of QCD backgrounds for jet-substructure observables from other MC programs, see Fig.~\ref{f.cutchaincompare}. 
}
\label{f.cutflow.light}
\end{figure*}

%%%%%%%%%%%%%%%%%%%%%%%%%%%%%%%%%%%%%%%%%%%%%%%%%%%%%%%%%%%%%%%%%%%%%%
%%%%%%%%%%%%%%%%%%%%%%%%%%%%%%%%%%%%%%%%%%%%%%%%%%%%%%%%%%%%%%%%%%%%%%%
\subsection{Results for $m_{\tilde g} \sim m_t$}
\label{ss.topmass} 
%%%%%%%%%%%%%%%%%%%%%%%%%%%%%%%%%%%%%%%%%%%%%%%%%%%%%%%%%%%%%%%%%%%%%%
%%%%%%%%%%%%%%%%%%%%%%%%%%%%%%%%%%%%%%%%%%%%%%%%%%%%%%%%%%%%%%%%%%%%%

While the recent ATLAS search for RPV gluinos excluded the mass range  $m_{\tilde g}\sim140-200$~GeV \cite{ATLASrpvgluinobound5}, we still consider gluinos in this mass window as they provide a large sample of highly boosted signal events with which we can demonstrate the full discriminating power of our color-flow variables. Furthermore, we show that a dramatically higher signal purity than in \cite{ATLASrpvgluinobound5} can be achieved, which translates to a correspondingly higher cross section exclusion. 

We find that the trigger requiring six jets with $\pt>60$~GeV yields the highest signal efficiency and tends to give boosted gluinos. Due to the abundance of signal we can cut hard on the substructure and color-flow variables. The cuts are outlined in Table~\ref{f.LIGHTcuts} and illustrated in \fref{cutflow.light}. While axis contraction was better suited for the soft cuts on color flow required by low signal numbers in the heavy gluino analysis, radial pull is more effective when we have enough signal to cut more aggressively. The result is an extremely powerful color-flow cut that keeps 12\% of the combinatorially correct signal (of events surviving the previous cut step) but only 2\% of combinatorially incorrect signal or $t\bar{t}$ background, and only 0.2\% of \texttt{Sherpa} QCD background. The resulting mass distribution \fref{cutflow.light}(g) is very signal-dominated and displays a  clean resonance at 175 GeV \footnote{We repeated this analysis for a 145 GeV gluino and found that the peak is indeed shifted appropriately to the left. This was to ensure that we are actually seeing a resonance, and not a falling jet-mass distribution shaped by the trigger and cuts.}. Applying a $b$-veto (assuming 70\% tagging efficiency for $b$ quarks and  a 1\% light-jet mistag rate \cite{ATLASbtagger}) yields \fref{cutflow.light}(h), with a resonance peak of $\sim 100$ gluino events and only about 1 background event in total. For other MC programs, as described in Appendix \ref{s.MCcomparison}, $S/B\sim10$, which is smaller than \texttt{Sherpa} but still shows the discriminatory power of color-flow observables. 

It is instructive to compare the radial-pull distribution after the kinematic substructure cuts \fref{cutflow.light}(f) to \fref{color} (left), which shows the same distribution before many of the kinematic and substructure cuts are applied. The similarity in the distributions (accounting for increased statistical uncertainty of the MC predictions after more cuts) suggests that  radial pull is relatively uncorrelated with other substructure variables, and provides genuinely new discriminating power that cannot be accessed by simply making another cut more aggressive. 

The counting experiment performed by ATLAS at LHC7 with 5 $\ifb$ has excluded top-mass gluinos with $\mathrm{Br}(\tilde g\rightarrow q q q)\lesssim 0.25$. Treating our nearly background-free sample in a similar way would exclude $\mathrm{Br}(\tilde g\rightarrow q q q)\lesssim0.15$ at LHC8 with the same luminosity, decreasing to $\mathrm{Br}(\tilde g\rightarrow q q q)\lesssim0.05$ at 20 $\ifb$. Apart from excluding other possible RPV spectra, this can be relevant for different color representations decaying to three jets, which have different production cross sections.  
The shape analysis serves as a useful check on any discovered signal, since a resonance is clearly constructed. Furthermore, if a three-jet resonance in this mass range were to be found, one could use the alternative version of radial pull, \eref{radialpull2}, to distinguish a hadronized color-octet from some other state, as illustrated in \fref{had.vs.nohad}.

%%%%%%%%%%%%%%%%%%%%%%%%%%%%%%%%%%%%%%%%%%%%%%%%%%%%%%%%%%%%%%%%%%%%%%
%%%%%%%%%%%%%%%%%%%%%%%%%%%%%%%%%%%%%%%%%%%%%%%%%%%%%%%%%%%%%%%%%%%%%%%
\section{Conclusions}
\label{s.conclusions}
%%%%%%%%%%%%%%%%%%%%%%%%%%%%%%%%%%%%%%%%%%%%%%%%%%%%%%%%%%%%%%%%%%%%%%
%%%%%%%%%%%%%%%%%%%%%%%%%%%%%%%%%%%%%%%%%%%%%%%%%%%%%%%%%%%%%%%%%%%%%

Jet-substructure variables are excellent tools in searches for boosted resonances. We have shown that such techniques can also be useful when applied to the boosted fraction of very heavy particles, such as RPV gluinos. Although this has recently been demonstrated in an LHC search for top-mass gluinos~\cite{ATLASrpvgluinobound5}, we propose such search strategies also for heavier masses at LHC8. We use existing variables, such as $N$-subjettiness, jet-mass symmetry, and subjet hierarchy, as well as our new color-flow variables, radial pull and axis contraction, to isolate a high-purity signal sample. Our suggested analysis strategy with aggressive cuts on these variables can be competitive with existing search strategies~\cite{rouventhesis,TEVATRONrpvgluinobound,CMSrpvgluinobound36, CMSrpvgluinobound5,ATLASrpvgluinobound5}, 
while providing nearly independent systematic uncertainties.  
With 20 $\ifb$ (5 $\ifb$) at LHC8, it should be possible to probe boosted gluino-like resonances around the top mass with $\sigma \times \mathrm{Br} \lesssim 0.3 \ (0.1) \  \mathrm{nb}$, corresponding to $\sim 0.05~(0.15)$ gluino branching fractions to three jets. 
Heavy gluinos can be excluded for $m_{\tilde g}\lesssim750$ $(650)$ GeV. 

Additionally, we have shown that radial pull and axis contraction~\footnote{While working on this paper we learned that M.~Freytsis, T.~Volansky, and J.~Walsh are independently exploring variables similar to axis shift, but in the context of boosted top tagging. Their results will appear shortly.} appear to be powerful variables for distinguishing processes with different color flows, particularly for highly boosted events. We anticipate that they have applications in studies of other boosted colored objects, and may prove useful in distinguishing different models in the case that a boosted resonance is discovered.

\subsection*{Acknowledgements}

We are grateful to T.~Cohen, M.~Freytsis, E.~Izaguirre, and M.~Lisanti for helpful discussions and 
comments on a draft version of this paper. 
We also thank M.~Begel, E.~Halkiadakis, J.~Hobbs, D.~Tsybychev, P.~Meade, D.~Miller, 
G.~Salam, J.~Shelton, G.~Sterman, J.~Thaler, and
J.~Wacker for helpful conversations.  We especially thank P.~Skands and T.~Sj\"ostrand for their help with \texttt{Pythia}, and 
S.~H\"oche and S.~Schumann for their help with \texttt{Sherpa}. 
We also thank an anonymous referee for several useful suggestions.
The work of D.C. was supported in part by the National Science Foundation under Grant PHY-0969739. 
RE is supported by the Department of Energy Early Career research program under Award Number DE-SC0008061. 
The work of B.S. was supported in part by the Harvard Center for the Fundamental Laws of Nature, the National Science Foundation under Grant PHY-0855591, and the Canadian Institute of Particle Physics. Research at the Perimeter Institute
is supported in part by the Government of Canada through Industry Canada, and by the
Province of Ontario through the Ministry of Research and Information (MRI).
The work of D.C. and R.E. was conducted in part at the Aspen Center for Physics, supported by the National Science Foundation under Grant No.~PHY-1066293.
Some of the numerical calculations in this paper were performed on
the Odyssey cluster supported by the FAS Research Group at Harvard University, as well as the facilities of the Shared Hierarchical Academic Research Computing Network 
(SHARCNET) and Compute/Calcul Canada.

\appendix

%%%%%%%%%%%%%%%%%%%%%%%%%%%%%%%%%%%%%%%%%%%%%%%%%%%%%%%%%%%%%%%%%%%%%%
%%%%%%%%%%%%%%%%%%%%%%%%%%%%%%%%%%%%%%%%%%%%%%%%%%%%%%%%%%%%%%%%%%%%%%%
\section{Review of RPV Gluinos}
\label{s.review} \setcounter{equation}{0} 
%%%%%%%%%%%%%%%%%%%%%%%%%%%%%%%%%%%%%%%%%%%%%%%%%%%%%%%%%%%%%%%%%%%%%%
%%%%%%%%%%%%%%%%%%%%%%%%%%%%%%%%%%%%%%%%%%%%%%%%%%%%%%%%%%%%%%%%%%%%%

In addition to the superpotential terms of the Minimal Supersymmetric Standard Model, Standard Model gauge invariance allows 
terms that violate baryon ($B$) and lepton ($L$) number.
$B$ and $L$ violation are strongly constrained by limits on proton decay and neutrino masses~\cite{RPVreview}. Often, $R$-parity 
is imposed to remove these dangerous terms.  However, if lepton number is conserved, the $B$-violating RPV term
\begin{equation}\label{e.W-rpv}
W_{\mathrm{RPV}}=\frac{1}{2}\lambda_{ijk}''\bar u_i\bar d_j\bar d_k
\end{equation}
 is allowed and only loosely constrained. Many theories accommodate RPV~\cite{Ruderman:2012jd,FileviezPerez:2009gr,Goity:1994dq,Chen:2010ss,MFVsusy}.   

Models with $\lambda''\neq0$ can be challenging to observe, since the hadronic decay of supersymmetric particles suffer large QCD backgrounds.
For RPV operators with one or more heavy-flavor quarks, multiple $b$-tags, leptons, and missing energy can help distinguish the signal from 
backgrounds~\cite{Brust:2012uf,Allanach:2012vj,Evans:2012bf}.  If the quarks are all light-flavored, however, no easy distinguishing 
property exists.

\renewcommand{\tilt}[1]{\begin{sideways}#1\end{sideways}}
\renewcommand{\cutentryfirst}[2]{\begin{tabular}{r}$#1$\\$#2$\end{tabular}}
\renewcommand{\cutentry}[2]{\begin{tabular}{r}$#1$\\$#2$\%\end{tabular}}
\renewcommand{\cutentryTF}[6]{\begin{tabular}{r}$#1$ \\ $(#2)$ \\$#4 \%$ \\ $(#5  \%)$\end{tabular}}
\renewcommand{\cutentrylast}[3]{\begin{tabular}{r} $#1$ \\ $#2$\% \\ $S/B = #3$ \end{tabular}}
\renewcommand{\cutentrylastTF}[7]{\begin{tabular}{r}$#1$\\ $(#2)$ \\$#4 \%$ \\ $(#5  \%$) \\ $S/B = #7$ \end{tabular}}
\renewcommand{\gluinoheader}[1]{\begin{tabular}{r}Gluinos\\ $[#1]$\end{tabular}}
\begin{table*}[tb!]
\footnotesize
\begin{center}
\begin{tabular}{|l|r||r|r|r|r|r||r|r|}
\hline
\multicolumn{2}{|m{1cm}||}{Generator}
&
\tilt{Preselection:} \tilt{6 thin jets with} \tilt{$p_T > 60 \gev$} \tilt{ and two fat jets} \tilt{with} \tilt{$p_T > 200 \gev$}
&
\tilt{$\mathrm{max}(\tau_{32}) < 0.5$}
&
\tilt{$\mathrm{max}(s_{\rm m}) < 0.1$}
&
\tilt{$\mathrm{min}(h_{31}) > 0.5$}
&
\tilt{radial pull cut} 
\tilt{$\mathrm{max}(t_{\rm r}) < -0.6$}
&
\tilt{$b$-veto}
\\ \hline \hline

\multirow{3}{*}{\tilt{Gluino}}
& \texttt{Pythia 8} 
& \cutentryfirst{9.5 \times 10^5}{-}
& \cutentry{3.3 \times 10^5}{34}
& \cutentry{6.1 \times 10^3}{30}
& \cutentry{1.5\times10^3}{24}
& \cutentry{128}{8.7}
& \cutentry{120 \pm 17}{94}
\\ &&&&&&& \\ &&&&&&& \\  \hline \hline

\multirow{3}{*}{\tilt{QCD}}
& \texttt{Sherpa} 
& \cutentryfirst{1.7 \times 10^6}{-}
& \cutentry{2.0 \times 10^4}{1.2}
& \cutentry{4.5 \times 10^3}{24}
& \cutentry{488}{11}
& \cutentry{0.86}{0.18}
& \cutentry{0.81 \pm 0.24}{94}
\\ 

\cline{2-8}
& \texttt{POWHEG + Pythia 6}
& \cutentryfirst{1.7 \times 10^6}{(normalized)}
& \cutentry{1.7 \times 10^4}{0.98}
& \cutentry{3.7\times10^3}{22}
& \cutentry{696}{19}
& \cutentry{8.4}{1.2}
& \cutentry{7.9^{+20}_{-6.4}}{94}
\\

\cline{2-8}
& \texttt{POWHEG + Pythia 8}
& \cutentryfirst{1.7 \times 10^6}{(normalized)}
& \cutentry{2.0 \times 10^4}{1.2}
& \cutentry{4.5 \times 10^3}{22}
& \cutentry{808}{18}
& \cutentry{8.9}{1.1}
& \cutentry{8.8^{+12}_{-5.5}}{94}
\\ \hline \hline

\multirow{3}{*}{\tilt{Top}}
& \texttt{Sherpa} 
& \cutentryfirst{2.3 \times 10^4}{-}
& \cutentry{2.2 \times 10^3}{9.7}
& \cutentry{771}{36}
& \cutentry{253}{33}
& \cutentry{5.6}{2.2}
& \cutentry{0.48 \pm 0.25}{8.6}
\\ 

\cline{2-8}
& \texttt{Madgraph 5 + Pythia 6}
& \cutentryfirst{2.3 \times 10^4}{(normalized)}
& \cutentry{2.0 \times 10^3}{8.9}
& \cutentry{900}{44}
& \cutentry{270}{30}
& \cutentry{14}{5.0}
& \cutentry{1.2\pm0.4}{8.6}
\\

\cline{2-8}
& \texttt{Madgraph 5 + Pythia 8}
& \cutentryfirst{2.3 \times 10^4}{(normalized)}
& \cutentry{1.2 \times 10^3}{5.3}
& \cutentry{451}{37}
& \cutentry{113}{25}
& \cutentry{5.1}{4.5}
& \cutentry{0.44\pm0.11}{8.6}
\\ \hline \hline

\end{tabular}
\end{center} \vspace{-3mm}
\caption{Comparison of expected events and cut efficiencies at LHC8 with 20 $\ifb$ for signal and background for different event generators at each stage of the cuts from Table \ref{f.LIGHTcuts}. The number of background events after preselection cuts are normalized to the corresponding \texttt{Sherpa} values to facilitate comparison. The first row for each sample shows the expected number of events at LHC8 with 20 $\ifb$, while the second row gives the cut efficiency. In the final column, the error bars are derived from the statistical uncertainty of the MC sample. 
}
\label{f.LIGHTcompare}
\end{table*}

The couplings $\lambda_{ijk}''$ are subject to various constraints for different $\{ijk\}$.  The dominant constraint on $\lambda_{112}''$ comes from nucleon-antinucleon oscillation through an intermediate strange squark and gluino. The precise bound depends on the off-diagonal entries of the unknown left-right strange squark mixing matrix. If there is no mixing suppression, the bound is $|\lambda_{112}''|\lesssim10^{-6}$ for  $m_{\tilde g}\sim$ TeV and $m_{\tilde s_{\rm R}}\sim5$ TeV \cite{RPVreview}.
The bound is greatly relaxed if the mixing is suppressed. 

The coupling $\lambda_{121}''$ induces strangeness-violating nucleon-antinucleon oscillation, leading to double nucleon decay processes such as $pp\rightarrow K^+K^+$ and $nn\rightarrow K^0K^0$. Bounds extracted from this process are highly sensitive to hadronic and nuclear matrix elements, and range from $|\lambda_{121}''|\lesssim10^{-7}$ to $1$ \cite{Goity:1994dq}, with  \cite{RPVreview} quoting a value of $|\lambda_{121}''|\lesssim10^{-6}$ for $m_{\tilde s_{\rm R}}=m_{\tilde g}=300$ GeV.

The couplings $\lambda_{212}''$ and $\lambda_{221}''$ are even less constrained. Bounds can be imposed by requiring that the RPV couplings remain perturbative up to the GUT scale, giving $|\lambda_{212}''|,|\lambda_{221}''|\lesssim1.25$.

We focus on models with light gluinos, which are pair-produced and each decay through an off-shell squark to three jets. Models exist in which the  gauginos are naturally lighter than the scalars, including ``split SUSY''~\cite{ArkaniHamed:2004fb,Giudice:2004tc}), leading to the possibility of a gluino LSP. 
The collider phenomenology depends on the gluino lifetime. Gluinos decaying through an off-shell squark and RPV couplings {\it always hadronize before decaying}, and depending on the magnitude of $\lambda''$, may also give rise to displaced vertices. To see this, we estimate the gluino width in the limit $m_{\tilde q}\gg m_{\tilde g}$:
\ben
\Gamma_{\tilde g} \sim \frac{\alpha_{\rm s}|\lambda''|^2\,m_{\tilde g}^5}{384\pi^2\,m_{\tilde q}^4}.
\een
Using a conservative value $|\lambda''|=10^{-6}$, then $\Gamma_{\tilde g}\sim 10^{-17}$~GeV with $m_{\tilde g}=1$~TeV and $m_{\tilde q}=5$~TeV, while $\Gamma_{\tilde g}\sim10^{-5}$~GeV with the most relaxed bound $|\lambda''|=1$. Since these are both well below $\Lambda_{\rm QCD}$, the gluino forms an $R$-hadron prior to decay whenever $m_{\tilde q}>m_{\tilde g}$, even when the RPV couplings are $\mathcal O(1)$. 
We consider only prompt decays after hadronization; for long-lived and stopped gluinos, see e.g.~\cite{Khachatryan:2010uf,Graham:2012th}.

%%%%%%%%%%%%%%%%%%%%%%%%%%%%%%%%%%%%%%%%%%%%%%%%%%%%%%%%%%%%%%%%%%%%%%
%%%%%%%%%%%%%%%%%%%%%%%%%%%%%%%%%%%%%%%%%%%%%%%%%%%%%%%%%%%%%%%%%%%%%%%
\section{Monte Carlo Comparison}
\label{s.MCcomparison} \setcounter{equation}{0} 
%%%%%%%%%%%%%%%%%%%%%%%%%%%%%%%%%%%%%%%%%%%%%%%%%%%%%%%%%%%%%%%%%%%%%%
%%%%%%%%%%%%%%%%%%%%%%%%%%%%%%%%%%%%%%%%%%%%%%%%%%%%%%%%%%%%%%%%%%%%%

\begin{figure*}
\newcommand{\upshift}{\vspace*{-26mm}}
\vspace*{-4mm}
\begin{center}
\hspace*{-3mm}
\begin{tabular}{p{5mm}p{1mm}p{\fval\textwidth}p{4mm}p{5mm}p{1mm}p{\fval\textwidth}}
\upshift (a)&&
\includegraphics[width=\fval\textwidth]{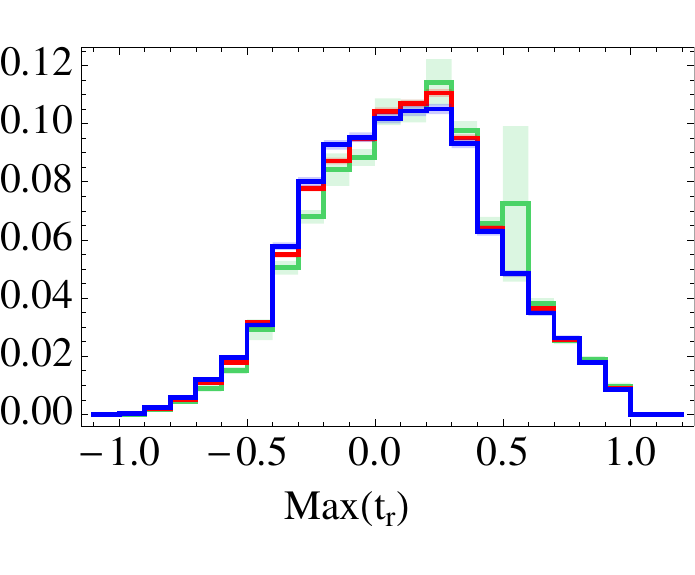}
&&
\upshift(b) &&
\includegraphics[width=\fval\textwidth]{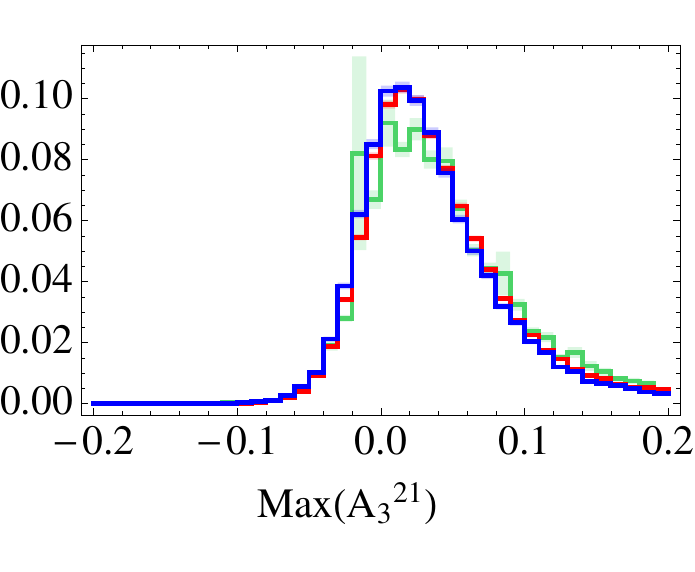}
 \vspace{2mm}
 
\end{tabular}
\end{center}
\vskip -0.8cm
\caption{
Distributions of QCD samples from \texttt{POWHEG + Pythia 6} (red), \texttt{POWHEG + Pythia 8} (blue), and \texttt{Sherpa} (green) after requiring six thin jets with $p_{\rm T}>60$ GeV and two fat jets with $p_{\rm T}>200$ GeV. (a) Radial pull $\mathrm{Max}(t_{\rm r})$. (b) Axis contraction $\mathrm{Max}(A_{3}^{21})$.
}
\label{f.colorflowcompare}
\end{figure*}

\begin{figure*}
\newcommand{\upshift}{\vspace*{-26mm}}
\vspace*{-4mm}
\begin{center}
\hspace*{-3mm}
\begin{tabular}{p{5mm}p{1mm}p{\fval\textwidth}p{4mm}p{5mm}p{1mm}p{\fval\textwidth}}
\upshift (a)&&
\includegraphics[width=\fval\textwidth]{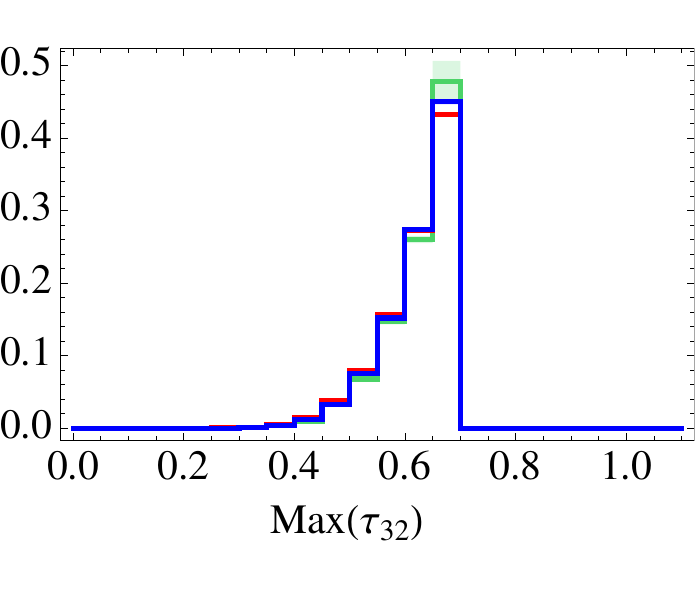}
&&
\upshift(b) &&
\includegraphics[width=\fval\textwidth]{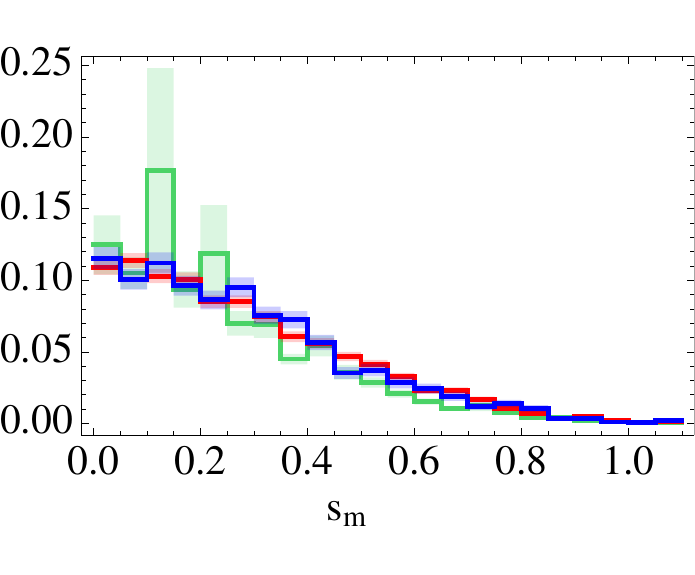}
\\
\upshift(c) &&
\includegraphics[width=\fval\textwidth]{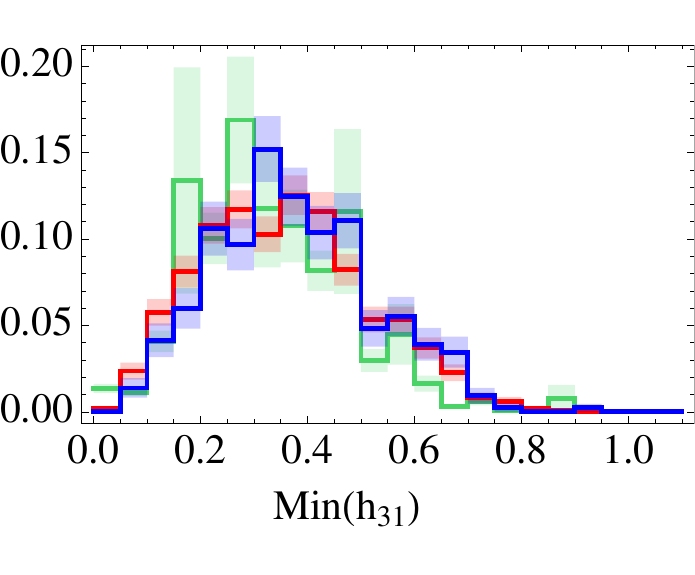}
&& 
\upshift(d) &&
\includegraphics[width=\fval\textwidth]{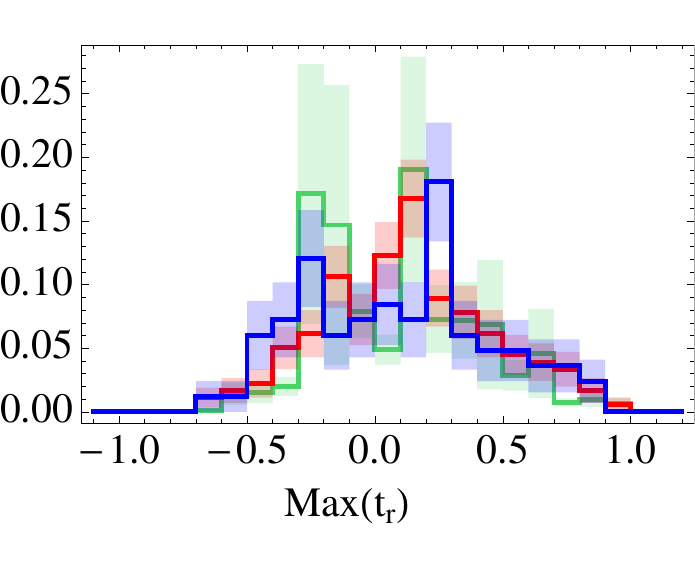}
 \vspace{1mm}
 
\end{tabular}
\end{center}
\vskip -0.8cm
\caption{
Distributions of QCD samples from \texttt{POWHEG + Pythia 6} (red), \texttt{POWHEG + Pythia 8} (blue), and \texttt{Sherpa} (green), following the cuts in Table \ref{f.LIGHTcuts}. (a) Maximum $N$-subjettiness ($\tau_{32}$) distribution prior to cut $\tau_{32}<0.5$. (b) Jet mass symmetry ($s_{\rm m}$) distribution prior to cut $s_{\rm m}<0.2$. (c) Subjet hierarchy ($h_{31}$) distribution prior to cut $h_{31}>0.5$. (d) Radial Pull ($\mathrm{Max}(t_r)$) distribution before the cut $\mathrm{Max}(t_r) < -0.6$.
}
\label{f.cutchaincompare}
\end{figure*}

For the results presented in the main body of the paper, we have exclusively used \texttt{Sherpa 1.4.0} to generate background event samples. Since our analysis includes a study of color-flow observables that may be sensitive to modeling of the parton shower and hadronization of colored objects, it is important to check how distributions of observables vary across Monte Carlo programs. Such a comparison also provides useful guidance for future substructure studies. We focus on QCD backgrounds, since these are dominant after all cuts, but also check $t\bar t$ distributions. We generated multijet backgrounds at parton-level with the \texttt{Dijet} package of \texttt{POWHEG 1.0} \cite{Nason:2004rx,Frixione:2007vw,Alioli:2010xa}, and the events were showered with the AUET2B tune \cite{ATLAS:2011gmi,ATLAS:2011zja}  of \texttt{Pythia 6.4.27} \cite{pythia6manual}. \texttt{POWHEG} is an NLO generator, and the \texttt{Dijet} package generates 3-parton final states. We chose the combination of \texttt{POWHEG + Pythia 6} because ATLAS found this provided a better detector-level description of the internal structure of high-$p_{\rm T}$ jets than \texttt{Pythia 6} alone \cite{ATLASrpvgluinobound5}. As a further check, we also showered the parton-level events with the default tune of \texttt{Pythia 8.1.65} \cite{pythia8}.  Matched top backgrounds were generated in \texttt{Madgraph 5.1.5} \cite{Alwall:2011uj} and showered in \texttt{Pythia 6} and \texttt{Pythia 8}. Since the hadronization of gluinos and their subsequent RPV decay is only implemented in \texttt{Pythia 8}, we are unable to check the color-flow distributions of the signal against another MC program.

We compare  distributions of the radial pull, axis contraction, $\tau_{32}$, $s_{\rm m}$, and $h_{31}$ variables defined in the text at various stages of the cut chain. We use the cuts from our top-mass gluino analysis (Table \ref{f.LIGHTcuts}), as some of these cuts were sensitive to the tails of certain observables such as radial pull. We summarize our results in Table \ref{f.LIGHTcompare}, in which we show the cut efficiencies for QCD and top backgrounds generated by each MC program. In the table, we normalize the number of events after preliminary cuts to the \texttt{Sherpa} value, both because that event sample was normalized to data and to allow a direct comparison of each cut's efficiency between MC generators. Furthermore, the fact that \texttt{Sherpa} generates matched samples with up to six partons suggests that it may give the best estimate of the number of events following the trigger and fat jet cuts. 

Table \ref{f.LIGHTcompare} shows that, overall, the MC programs are consistent in their modeling of jet substructure, although the tails of the subjet hierarchy and radial pull distributions can differ somewhat between programs; the total efficiency after all cuts can vary by about a factor of 10 for QCD. In spite of the more pessimistic estimates from \texttt{POWHEG + Pythia 6} and \texttt{POWHEG + Pythia 8}, however, there remains a gain in $S/B$ of $\mathcal{O}(10)$ from the radial pull cut, and $S/B\gtrsim10$ for the mean number of events after all cuts for QCD. Furthermore, an optimization of the cut with the new MC program allows for some recovery of the gain in $S/B$. For top backgrounds, the agreement is better, with a gain of $\sim2$ for the radial pull cut, and with the number of events passing all cuts agreeing within a factor of 2-3. All MC generators studied thus show that radial pull, in conjunction with other jet substructure cuts, is extremely effective at isolating a pure signal sample for top-mass gluinos, although the degree of improvement differs between programs. Experimental study is needed to determine the precise gain anticipated from such cuts.

\begin{figure*}[th!]
\newcommand{\upshift}{\vspace*{-26mm}}
\vspace*{-4mm}
\begin{center}
\hspace*{-3mm}
\begin{tabular}{p{5mm}p{1mm}p{\fval\textwidth}p{4mm}p{5mm}p{1mm}p{\fval\textwidth}}
\upshift (a)&&
\includegraphics[width=\fval\textwidth]{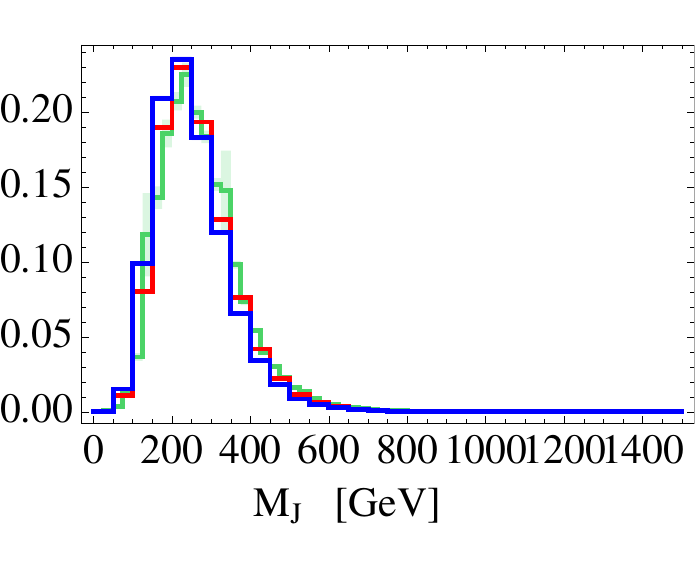}
&&
\upshift(b) &&
\includegraphics[width=\fval\textwidth]{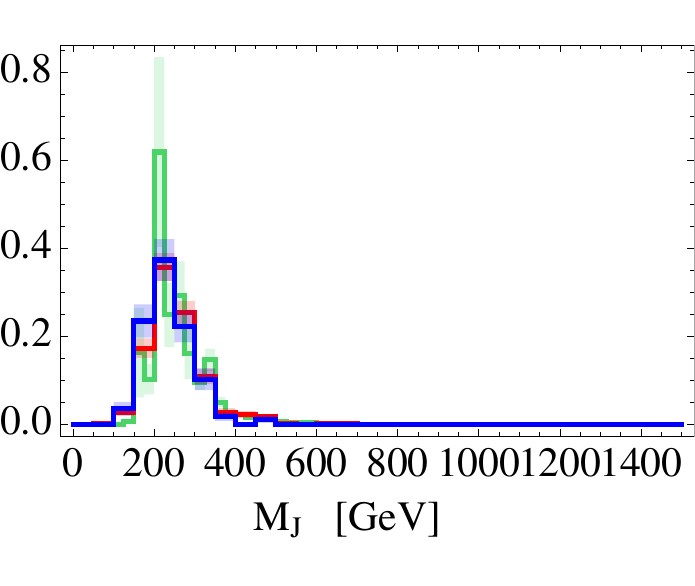}
\vspace{1mm}
\end{tabular}
\end{center}
\vskip -0.8cm
\caption{
Fat jet mass distributions of QCD samples from \texttt{POWHEG + Pythia 6} (red), \texttt{POWHEG + Pythia 8} (blue), and \texttt{Sherpa} (green). (a) Jet mass distribution after requiring six thin jets with $p_{\rm T}>60$ GeV and two fat jets with $p_{\rm T}>200$ GeV. (b) Jet mass distribution after all cuts in Table \ref{f.LIGHTcuts} prior to the cut on $t_{\rm r}$.
}
\label{f.jetmasscompare}
\end{figure*}

We now provide details of the MC comparison for the QCD Background. The \texttt{Dijet} package of \texttt{POWHEG} is an NLO parton-level event generator. It generates dijet events at leading (Born) order, and then includes NLO corrections to generate three-parton events. Higher multiplicity jets are generated by the subsequent parton shower, making the efficiency of passing the six-jet trigger extremely low. The only cut that can be placed on QCD events is the $p_{\rm T}$ of the dijet system prior to NLO emission (called the Born $p_{\rm T}$) \footnote{It is possible to generate weighted events in $p_{\rm T}$, but we find that the majority of events passing all cuts are those which marginally pass the six-jet trigger, and therefore generating many low-weight, high-$p_{\rm T}$ events does not improve the statistics after all cuts are imposed.}. This is in contrast with \texttt{Sherpa}, where six-parton final states could be generated directly with $p_{\rm T}$ cuts on each parton. To choose the value of the \texttt{POWHEG} Born $p_{\rm T}$ cut, we use the fact that parton emissions beyond LO are typically softer than the original hard scale; therefore, if applying a fat dijet cut of $p_{\rm T}>200$ GeV to the showered sample (as in our top-mass gluino analysis), the corresponding cut on the Born $p_{\rm T}$ should be similar but somewhat softer than this. 
\footnote{
This can be understood physically. A configuration with two hard three-pronged fat jets would typically arise from a parton-level configuration with two partially-aligned partons recoiling off a third parton, with the shower providing additional hard subjets.  In particular, most of one fat jet's $p_{\rm T}$ comes from a single hard parton, which has a momentum roughly bounded by the Born $p_{\rm T}$ cut.
}
We use the Born cut $p_{\rm T}>140$ GeV, and we confirmed its validity by checking that the cross section passing  trigger and fat jet cuts is the same for other generator-level cuts up to 140~GeV. With this cut, only 0.05\% of MC events pass the preliminary six-jet trigger and fat jet cuts; even though we generated 500 million QCD events, we still suffer from small statistics for aggressive values of our substructure cuts, and only 1-2 such events pass the final radial pull cut from Section \ref{ss.topmass}. The results for \texttt{POWHEG+Pythia}  therefore have a much higher uncertainty than for \texttt{Sherpa}.

In Fig.~\ref{f.colorflowcompare}, we compare the shapes of radial pull and axis contraction distributions for QCD samples generated with each MC program. The distributions are shown after the preliminary cuts of the top-mass gluino analysis (six-jet trigger and two fat jets $p_{\rm T}>200$ GeV), and the shapes are similar for all generators. We also plot the distributions in Fig.~\ref{f.cutchaincompare} of each substructure variable before the cut on this variable as listed in Table \ref{f.LIGHTcuts}. These distributions are largely consistent across different MC programs; the least-well-modeled variables are the subjet hierarchy $h_{31}$ and the radial pull $\mathrm{Max}(t_{\rm r})$, where we find that the cut efficiency of $\mathrm{min}(h_{31})>0.5$ varies by at most a factor of two between MC descriptions, while the cuts on $\mathrm{Max}(t_{\rm r})$ have efficiencies in the \texttt{Pythia} samples that are approximately six times higher than the \texttt{Sherpa} prediction. Some of this variation, especially for the $h_{31}$ cut,  may be due to \texttt{Sherpa} generating up to 6 jets at the matrix-element level while \texttt{POWHEG} must rely on the shower. This suggests that \texttt{Sherpa}'s predictions might be more trustworthy. At any rate, the statistical uncertainties are very large because only one or two MC events pass cuts from the \texttt{Pythia} samples. For completeness, we also plot the fat-jet mass distributions before and after jet substructure cuts in Fig.~\ref{f.jetmasscompare}, 

Figs.~\ref{f.colorflowcompare}-\ref{f.cutchaincompare} therefore demonstrate that our substructure analysis is generally robust across different MC event generators for QCD backgrounds. The most sensitive observables are the tails of the radial pull and axis contraction variables, which can change background cut efficiencies by up to a factor of $\sim 10$, but $S/B$ is so large that the results are not invalidated.  Together with the already outlined comparison of top backgrounds, which are anyway subdominant to QCD after a $b$-veto, this provides evidence that the \texttt{Sherpa} analysis is reliable.


\begin{thebibliography}{99}





\bibitem{ATLASnote:2012a} 
[ATLAS Collaboration],
  %``Measurement of Jet Mass and Substructure for Inclusive Jets in $\sqrt{s} = 7$ TeV pp Collisions with the ATLAS Experiment''
ATLAS-CONF-2012-109.
  %%CITATION = ATL-PHYS-CONF-2008-008;%%
  
\bibitem{2012mfa} 
  S.~Chatrchyan {\it et al.}  [CMS Collaboration],
  %``Search for new physics in the multijet and missing transverse momentum final state in proton-proton collisions at $\sqrt{s} = 7$ TeV,''
  arXiv:1207.1898 [hep-ex].
  %%CITATION = ARXIV:1207.1898;%%

\bibitem{ATLASnote:2012b} 
[ATLAS Collaboration],
  %``Measurement of Jet Mass and Substructure for Inclusive Jets in $\sqrt{s} = 7$ TeV pp Collisions with the ATLAS Experiment''
ATLAS-CONF-2012-103.
  %%CITATION = ATL-PHYS-CONF-2008-008;%%

\bibitem{CMSnote:2011b} 
[CMS Collaboration],
  %``Measurement of Jet Mass and Substructure for Inclusive Jets in $\sqrt{s} = 7$ TeV pp Collisions with the ATLAS Experiment''
CMS PAS SUS-11-006.
  %%CITATION = ATL-PHYS-CONF-2008-008;%%
  
\bibitem{2012ar} 
  G.~Aad {\it et al.}  [ATLAS Collaboration],
  %``Search for direct top squark pair production in final states with one isolated lepton, jets, and missing transverse momentum in sqrt(s) = 7 TeV pp collisions using 4.7 fb-1 of ATLAS data,''
  arXiv:1208.2590 [hep-ex].
  %%CITATION = ARXIV:1208.2590;%%
  
\bibitem{CMSnote:2011c} 
[CMS Collaboration],
  %``Measurement of Jet Mass and Substructure for Inclusive Jets in $\sqrt{s} = 7$ TeV pp Collisions with the ATLAS Experiment''
CMS PAS SUS-11-022.
  %%CITATION = ATL-PHYS-CONF-2008-008;%%
  
\bibitem{Chatrchyan:2012sa} 
  S.~Chatrchyan {\it et al.}  [CMS Collaboration],
  %``Search for new physics in events with same-sign dileptons and b-tagged jets in pp collisions at sqrt(s) = 7 TeV,''
  arXiv:1205.3933 [hep-ex].
  %%CITATION = ARXIV:1205.3933;%%
  
\bibitem{Aad:2011cw} 
  G.~Aad {\it et al.}  [ATLAS Collaboration],
  %``Search for scalar bottom pair production with the ATLAS detector in pp Collisions at sqrt{s} = 7 TeV,''
  Phys.\ Rev.\ Lett.\  {\bf 108}, 181802 (2012)
  [arXiv:1112.3832 [hep-ex]].
  %%CITATION = ARXIV:1112.3832;%%
  
\bibitem{2012he} 
  G.~Aad {\it et al.}  [ATLAS Collaboration],
  %``Further search for supersymmetry at sqrt(s) = 7 TeV in final states with jets, missing transverse momentum and isolated leptons with the ATLAS detector,''
  arXiv:1208.4688 [hep-ex].
  %%CITATION = ARXIV:1208.4688;%%
  

\bibitem{Sterman:1977wj}
G.~F.~Sterman and S.~Weinberg,
``Jets from Quantum Chromodynamics,''
Phys.\ Rev.\ Lett.\ {\bf 39} (1977) 1436.
%%CITATION = PRLTA,39,1436;%%

\bibitem{Chivukula:1990di}
R.~S.~Chivukula, M.~Golden and E.~H.~Simmons,
``Six Jet Signals of Highly Colored Fermions,''
Phys.\ Lett.\ B {\bf 257} (1991) 403.
%%CITATION = PHLTA,B257,403;%%

\bibitem{Chivukula:1991zk}
R.~S.~Chivukula, M.~Golden and E.~H.~Simmons,
``Multi - Jet Physics at Hadron Colliders,''
Nucl.\ Phys.\ B {\bf 363} (1991) 83.
%%CITATION = NUPHA,B363,83;%%

\bibitem{Farhi:1979zx}
E.~Farhi and L.~Susskind,
``A Technicolored G.U.T.,''
Phys.\ Rev.\ D {\bf 20} (1979) 3404.
%%CITATION = PHRVA,D20,3404;%%

\bibitem{Marciano:1980zf}
W.~J.~Marciano,
%``Exotic New Quarks and Dynamical Symmetry Breaking,''
Phys.\ Rev.\ D {\bf 21} (1980) 2425.
%%CITATION = PHRVA,D21,2425;%%

\bibitem{Frampton:1987ut}
P.~H.~Frampton and S.~L.~Glashow,
%``Unifiable Chiral Color with Natural Gim Mechanism,''
Phys.\ Rev.\ Lett.\ {\bf 58} (1987) 2168.
%%CITATION = PRLTA,58,2168;%%

\bibitem{Frampton:1987dn}
P.~H.~Frampton and S.~L.~Glashow,
%``Chiral Color: an Alternative to the Standard Model,''
Phys.\ Lett.\ B {\bf 190} (1987) 157.
%%CITATION = PHLTA,B190,157;%%

\bibitem{rouventhesis}
Based on work by S-H.~Chuang, R.~Essig, E.~Halkiadakis, A.~Lath, S.Thomas; see Ch.~6 in R.~Essig,
%``Physics Beyond the Standard Model: Supersymmetry, Dark Matter, and LHC Phenomenology'',
AAT-3349692, PhD Thesis, Rutgers University, 2008, {\footnotesize http://hdl.rutgers.edu/1782.2/rucore10001600001.ETD.17462}.
%%CITATION = AAT-3349692;%%

\bibitem{RPVoriginal} 
  L.~J.~Hall and M.~Suzuki,
  %``Explicit R-Parity Breaking in Supersymmetric Models,''
  Nucl.\ Phys.\ B {\bf 231}, 419 (1984).
  %%CITATION = NUPHA,B231,419;%%
  
\bibitem{RPVreview} 
%  R.~Barbier, C.~Berat, M.~Besancon, M.~Chemtob, A.~Deandrea, E.~Dudas, P.~Fayet and S.~Lavignac {\it et al.},
    R.~Barbier {\it et al.},
  %``R-parity violating supersymmetry,''
  Phys.\ Rept.\  {\bf 420}, 1 (2005)
  [hep-ph/0406039].
  %%CITATION = HEP-PH/0406039;%%


\bibitem{Alwall:2008ve} 
  J.~Alwall, M.~-P.~Le, M.~Lisanti and J.~G.~Wacker,
  %``Searching for Directly Decaying Gluinos at the Tevatron,''
  Phys.\ Lett.\ B {\bf 666}, 34 (2008)
  [arXiv:0803.0019 [hep-ph]].
  %%CITATION = ARXIV:0803.0019;%%

\bibitem{Kilic:2008pm}
C.~Kilic, T.~Okui and R.~Sundrum,
%``Colored Resonances at the Tevatron: Phenomenology and Discovery Potential in Multijets,''
JHEP {\bf 0807} (2008) 038
[arXiv:0802.2568 [hep-ph]].
%%CITATION = ARXIV:0802.2568;%%

\bibitem{Kilic:2010et}
C.~Kilic and T.~Okui,
%``The Lhc Phenomenology of Vectorlike Confinement,''
JHEP {\bf 1004} (2010) 128
[arXiv:1001.4526 [hep-ph]].
%%CITATION = ARXIV:1001.4526;%%

\bibitem{Alves:2011wf}
D.~Alves {\it et al.} [LHC New Physics Working Group Collaboration],
%``Simplified Models for Lhc New Physics Searches,''
J.\ Phys.\ G {\bf 39} (2012) 105005
[arXiv:1105.2838 [hep-ph]].
%%CITATION = ARXIV:1105.2838;%%

\bibitem{Tavares:2011zg}
G.~Marques Tavares and M.~Schmaltz,
%``Explaining the T-Tbar Asymmetry with a Light Axigluon,''
Phys.\ Rev.\ D {\bf 84} (2011) 054008
[arXiv:1107.0978 [hep-ph]].
%%CITATION = ARXIV:1107.0978;%%

\bibitem{Gross:2012bz}
C.~Gross, G.~M.~Tavares, C.~Spethmann and M.~Schmaltz,
%``Light Axigluon Explanation of the Tevatron t-tbar Asymmetry and Multijet Signals at the LHC,''
arXiv:1209.6375 [hep-ph].
%%CITATION = ARXIV:1209.6375;%% 
 
\bibitem{Hook:2012fd} 
  A.~Hook, E.~Izaguirre, M.~Lisanti and J.~G.~Wacker,
  %``High Multiplicity Searches at the LHC Using Jet Masses,''
  Phys.\ Rev.\ D {\bf 85}, 055029 (2012)
  [arXiv:1202.0558 [hep-ph]].
  %%CITATION = ARXIV:1202.0558;%%

\bibitem{rpvmsugra} 
  M.~Asano, K.~Rolbiecki and K.~Sakurai,
  %``Can R-parity violation hide vanilla supersymmetry at the LHC?,''
  arXiv:1209.5778 [hep-ph].
  %%CITATION = ARXIV:1209.5778;%%

\bibitem{LEPrpvgluinobound} 
  D.~E.~Kaplan and M.~D.~Schwartz,
  %``Constraining Light Colored Particles with Event Shapes,''
  Phys.\ Rev.\ Lett.\  {\bf 101}, 022002 (2008)
  [arXiv:0804.2477 [hep-ph]].
  %%CITATION = ARXIV:0804.2477;%%
  
\bibitem{TEVATRONrpvgluinobound} 
  T.~Aaltonen {\it et al.}  [CDF Collaboration],
  %``First Search for Multijet Resonances in $\sqrt{s} = 1.96$ TeV $ p\bar{p}$ Collisions,''
  Phys.\ Rev.\ Lett.\  {\bf 107}, 042001 (2011)
  [arXiv:1105.2815 [hep-ex]].
  %%CITATION = ARXIV:1105.2815;%%

\bibitem{CMSrpvgluinobound36} 
  S.~Chatrchyan {\it et al.}  [CMS Collaboration],
  %``Search for Three-Jet Resonances in pp Collisions at sqrt(s) = 7 TeV,''
  Phys.\ Rev.\ Lett.\  {\bf 107}, 101801 (2011)
  [arXiv:1107.3084 [hep-ex]].
  %%CITATION = ARXIV:1107.3084;%%
  
\bibitem{CMSrpvgluinobound5} 
S.~Chatrchyan {\it et al.} [CMS Collaboration],
%``Search for Three-Jet Resonances in PP Collisions at Sqrt(S) = 7 Tev,''
arXiv:1208.2931 [hep-ex].
%%CITATION = ARXIV:1208.2931;%%

\bibitem{ATLASrpvgluinobound5} 
  G.~Aad {\it et al.}  [ATLAS Collaboration],
  arXiv:1210.4813 [hep-ex].
  %%CITATION = ARXIV:1210.4813;%%
  
\bibitem{Allanach:2012vj} 
  B.~C.~Allanach and B.~Gripaios,
  %``Hide and Seek With Natural Supersymmetry at the LHC,''
  JHEP {\bf 1205}, 062 (2012)
  [arXiv:1202.6616 [hep-ph]].
  %%CITATION = ARXIV:1202.6616;%%

\bibitem{Brust:2011tb} 
  C.~Brust, A.~Katz, S.~Lawrence and R.~Sundrum,
  %``SUSY, the Third Generation and the LHC,''
  JHEP {\bf 1203}, 103 (2012)
  [arXiv:1110.6670 [hep-ph]].
  %%CITATION = ARXIV:1110.6670;%%pe
  
\bibitem{Brust:2012uf} 
  C.~Brust, A.~Katz and R.~Sundrum,
  %``SUSY Stops at a Bump,''
  JHEP {\bf 1208}, 059 (2012)
  [arXiv:1206.2353 [hep-ph]].
  %%CITATION = ARXIV:1206.2353;%%

\bibitem{Evans:2012bf} 
  J.~A.~Evans and Y.~Kats,
  %``LHC Coverage of RPV MSSM with Light Stops,''
  arXiv:1209.0764 [hep-ph].
  %%CITATION = ARXIV:1209.0764;%%
    
\bibitem{nsubjettiness} 
  J.~Thaler and K.~Van Tilburg,
  %``Identifying Boosted Objects with N-subjettiness,''
  JHEP {\bf 1103}, 015 (2011)
  [arXiv:1011.2268 [hep-ph]].
  %%CITATION = ARXIV:1011.2268;%%

\bibitem{Stewart:2010tn}
I.~W.~Stewart, F.~J.~Tackmann and W.~J.~Waalewijn,
%``N-Jettiness: an Inclusive Event Shape to Veto Jets,''
Phys.\ Rev.\ Lett.\ {\bf 105} (2010) 092002
[arXiv:1004.2489 [hep-ph]].
%%CITATION = ARXIV:1004.2489;%%
  
\bibitem{nsubjettinessminaxes} 
  J.~Thaler and K.~Van Tilburg,
  %``Maximizing Boosted Top Identification by Minimizing N-subjettiness,''
  JHEP {\bf 1202}, 093 (2012)
  [arXiv:1108.2701 [hep-ph]].
  %%CITATION = ARXIV:1108.2701;%%

\bibitem{pull} 
  J.~Gallicchio and M.~D.~Schwartz,
  %``Seeing in Color: Jet Superstructure,''
  Phys.\ Rev.\ Lett.\  {\bf 105}, 022001 (2010)
  [arXiv:1001.5027 [hep-ph]].
  %%CITATION = ARXIV:1001.5027;%%
  
\bibitem{Bassetto:1984ik} 
  A.~Bassetto, M.~Ciafaloni and G.~Marchesini,
  %``Jet Structure and Infrared Sensitive Quantities in Perturbative QCD,''
  Phys.\ Rept.\  {\bf 100}, 201 (1983).
  %%CITATION = PRPLC,100,201;%%
  
\bibitem{dipolarity} 
  A.~Hook, M.~Jankowiak and J.~G.~Wacker,
  %``Jet Dipolarity: Top Tagging with Color Flow,''
  JHEP {\bf 1204}, 007 (2012)
  [arXiv:1102.1012 [hep-ph]].
  %%CITATION = ARXIV:1102.1012;%%
  
\bibitem{girth} 
  J.~Gallicchio, J.~Huth, M.~Kagan, M.~D.~Schwartz, K.~Black and B.~Tweedie,
  %``Multivariate discrimination and the Higgs + W/Z search,''
  JHEP {\bf 1104}, 069 (2011)
  [arXiv:1010.3698 [hep-ph]].
  %%CITATION = ARXIV:1010.3698;%%
 
\bibitem{WtagRcores} 
  Y.~Cui, Z.~Han and M.~D.~Schwartz,
  %``W-jet Tagging: Optimizing the Identification of Boosted Hadronically-Decaying W Bosons,''
  Phys.\ Rev.\ D {\bf 83}, 074023 (2011)
  [arXiv:1012.2077 [hep-ph]].
  %%CITATION = ARXIV:1012.2077;%%

\bibitem{quarkgluontag} 
  J.~Gallicchio and M.~D.~Schwartz,
  %``Quark and Gluon Tagging at the LHC,''
  Phys.\ Rev.\ Lett.\  {\bf 107}, 172001 (2011)
  [arXiv:1106.3076 [hep-ph]].
  %%CITATION = ARXIV:1106.3076;%%
  
\bibitem{planarflowtoptag} 
  L.~G.~Almeida, S.~J.~Lee, G.~Perez, I.~Sung and J.~Virzi,
  %``Top Jets at the LHC,''
  Phys.\ Rev.\ D {\bf 79}, 074012 (2009)
  [arXiv:0810.0934 [hep-ph]].
  %%CITATION = ARXIV:0810.0934;%%

\bibitem{BDRStagger} 
  J.~M.~Butterworth, A.~R.~Davison, M.~Rubin and G.~P.~Salam,
  %``Jet substructure as a new Higgs search channel at the LHC,''
  Phys.\ Rev.\ Lett.\  {\bf 100}, 242001 (2008)
  [arXiv:0802.2470 [hep-ph]].
  %%CITATION = ARXIV:0802.2470;%%
  
\bibitem{jetpruning} 
  S.~D.~Ellis, C.~K.~Vermilion and J.~R.~Walsh,
  %``Techniques for improved heavy particle searches with jet substructure,''
  Phys.\ Rev.\ D {\bf 80}, 051501 (2009)
  [arXiv:0903.5081 [hep-ph]].
  %%CITATION = ARXIV:0903.5081;%%
  
\bibitem{filteringfortag} 
  T.~Plehn, G.~P.~Salam and M.~Spannowsky,
  %``Fat Jets for a Light Higgs,''
  Phys.\ Rev.\ Lett.\  {\bf 104}, 111801 (2010)
  [arXiv:0910.5472 [hep-ph]].
  %%CITATION = ARXIV:0910.5472;%%
  
\bibitem{trimming} 
  D.~Krohn, J.~Thaler and L.~-T.~Wang,
  %``Jet Trimming,''
  JHEP {\bf 1002}, 084 (2010)
  [arXiv:0912.1342 [hep-ph]].
  %%CITATION = ARXIV:0912.1342;%%
  
\bibitem{filterforhiggs} 
  G.~D.~Kribs, A.~Martin, T.~S.~Roy and M.~Spannowsky,
  %``Discovering the Higgs Boson in New Physics Events using Jet Substructure,''
  Phys.\ Rev.\ D {\bf 81}, 111501 (2010)
  [arXiv:0912.4731 [hep-ph]].
  %%CITATION = ARXIV:0912.4731;%%
  
\bibitem{YsplitterATLAS} 
  G.~Brooijmans,
  %``High p**T hadronic top quark identification. Part I: Jet mass and Ysplitter,''
  ATL-PHYS-CONF-2008-008.
  %%CITATION = ATL-PHYS-CONF-2008-008;%%
  
\bibitem{Soper:2011cr} 
  D.~E.~Soper and M.~Spannowsky,
  %``Finding physics signals with shower deconstruction,''
  Phys.\ Rev.\ D {\bf 84}, 074002 (2011)
  [arXiv:1102.3480 [hep-ph]].
  %%CITATION = ARXIV:1102.3480;%%
  
\bibitem{Butterworth:2007ke} 
  J.~M.~Butterworth, J.~R.~Ellis and A.~R.~Raklev,
  %``Reconstructing sparticle mass spectra using hadronic decays,''
  JHEP {\bf 0705}, 033 (2007)
  [hep-ph/0702150 [hep-ph]].
  %%CITATION = HEP-PH/0702150;%%
  
\bibitem{Almeida:2010pa} 
  L.~G.~Almeida, S.~J.~Lee, G.~Perez, G.~Sterman and I.~Sung,
  %``Template Overlap Method for Massive Jets,''
  Phys.\ Rev.\ D {\bf 82}, 054034 (2010)
  [arXiv:1006.2035 [hep-ph]].
  %%CITATION = ARXIV:1006.2035;%%
  
\bibitem{Jankowiak:2011qa} 
  M.~Jankowiak and A.~J.~Larkoski,
  %``Jet Substructure Without Trees,''
  JHEP {\bf 1106}, 057 (2011)
  [arXiv:1104.1646 [hep-ph]].
  %%CITATION = ARXIV:1104.1646;%%
    
\bibitem{Ellis:2012sn} 
  S.~D.~Ellis, A.~Hornig, T.~S.~Roy, D.~Krohn and M.~D.~Schwartz,
  %``Qjets: A Non-Deterministic Approach to Tree-Based Jet Substructure,''
  Phys.\ Rev.\ Lett.\  {\bf 108}, 182003 (2012)
  [arXiv:1201.1914 [hep-ph]].
  %%CITATION = ARXIV:1201.1914;%%

\bibitem{Butterworth:2002tt} 
  J.~M.~Butterworth, B.~E.~Cox and J.~R.~Forshaw,
  %``$W W$ scattering at the CERN LHC,''
  Phys.\ Rev.\ D {\bf 65}, 096014 (2002)
  [hep-ph/0201098].
  %%CITATION = HEP-PH/0201098;%%
  
\bibitem{Jankowiak:2012na} 
  M.~Jankowiak and A.~J.~Larkoski,
  %``Angular Scaling in Jets,''
  JHEP {\bf 1204}, 039 (2012)
  [arXiv:1201.2688 [hep-ph]].
  %%CITATION = ARXIV:1201.2688;%%
  
  
\bibitem{Salam:2009jx} 
  G.~P.~Salam,
  %``Towards Jetography,''
  Eur.\ Phys.\ J.\ C {\bf 67}, 637 (2010)
  [arXiv:0906.1833 [hep-ph]].
  %%CITATION = ARXIV:0906.1833;%%
       
\bibitem{TWtoptagger} 
  J.~Thaler and L.~-T.~Wang,
  %``Strategies to Identify Boosted Tops,''
  JHEP {\bf 0807}, 092 (2008)
  [arXiv:0806.0023 [hep-ph]].
  %%CITATION = ARXIV:0806.0023;%%
  
  
\bibitem{Hopkinstoptagger} 
  D.~E.~Kaplan, K.~Rehermann, M.~D.~Schwartz and B.~Tweedie,
  %``Top Tagging: A Method for Identifying Boosted Hadronically Decaying Top Quarks,''
  Phys.\ Rev.\ Lett.\  {\bf 101}, 142001 (2008)
  [arXiv:0806.0848 [hep-ph]].
  %%CITATION = ARXIV:0806.0848;%%
  
\bibitem{Abazov:2011vh} 
  V.~M.~Abazov {\it et al.}  [D0 Collaboration],
  %``Measurement of color flow in $\mathbf{t\bar{t}}$ events from $\mathbf{p\bar{p}}$ collisions at $\mathbf{\sqrt{s}=1.96}$ TeV,''
  Phys.\ Rev.\ D {\bf 83}, 092002 (2011)
  [arXiv:1101.0648 [hep-ex]].
  %%CITATION = ARXIV:1101.0648;%%

\bibitem{ATLASsubstructurecomparison} 
  G.~Aad {\it et al.}  [ATLAS Collaboration],
  %``Jet mass and substructure of inclusive jets in sqrt(s) = 7 TeV pp collisions with the ATLAS experiment,''
  JHEP {\bf 1205}, 128 (2012)
  [arXiv:1203.4606 [hep-ex]].
  %%CITATION = ARXIV:1203.4606;%%
  
\bibitem{Aad:2011kq} 
  G.~Aad {\it et al.}  [Atlas Collaboration],
  %``Study of Jet Shapes in Inclusive Jet Production in pp Collisions at sqrt(s) = 7 TeV using the ATLAS Detector,''
  Phys.\ Rev.\ D {\bf 83}, 052003 (2011)
  [arXiv:1101.0070 [hep-ex]].
  %%CITATION = ARXIV:1101.0070;%%
  
  
  

\bibitem{Aad:2012jf} 
  G.~Aad {\it et al.}  [ATLAS Collaboration],
  %``ATLAS measurements of the properties of jets for boosted particle searches,''
  arXiv:1206.5369 [hep-ex].
  %%CITATION = ARXIV:1206.5369;%%


\bibitem{CMSnote:2011a} 
  [CMS Collaboration],
  %``Jet Substructure Algorithms,''
  CMS-PAS-JME-10-013.
  %%CITATION = CMS-PAS-JME-10-013;%%


\bibitem{Aaltonen:2011pg} 
  T.~Aaltonen {\it et al.}  [CDF Collaboration],
  %``Study of Substructure of High Transverse Momentum Jets Produced in Proton-Antiproton Collisions at $\sqrt{s}=1.96$ TeV,''
  Phys.\ Rev.\ D {\bf 85}, 091101 (2012)
  [arXiv:1106.5952 [hep-ex]].
  %%CITATION = ARXIV:1106.5952;%%
  
\bibitem{ATLASnote:2011a} 
[ATLAS Collaboration],
  %``Measurement of Jet Mass and Substructure for Inclusive Jets in $\sqrt{s} = 7$ TeV pp Collisions with the ATLAS Experiment''
 ATLAS-CONF-2011-073.
  %%CITATION = ATL-PHYS-CONF-2008-008;%%

\bibitem{Chatrchyan:2012zt} 
  S.~Chatrchyan {\it et al.}  [CMS Collaboration],
  %``Shape, transverse size, and charged hadron multiplicity of jets in pp collisions at 7 TeV,''
  JHEP {\bf 1206}, 160 (2012)
  [arXiv:1204.3170 [hep-ex]].
  %%CITATION = ARXIV:1204.3170;%%



\bibitem{Seymour:1993mx} 
  M.~H.~Seymour,
  %``Searches for new particles using cone and cluster jet algorithms: A Comparative study,''
  Z.\ Phys.\ C {\bf 62}, 127 (1994).
  %%CITATION = ZEPYA,C62,127;%%
  
  
  
  





  
  
  
  
\bibitem{boost2010} 
%  A.~Abdesselam, E.~B.~Kuutmann, U.~Bitenc, G.~Brooijmans, J.~Butterworth, P.~Bruckman de Renstrom, D.~Buarque Franzosi and R.~Buckingham {\it et al.},
    A.~Abdesselam {\it et al.},
  %``Boosted objects: A Probe of beyond the Standard Model physics,''
  Eur.\ Phys.\ J.\ C {\bf 71}, 1661 (2011)
  [arXiv:1012.5412 [hep-ph]].
  %%CITATION = ARXIV:1012.5412;%%
  
\bibitem{toptagreview} 
  T.~Plehn and M.~Spannowsky,
  %``Top Tagging,''
  J.\ Phys.\ G G {\bf 39}, 083001 (2012)
  [arXiv:1112.4441 [hep-ph]].
  %%CITATION = ARXIV:1112.4441;%%
  
\bibitem{boost2011} 
%  A.~Altheimer, S.~Arora, L.~Asquith, G.~Brooijmans, J.~Butterworth, M.~Campanelli, B.~Chapleau and A.~E.~Cholakian {\it et al.},
    A.~Altheimer {\it et al.},
  %``Jet Substructure at the Tevatron and LHC: New results, new tools, new benchmarks,''
  J.\ Phys.\ G G {\bf 39}, 063001 (2012)
  [arXiv:1201.0008 [hep-ph]].
  %%CITATION = ARXIV:1201.0008;%%

\bibitem{tevatronboostedgluinos} 
  G.~Brooijmans {\it et al.}  [New Physics Working Group Collaboration],
  %``New Physics at the LHC. A Les Houches Report: Physics at TeV Colliders 2009 - New Physics Working Group,''
  arXiv:1005.1229 [hep-ph];
  %%CITATION = ARXIV:1005.1229;%%
  A.~R.~Raklev, G.~P.~Salam and J.~G.~Wacker,
  %``Light gluinos in hiding: Reconstructing R-parity violating decays at the Tevatron,''
  SLAC-REPRINT-2012-045.
  %%CITATION = SLAC-REPRINT-2012-045;%%

\bibitem{futurework}
D.~Curtin, R.~Essig, B.~Shuve, In preparation. 
  
\bibitem{Butterworth:2009qa} 
  J.~M.~Butterworth, J.~R.~Ellis, A.~R.~Raklev and G.~P.~Salam,
  %``Discovering baryon-number violating neutralino decays at the LHC,''
  Phys.\ Rev.\ Lett.\  {\bf 103}, 241803 (2009)
  [arXiv:0906.0728 [hep-ph]].
  %%CITATION = ARXIV:0906.0728;%%

\bibitem{Bjorken:1991xr} 
  J.~D.~Bjorken,
  %``A Full Acceptance Detector for SSC Physics at Low and Intermediate Mass Scales: An Expression of Interest to the SSC,''
  Int.\ J.\ Mod.\ Phys.\ A {\bf 7}, 4189 (1992).
  %%CITATION = IMPAE,A7,4189;%%
  
\bibitem{Bjorken:1992er} 
  J.~D.~Bjorken,
  %``Rapidity gaps and jets as a new physics signature in very high-energy hadron hadron collisions,''
  Phys.\ Rev.\ D {\bf 47}, 101 (1993).
  %%CITATION = PHRVA,D47,101;%%
  
\bibitem{Fletcher:1993ij} 
  R.~S.~Fletcher and T.~Stelzer,
  %``Rapidity gap signals in Higgs production at the SSC,''
  Phys.\ Rev.\ D {\bf 48}, 5162 (1993)
  [hep-ph/9306253].
  %%CITATION = HEP-PH/9306253;%%
  
\bibitem{Barger:1994zq} 
  V.~D.~Barger, R.~J.~N.~Phillips and D.~Zeppenfeld,
  %``Mini - jet veto: A Tool for the heavy Higgs search at the LHC,''
  Phys.\ Lett.\ B {\bf 346}, 106 (1995)
  [hep-ph/9412276].
  %%CITATION = HEP-PH/9412276;%%
  
\bibitem{pythiaRhadrondiscussion} 
  M.~Fairbairn, A.~C.~Kraan, D.~A.~Milstead, T.~Sjostrand, P.~Z.~Skands and T.~Sloan,
  %``Stable massive particles at colliders,''
  Phys.\ Rept.\  {\bf 438}, 1 (2007)
  [hep-ph/0611040].
  %%CITATION = HEP-PH/0611040;%%
  
  \bibitem{georgeeventshape} 
  C.~F.~Berger, T.~Kucs and G.~F.~Sterman,
  %``Event shape / energy flow correlations,''
  Phys.\ Rev.\ D {\bf 68}, 014012 (2003)
  [hep-ph/0303051].
  %%CITATION = HEP-PH/0303051;%%
  
  
  
\bibitem{pythia6manual} 
  T.~Sjostrand, S.~Mrenna and P.~Z.~Skands,
  %``PYTHIA 6.4 Physics and Manual,''
  JHEP {\bf 0605}, 026 (2006)
  [hep-ph/0603175].
  %%CITATION = HEP-PH/0603175;%%

\bibitem{pythia8RPV} 
  N.~Desai and P.~Z.~Skands,
  %``Supersymmetry and Generic BSM Models in PYTHIA 8,''
  arXiv:1109.5852 [hep-ph].
  %%CITATION = ARXIV:1109.5852;%%

\bibitem{pythia6RPV} 
  T.~Sjostrand and P.~Z.~Skands,
  %``Baryon number violation and string topologies,''
  Nucl.\ Phys.\ B {\bf 659}, 243 (2003)
  [hep-ph/0212264].
  %%CITATION = HEP-PH/0212264;%%
  
  
  
  \bibitem{pythia8}
  T.~Sjostrand, S.~Mrenna and P.~Z.~Skands,
  %``A Brief Introduction to PYTHIA 8.1,''
  Comput.\ Phys.\ Commun.\  {\bf 178}, 852 (2008)
  [arXiv:0710.3820 [hep-ph]].
  %%CITATION = ARXIV:0710.3820;%%
  
  
  \bibitem{prospino}
    W.~Beenakker, R.~Hopker and M.~Spira,
  %``PROSPINO: A Program for the production of supersymmetric particles in next-to-leading order QCD,''
  hep-ph/9611232;
  %%CITATION = HEP-PH/9611232;%%
    W.~Beenakker, M.~Klasen, M.~Kramer, T.~Plehn, M.~Spira and P.~M.~Zerwas,
  %``The Production of charginos / neutralinos and sleptons at hadron colliders,''
  Phys.\ Rev.\ Lett.\  {\bf 83}, 3780 (1999)
  [Erratum-ibid.\  {\bf 100}, 029901 (2008)]
  [hep-ph/9906298].
  %%CITATION = HEP-PH/9906298;%%
    
\bibitem{sherpa} 
  T.~Gleisberg, S.~.Hoeche, F.~Krauss, M.~Schonherr, S.~Schumann, F.~Siegert and J.~Winter,
  %``Event generation with SHERPA 1.1,''
  JHEP {\bf 0902}, 007 (2009)
  [arXiv:0811.4622 [hep-ph]].
  %%CITATION = ARXIV:0811.4622;%%

\bibitem{sherpaother}
  %AMEGIC
  F.~Krauss, R.~Kuhn and G.~Soff,
  %``AMEGIC++ 1.0: A Matrix element generator in C++,''
  JHEP {\bf 0202}, 044 (2002)
  [hep-ph/0109036];
  %%CITATION = HEP-PH/0109036;%%
%The Catani-Seymour subtraction based shower
  S.~Schumann and F.~Krauss,
  %``A Parton shower algorithm based on Catani-Seymour dipole factorisation,''
  JHEP {\bf 0803}, 038 (2008)
  [arXiv:0709.1027 [hep-ph]];
  %%CITATION = ARXIV:0709.1027;%%
  %COMIX
  T.~Gleisberg and S.~Hoeche,
  %``Comix, a new matrix element generator,''
  JHEP {\bf 0812}, 039 (2008)
  [arXiv:0808.3674 [hep-ph]];
  %%CITATION = ARXIV:0808.3674;%%
  %Matrix element merging with truncated showers
  S.~Hoeche, F.~Krauss, S.~Schumann and F.~Siegert,
  %``QCD matrix elements and truncated showers,''
  JHEP {\bf 0905}, 053 (2009)
  [arXiv:0903.1219 [hep-ph]].
  %%CITATION = ARXIV:0903.1219;%%
  

\bibitem{private.communication.1}
M.~Begel and D.~Tsybychev, private communication.

\bibitem{fastjet}
    M.~Cacciari and G.~P.~Salam,
  %``Dispelling the $N^{3}$ myth for the $k_t$ jet-finder,''
  Phys.\ Lett.\ B {\bf 641}, 57 (2006)
  [hep-ph/0512210];
  %%CITATION = HEP-PH/0512210;%%
    M.~Cacciari, G.~P.~Salam and G.~Soyez,
  %``FastJet user manual,''
  Eur.\ Phys.\ J.\ C {\bf 72}, 1896 (2012)
  [arXiv:1111.6097 [hep-ph]].
  %%CITATION = ARXIV:1111.6097;%%

\bibitem{Cacciari:2008gp} 
  M.~Cacciari, G.~P.~Salam and G.~Soyez,
  %``The Anti-k(t) jet clustering algorithm,''
  JHEP {\bf 0804}, 063 (2008)
  [arXiv:0802.1189 [hep-ph]].
  %%CITATION = ARXIV:0802.1189;%%

\bibitem{Falkowski:2010hi} 
  A.~Falkowski, D.~Krohn, L.~-T.~Wang, J.~Shelton and A.~Thalapillil,
  %``Unburied Higgs boson: Jet substructure techniques for searching for Higgs' decay into gluons,''
  Phys.\ Rev.\ D {\bf 84}, 074022 (2011)
  [arXiv:1006.1650 [hep-ph]].
  %%CITATION = ARXIV:1006.1650;%%
  
  \bibitem{planarflow} 
  L.~G.~Almeida, S.~J.~Lee, G.~Perez, G.~F.~Sterman, I.~Sung and J.~Virzi,
  %``Substructure of high-p_T Jets at the LHC,''
  Phys.\ Rev.\ D {\bf 79}, 074017 (2009)
  [arXiv:0807.0234 [hep-ph]].
  %%CITATION = ARXIV:0807.0234;%%



  
  
  
 
	  
\bibitem{Ellis:2009me} 
  S.~D.~Ellis, C.~K.~Vermilion and J.~R.~Walsh,
  %``Recombination Algorithms and Jet Substructure: Pruning as a Tool for Heavy Particle Searches,''
  Phys.\ Rev.\ D {\bf 81}, 094023 (2010)
  [arXiv:0912.0033 [hep-ph]].
  %%CITATION = ARXIV:0912.0033;%%



 \bibitem{pileupboost}
 G.~Soyez, ``Pile-up subtraction for jet $\pt$, masses, and shapes'', BOOST 2012 talk.
  
\bibitem{CLsmethod}
A.~L.~Read, 
%``Presentation of search results: the CLS technique'', 
J. Phys. G {\bf 28} (2002) 2693, \texttt{doi:10.1088/0954-3899/28/10/313}.
  
\bibitem{Cowan:2010js} 
  G.~Cowan, K.~Cranmer, E.~Gross and O.~Vitells,
  %``Asymptotic formulae for likelihood-based tests of new physics,''
  Eur.\ Phys.\ J.\ C {\bf 71}, 1554 (2011)
  [arXiv:1007.1727 [physics.data-an]].
  %%CITATION = ARXIV:1007.1727;%%
  
\bibitem{ATLASbtagger} 
The ATLAS Collaboration, 
%``Commissioning of the ATLAS high-performance $b$-tagging	 algorithms in the 7 TeV collision data'', 
ATLAS-CONF-2011-102, July 2011.
  
  
  
\bibitem{Ruderman:2012jd} 
  J.~T.~Ruderman, T.~R.~Slatyer and N.~Weiner,
  %``A Collective Breaking of R-Parity,''
  arXiv:1207.5787 [hep-ph].
  %%CITATION = ARXIV:1207.5787;%%
  
  \bibitem{FileviezPerez:2009gr} 
  P.~Fileviez Perez and S.~Spinner,
  %``Spontaneous R-Parity Breaking in SUSY Models,''
  Phys.\ Rev.\ D {\bf 80}, 015004 (2009)
  [arXiv:0904.2213 [hep-ph]].
  %%CITATION = ARXIV:0904.2213;%%
  
\bibitem{Goity:1994dq} 
  J.~L.~Goity and M.~Sher,
  %``Bounds on delta B = 1 couplings in the supersymmetric standard model,''
  Phys.\ Lett.\ B {\bf 346}, 69 (1995)
  [Erratum-ibid.\ B {\bf 385}, 500 (1996)]
  [hep-ph/9412208].
  %%CITATION = HEP-PH/9412208;%%

\bibitem{Chen:2010ss} 
  S.~-L.~Chen, D.~K.~Ghosh, R.~N.~Mohapatra and Y.~Zhang,
  %``Dynamical R-parity Breaking at the LHC,''
  JHEP {\bf 1102}, 036 (2011)
  [arXiv:1011.2214 [hep-ph]].
  %%CITATION = ARXIV:1011.2214;%%
  
\bibitem{MFVsusy} 
  C.~Csaki, Y.~Grossman and B.~Heidenreich,
  %``MFV SUSY: A Natural Theory for R-Parity Violation,''
  Phys.\ Rev.\ D {\bf 85}, 095009 (2012)
  [arXiv:1111.1239 [hep-ph]].
  %%CITATION = ARXIV:1111.1239;%%

\bibitem{ArkaniHamed:2004fb} 
  N.~Arkani-Hamed and S.~Dimopoulos,
  %``Supersymmetric unification without low energy supersymmetry and signatures for fine-tuning at the LHC,''
  JHEP {\bf 0506}, 073 (2005)
  [hep-th/0405159].
  %%CITATION = HEP-TH/0405159;%%
  
\bibitem{Giudice:2004tc} 
  G.~F.~Giudice and A.~Romanino,
  %``Split supersymmetry,''
  Nucl.\ Phys.\ B {\bf 699}, 65 (2004)
  [Erratum-ibid.\ B {\bf 706}, 65 (2005)]
  [hep-ph/0406088].
  %%CITATION = HEP-PH/0406088;%%

\bibitem{Khachatryan:2010uf} 
  V.~Khachatryan {\it et al.}  [CMS Collaboration],
  %``Search for Stopped Gluinos in $pp$ collisions at $\sqrt{s}=7$ TeV,''
  Phys.\ Rev.\ Lett.\  {\bf 106}, 011801 (2011)
  [arXiv:1011.5861 [hep-ex]].
  %%CITATION = ARXIV:1011.5861;%%

\bibitem{Graham:2012th} 
  P.~W.~Graham, D.~E.~Kaplan, S.~Rajendran and P.~Saraswat,
  %``Displaced Supersymmetry,''
  JHEP {\bf 1207}, 149 (2012)
  [arXiv:1204.6038 [hep-ph]].
  %%CITATION = ARXIV:1204.6038;%%
  
  \bibitem{Nason:2004rx} 
  P.~Nason,
  %``A New method for combining NLO QCD with shower Monte Carlo algorithms,''
  JHEP {\bf 0411}, 040 (2004)
  [hep-ph/0409146].
  %%CITATION = HEP-PH/0409146;%%
  %440 citations counted in INSPIRE as of 17 May 2013
  
\bibitem{Frixione:2007vw} 
  S.~Frixione, P.~Nason and C.~Oleari,
  %``Matching NLO QCD computations with Parton Shower simulations: the POWHEG method,''
  JHEP {\bf 0711}, 070 (2007)
  [arXiv:0709.2092 [hep-ph]].
  %%CITATION = ARXIV:0709.2092;%%
  %590 citations counted in INSPIRE as of 17 May 2013
  
\bibitem{Alioli:2010xa} 
  S.~Alioli, K.~Hamilton, P.~Nason, C.~Oleari and E.~Re,
  %``Jet pair production in POWHEG,''
  JHEP {\bf 1104}, 081 (2011)
  [arXiv:1012.3380 [hep-ph]].
  %%CITATION = ARXIV:1012.3380;%%
  %83 citations counted in INSPIRE as of 17 May 2013
  
\bibitem{ATLAS:2011gmi} 
  [ATLAS Collaboration],
  %``New ATLAS event generator tunes to 2010 data,''
  ATL-PHYS-PUB-2011-008.
  %%CITATION = ATL-PHYS-PUB-2011-008;%%
  %46 citations counted in INSPIRE as of 21 May 2013
  
\bibitem{ATLAS:2011zja} 
  [ATLAS Collaboration],
  %``ATLAS tunes of PYTHIA 6 and Pythia 8 for MC11,''
  ATL-PHYS-PUB-2011-009.
  %%CITATION = ATL-PHYS-PUB-2011-009;%%
  %72 citations counted in INSPIRE as of 21 May 2013
  
\bibitem{Alwall:2011uj} 
  J.~Alwall, M.~Herquet, F.~Maltoni, O.~Mattelaer and T.~Stelzer,
  %``MadGraph 5 : Going Beyond,''
  JHEP {\bf 1106}, 128 (2011)
  [arXiv:1106.0522 [hep-ph]].
  %%CITATION = ARXIV:1106.0522;%%
  %602 citations counted in INSPIRE as of 23 May 2013
  
  
  
\end{thebibliography}
\end{document}